\documentclass[ALICE,manyauthors]{cernphprep}
\usepackage[comma,square,numbers,sort&compress]{natbib}
\usepackage{hyperref}
\usepackage{lineno}
\usepackage{xspace}
\usepackage[dvipsnames]{xcolor}
\usepackage{afterpage}

\usepackage[T1]{fontenc} 
\usepackage{orcidlink}

\begin{document}
%

\newcommand{\todo}[1]{\textit{\color{red}[TODO: #1]}\xspace}
\newcommand{\alice}[1]{\textit{\color{Purple}[Alice's comment: #1]}\xspace}

\newcommand{\pp}           {pp\xspace}
\newcommand{\ppbar}        {\mbox{$\mathrm {p\overline{p}}$}\xspace}
\newcommand{\XeXe}         {\mbox{Xe--Xe}\xspace}
\newcommand{\PbPb}         {\mbox{Pb--Pb}\xspace}
\newcommand{\pA}           {\mbox{pA}\xspace}
\newcommand{\pPb}          {\mbox{p--Pb}\xspace}
\newcommand{\AuAu}         {\mbox{Au--Au}\xspace}
\newcommand{\dAu}          {\mbox{d--Au}\xspace}
\newcommand{\eee}          {\mbox{e$^{+}$e$^{-}$}\xspace}

\newcommand{\s}            {\ensuremath{\sqrt{s}}\xspace}
\newcommand{\snn}          {\ensuremath{\sqrt{s_{\mathrm{NN}}}}\xspace}
\newcommand{\pt}           {\ensuremath{p_{\rm T}}\xspace}
\newcommand{\meanpt}       {$\langle p_{\mathrm{T}}\rangle$\xspace}
\newcommand{\ycms}         {\ensuremath{y_{\rm CMS}}\xspace}
\newcommand{\ylab}         {\ensuremath{y_{\rm lab}}\xspace}
\newcommand{\etarange}[1]  {\mbox{$\left | \eta \right |~<~#1$}}
\newcommand{\yrange}[1]    {\mbox{$\left | y \right |~<~#1$}}
\newcommand{\ptrange}[2] {\mbox{$#1 < \pt < #2$~\GeVc}}
\newcommand{\dndy}         {\ensuremath{\mathrm{d}N_\mathrm{ch}/\mathrm{d}y}\xspace}
\newcommand{\dndeta}       {\ensuremath{\mathrm{d}N_\mathrm{ch}/\mathrm{d}\eta}\xspace}
\newcommand{\avdndeta}     {\ensuremath{\langle\dndeta\rangle}\xspace}
\newcommand{\dNdy}         {\ensuremath{\mathrm{d}N_\mathrm{ch}/\mathrm{d}y}\xspace}
\newcommand{\Npart}        {\ensuremath{N_\mathrm{part}}\xspace}
\newcommand{\Ncoll}        {\ensuremath{N_\mathrm{coll}}\xspace}
\newcommand{\dEdx}         {\ensuremath{\textrm{d}E/\textrm{d}x}\xspace}
\newcommand{\RpPb}         {\ensuremath{R_{\rm pPb}}\xspace}

\newcommand{\nineH}        {$\sqrt{s}~=~0.9$~Te\kern-.1emV\xspace}
\newcommand{\seven}        {$\sqrt{s}~=~7$~Te\kern-.1emV\xspace}
\newcommand{\twoH}         {$\sqrt{s}~=~0.2$~Te\kern-.1emV\xspace}
\newcommand{\twosevensix}  {$\sqrt{s}~=~2.76$~Te\kern-.1emV\xspace}
\newcommand{\five}         {$\sqrt{s}~=~5.02$~Te\kern-.1emV\xspace}
\newcommand{\twosevensixnn}{$\sqrt{s_{\mathrm{NN}}}~=~2.76$~Te\kern-.1emV\xspace}
\newcommand{\fivenn}       {$\sqrt{s_{\mathrm{NN}}}~=~5.02$~Te\kern-.1emV\xspace}
\newcommand{\LT}           {L{\'e}vy-Tsallis\xspace}
\newcommand{\GeVc}         {Ge\kern-.1emV/$c$\xspace}
\newcommand{\MeVc}         {Me\kern-.1emV/$c$\xspace}
\newcommand{\TeV}          {Te\kern-.1emV\xspace}
\newcommand{\GeV}          {Ge\kern-.1emV\xspace}
\newcommand{\MeV}          {Me\kern-.1emV\xspace}
\newcommand{\GeVmass}      {Ge\kern-.1emV/$c^2$\xspace}
\newcommand{\MeVmass}      {Me\kern-.1emV/$c^2$\xspace}
\newcommand{\lumi}         {\ensuremath{\mathcal{L}}\xspace}

\newcommand{\ITS}          {\rm{ITS}\xspace}
\newcommand{\TOF}          {\rm{TOF}\xspace}
\newcommand{\ZDC}          {\rm{ZDC}\xspace}
\newcommand{\ZDCs}         {\rm{ZDCs}\xspace}
\newcommand{\ZNA}          {\rm{ZNA}\xspace}
\newcommand{\ZNC}          {\rm{ZNC}\xspace}
\newcommand{\SPD}          {\rm{SPD}\xspace}
\newcommand{\SDD}          {\rm{SDD}\xspace}
\newcommand{\SSD}          {\rm{SSD}\xspace}
\newcommand{\TPC}          {\rm{TPC}\xspace}
\newcommand{\TRD}          {\rm{TRD}\xspace}
\newcommand{\VZERO}        {\rm{V0}\xspace}
\newcommand{\VZEROA}       {\rm{V0A}\xspace}
\newcommand{\VZEROC}       {\rm{V0C}\xspace}
\newcommand{\Vdecay} 	   {\ensuremath{V^{0}}\xspace}

\newcommand{\ee}           {\ensuremath{e^{+}e^{-}}} 
\newcommand{\pip}          {\ensuremath{\pi^{+}}\xspace}
\newcommand{\pim}          {\ensuremath{\pi^{-}}\xspace}
\newcommand{\kap}          {\ensuremath{\mathrm{K}^{+}}\xspace}
\newcommand{\kam}          {\ensuremath{\mathrm{K}^{-}}\xspace}
\newcommand{\pbar}         {\ensuremath{\rm\overline{p}}\xspace}
\newcommand{\kzero}        {\ensuremath{{\rm K}^{0}_{\rm{S}}}\xspace}
\newcommand{\lmb}          {\ensuremath{\Lambda}\xspace}
\newcommand{\almb}         {\ensuremath{\overline{\Lambda}}\xspace}
\newcommand{\Om}           {\ensuremath{\Omega^-}\xspace}
\newcommand{\Mo}           {\ensuremath{\overline{\Omega}^+}\xspace}
\newcommand{\X}            {\ensuremath{\Xi^-}\xspace}
\newcommand{\Ix}           {\ensuremath{\overline{\Xi}^+}\xspace}
\newcommand{\Xis}          {\ensuremath{\Xi^{\pm}}\xspace}
\newcommand{\Oms}          {\ensuremath{\Omega^{\pm}}\xspace}
\newcommand{\degree}       {\ensuremath{^{\rm o}}\xspace}

\newcommand{\pythia}          {\textsc{Pythia~8}}
\newcommand{\epos}        {\textsc{Epos}}
\newcommand{\eposlhc}        {\textsc{Epos-lhc}}
\newcommand{\herwig}        {\textsc{Herwig~7}}
\newcommand{\geant}        {\textsc{Geant}}

\begin{titlepage}
\PHyear{2023}       
\PHnumber{198}      
\PHdate{31 August}  

\title{Studying strangeness and baryon production mechanisms through angular correlations between charged $\Xi$ baryons and identified hadrons in pp collisions at $\mathbf{\sqrt{\textit s}}$ = 13 TeV}
\ShortTitle{$\Xi$--hadron correlations in pp collisions}   

\Collaboration{ALICE Collaboration\thanks{See Appendix~\ref{app:collab} for the list of collaboration members}}
\ShortAuthor{ALICE Collaboration} 

\begin{abstract}
The angular correlations between charged $\Xi$ baryons and associated identified hadrons (pions, kaons, protons, $\Lambda$ baryons, and $\Xi$ baryons) are measured in \pp{} collisions at $\s = 13$ TeV with the ALICE detector to give insight into the particle production mechanisms and balancing of quantum numbers on the microscopic level.  In particular, the distribution of strangeness is investigated in the correlations between the doubly-strange $\Xi$ baryon and mesons and baryons that contain a single strange quark, K and $\Lambda$.  As a reference, the results are compared to $\Xi\pi$ and $\Xi\mathrm{p}$ correlations, where the associated mesons and baryons do not contain a strange valence quark.  These measurements are expected to be sensitive to whether strangeness is produced through string breaking or in a thermal production scenario.  
Furthermore, the multiplicity dependence of the correlation functions is measured to look for the turn-on of additional particle production mechanisms with event activity.  The results are compared to predictions from the string-breaking model \pythia{}, including tunes with baryon junctions and rope hadronisation enabled, the cluster hadronisation model \herwig{}, and the core--corona model \eposlhc{}.  While some aspects of the experimental data are described quantitatively or qualitatively by the Monte Carlo models, no model can match all features of the data.  These results provide stringent constraints on the strangeness and baryon number production mechanisms in \pp{} collisions.  

\end{abstract}
\end{titlepage}

\setcounter{page}{2} 


\section{Introduction} 

Studies of hadrons containing strange quarks in high energy elementary and nuclear collisions give insight into the production processes of both partons and hadrons. The production of strangeness is of particular interest in proton--proton and nucleus--nucleus collisions because there are no strange valence quarks in the incoming colliding particles, and thus all strangeness observed in the final state must have been produced in the collision (or later, through decays of heavier quarks). Furthermore, the enhancement of strangeness production was one of the first proposed signatures for the creation of a deconfined quark--gluon plasma in heavy nucleus–nucleus collisions~\cite{Rafelski:1982pu,Koch:1986ud}, based on the hypothesis that strange quarks are light enough that they can be produced in thermal processes if a QGP phase is present.  For a review of experimental results on QCD physics in \pp{} and heavy-ion collisions by the ALICE Collaboration, see Ref.~\cite{ALICE:2022wpn}.  

However, recent experimental measurements demonstrate a smooth turn-on of strangeness enhancement as a function of final-state-particle multiplicity from \pp collisions, to \pPb collisions, to \PbPb collisions~\cite{ALICE:2016fzo}.  This observation calls into question the uniqueness of strangeness enhancement as a signature of the creation of a thermalised QGP.  Furthermore, the strangeness enhancement is observed to vary with multiplicity even within the smallest hadronic systems, namely \pp collisions.  This challenges the long-standing concept of ``jet universality,'' the idea that while the underlying parton interactions depend on the colliding system and energy, the fragmentation and hadronisation of the resulting colour fields are independent of the system and should be universal from \eee to hadronic collisions~\cite{Field:1976ve}.  The multiplicity dependence of strange hadron yields has gathered tremendous attention in the \pp phenomenology community as it demonstrates that a \pp collision cannot be modelled as a sum of semi-independent parton--parton collisions, but that significant final-state interactions must be included.

Currently, there are several different explanations for the strangeness enhancement in small and large collision systems. 
In models based on Lund strings~\cite{lundstring,lundstring2}, such as \pythia{}~\cite{Bierlich:2022pfr}, particle production occurs through string breaking mainly in multiparton interactions (MPI) which contribute to the bulk of soft (low-momentum) particle production, and in hard scatterings and their subsequent fragmentation.  In recent versions of \pythia{}, additional string interactions can change the string topology via colour reconnection~\cite{Christiansen:2015yqa} in dense regions prior to hadronisation. The main new topologies are ropes~\cite{Bierlich:2014xba}, in which overlapping strings can interact coherently leading to higher string tensions and enhanced strangeness production~\cite{ALICE:2020nkc}, and junction formation~\cite{Bierlich:2015rha} that enhance the production of baryons.
\herwig{}~\cite{Bahr:2008pv,Bellm:2015jjp}, which also generates parton showers via string breaking but employs a cluster-based hadronisation mechanism, recently introduced a reconnection scheme for baryon ropes which leads to the enhancement of strange and multi-strange baryons with increasing multiplicity~\cite{ALICE:2020nkc}.  In core--corona~\cite{Werner:2007bf} models such as \eposlhc~\cite{Pierog:2013ria}, the system is subdivided into dilute corona regions, which are dominated by string-breaking processes, and dense QGP regions, which evolve hydrodynamically and then undergo cluster hadronisation. As the rates of strangeness production in these two regions are different, strangeness enhancement arises from the increase of the core region relative to the corona size with multiplicity~\cite{Aichelin:2008mi}.
In models based purely on a statistical thermal model, strangeness enhancement is alternatively viewed as the canonical suppression of strangeness in small collision systems, which is then lifted as the system size increases~\cite{Redlich:2001kb}.  

In low-multiplicity collisions, these three classes of models predict similar behaviour for the correlation between $\Xi^{-}(\overline{\Xi}^{+})$ baryons and strange hadrons, since strange and anti-strange quarks are produced in pairs and remain correlated through the evolution of the collision.  In the final state, the resultant strange hadrons are expected to be strongly correlated in angular space, since the system is too small for high-density effects to significantly alter the parton or hadron momentum distributions.  However, in high-multiplicity collisions, qualitatively different behaviour is expected from each model.  In the string-breaking picture, strange quarks and the resulting strange hadrons remain mostly correlated, and only minor modifications relative to the low-multiplicity case are expected due to the effects of colour reconnection, junctions, and ropes.  In \herwig{}, the newly-implemented baryonic rope mechanism may lead to a multiplicity-dependence of the correlations, particularly for strange baryons like the $\Lambda(\overline{\Lambda})$ and $\Xi^-(\overline{\Xi}^+)$.  On the contrary, in the core--corona and thermal models, a significant decorrelation of the strange and anti-strange quarks is expected, since a high-density region builds up in which partons may become deconfined or thermalised.  In these two pictures, strangeness is only conserved globally instead of locally.  

The goal of this work is to directly test the underlying assumptions of these qualitatively very different scenarios by investigating the angular correlations between the multi-strange $\Xi^{-}(\overline{\Xi}^{+})$ baryon and other hadron species ($\pi$, K, p, $\Lambda$, $\Xi$).  By measuring the relative distributions of ``associated'' hadrons of different electric charge ($Q$), strangeness ($S$), and baryon number ($B$) with respect to a ``trigger'' $\Xi^{-}$ baryon (quark content dss, with $S=-2$, $B=1$, $Q=-1$) or $\overline{\Xi}^{+}$ baryon ($\overline{\rm dss}$, with $S=2$, $B=-1$, $Q=1$), the distribution of the balancing quantum numbers in momentum space can be determined.  
Furthermore, by studying the multiplicity dependence of these correlations, the goal is to be able to determine experimentally if multiple production mechanisms are present and if there is an evolution from local to global strangeness conservation with system size.

In the following, measurements of the per-trigger yield of associated identified hadrons with respect to trigger $\Xi^{-}$ baryons will be shown.  The associated yield per trigger particle is defined as
\begin{equation}\label{eq:yield}
Y(\Delta y, \Delta\varphi)=\dfrac{1}{N_\mathrm{trig}}\dfrac{\mathrm{d^2}N_\mathrm{pairs}}{\mathrm{d}\Delta y\, \mathrm{d}\Delta\varphi},
\end{equation}
where $\Delta y = y_{\rm assoc} - y_{\rm trig}$ is the difference in rapidity and $\Delta\varphi = \varphi_{\rm assoc} - \varphi_{\rm trig}$ is the relative azimuthal angle between the trigger and associated particles, $N_\mathrm{pairs}$ is the number of trigger--associated particle pairs, and $N_\mathrm{trig}$ is the number of trigger particles.  In all cases, the charge conjugate pairs, i.e.\ the corresponding correlations with $\overline{\Xi}^{+}$ baryons, are also included in the correlation functions.  For the remainder of the paper, the notation ``$\Xi$'' will represent both the negatively-charged $\Xi^-$ and the positively-charged $\overline{\Xi}^+$, unless otherwise specified.  

Strangeness production is investigated through the correlations between the multi-strange $\Xi$ baryons and hadrons which contain only a single strange quark, kaons (\kap = u$\overline{\rm s}$, \kam = s$\overline{\rm u}$) and $\Lambda$ baryons ($\Lambda$ = uds, $\overline{\Lambda} = \overline{\rm uds}$).  These are compared with the correlations between $\Xi$ baryons and hadrons which do not carry any strangeness, pions (\pip = u$\overline{\rm d}$, \pim = d$\overline{\rm u}$) and protons (p = uud, $\overline{\rm p} = \overline{\rm uud}$). Furthermore, baryon production will be probed by comparing baryon--baryon correlations (such as $\Xi\mathrm{p}$ and $\Xi\Lambda$) with baryon--meson correlations ($\Xi\pi$ and $\Xi\mathrm{K}$).  Finally, the first measurement of $\Xi\Xi$ correlations will be shown.  

In order to isolate the quantum-number-dependent part of the correlation function and remove correlations due to (mini)jet fragmentation, flow, or the underlying event, the difference between the opposite-quantum-number (``opposite-sign'') and same-quantum-number (``same-sign'') correlations is also calculated (hereafter denoted as ``OS--SS;'' for the $\Xi$-baryon correlations ``OB--SB'' represents the opposite-baryon-number minus same-baryon-number differences).  Specifically, the difference between $\Xi^{-}\pip$ and $\Xi^{-}\pim$ correlations provides information on the distribution of electric charge, while strangeness correlations are measured by subtracting the distribution of pairs with the same strangeness (e.g.\ $\Xi^{-}\kam$) from those with opposite strangeness ($\Xi^{-}\kap$).  The balancing of baryon number is explored with the difference between $\Xi^{-}\mathrm{p}$ and $\Xi^{-}\overline{\rm p}$ correlations, and similarly in the strangeness sector with $\Xi^{-}\Lambda$ and $\Xi^{-}\overline{\Lambda}$.  Note that this difference between opposite-sign and same-sign per-trigger yields is closely related to the balance function~\cite{Bass:2000az,Pratt:2011bc}, differing only by a normalisation factor.  

The paper is organised as follows: Section~\ref{sec:exp} describes the relevant subsystems of the ALICE detector (\ref{sec:ALICE}), the data collection and event selection (\ref{sec:eventsel}), and the track reconstruction and particle identification (\ref{sec:piKp}--\ref{sec:lambdaxi}).  Section~\ref{sec:analysis} details the analysis method, experimental corrections (\ref{sec:corr}--\ref{sec:feeddown}), and systematic uncertainties (\ref{sec:systuncert}).  The experimental measurements and Monte Carlo model comparisons are presented and discussed in Section~\ref{sec:results}, and final conclusions are drawn in Section~\ref{sec:conclusions}.  

\section{Experimental setup, data selection}
\label{sec:exp}

\subsection{The ALICE detector}
\label{sec:ALICE}

ALICE is the detector designed to study heavy-ion collisions at the Large Hadron Collider (LHC), and is optimised for precise tracking and identification of particles over a wide momentum range, particularly in the high-multiplicity environment of heavy-ion collisions.  A full description of the ALICE detector and its performance can be found in Refs.~\cite{ALICE:2008ngc,ALICE:2014sbx}.  The detector subsystems utilised in this analysis were the Inner Tracking System (ITS), the Time Projection Chamber (TPC), the Time-Of-Flight detector (TOF), and the V0, which all provide full azimuthal coverage ($0 < \varphi < 2\pi$) around the beam line.  

The ITS, located closest to the beam pipe, consists of six cylindrical layers of silicon detectors, which contribute to the high resolution tracking of charged particles and vertex reconstruction. The two innermost Silicon Pixel Detector (SPD) layers are at radial distances of \mbox{3.9 cm} and \mbox{7.6 cm}, the two Silicon Drift Detector (SDD) layers are at \mbox{15.0 cm} and \mbox{23.9 cm}, and the outermost two Silicon Strip Detector (SSD) layers are at \mbox{38.0 cm} and \mbox{43.0 cm} from the beam line.  By providing track points close to the beam axis, the ITS makes it possible to reconstruct primary and displaced secondary decay vertices with high precision.  
The TPC is the principal tracking detector of ALICE. It is a \mbox{5 m}-long gaseous cylindrical detector with an active volume of \mbox{90 m$^3$}, and its inner and outer radii are \mbox{85 cm} and \mbox{250 cm}, respectively.
The TPC provides high-precision tracking and momentum determination for charged particles that traverse the active detector volume with transverse momentum $0.15 < \pt < 100$~\GeVc.  Through measurements of the specific energy loss, \dEdx{}, the TPC is also used for particle species identification.  In kinematic regions where \dEdx{} information does not give good species separation, complementary time-of-flight measurements with better than 90~ps timing resolution from the TOF detector are used for particle identification (PID).   The ITS, TPC, and TOF are used for tracking and PID in the midrapidity region, while the V0 comprises two scintillator arrays on either side of the interaction point which cover the forward pseudorapidity ranges $2.8 < \eta < 5.1$ (V0A) and $-3.7 < \eta < -1.7$ (V0C).  In this analysis, the V0 contributed to the event trigger and was used to classify events based on their activity, estimated through the sum of the energy deposited in the V0A and V0C (denoted as V0M)~\cite{ALICE:2018pal}. While the correlation between the final-state-particle multiplicities measured at forward rapidity and midrapidity is relatively broad in \pp{} collisions, a high event activity measured in the forward V0M is correlated with a high multiplicity at midrapidity, and vice versa.  The V0M amplitude distribution was divided into percentiles, with 0\% denoting the highest multiplicity events and 100\% the lowest.  

\subsection{Event selection}
\label{sec:eventsel}

The \pp collision data analysed here were collected with the ALICE detector from 2016 to 2018 during \mbox{Run 2} at the LHC.  Minimum bias \pp collision events were selected using a hardware trigger that required energy deposition in both the V0A and V0C. Events flagged as originating from pileup or beam--gas interactions based on the timing information from the V0 were rejected.  The primary collision vertex was reconstructed using full ITS+TPC tracks (see Section~\ref{sec:piKp}) which provide high vertex position resolution.  In each event, a second estimate of the primary vertex position was obtained from short track segments (tracklets) measured in the SPD, which provide higher efficiency but lower position resolution.  To ensure a high quality primary vertex, the estimates from the TPC+ITS vertex and the SPD vertex were compared and required to be within \mbox{0.5 cm} of each other along the direction of the beam line (the $z$-axis).  Furthermore, the primary collision vertices were required to lie within \mbox{10 cm} in the $z$-direction of the nominal interaction point at the centre of the ALICE detector so that the track acceptance was uniform for all events. In total, approximately 1.36 billion events passed the event selection. 

\subsection{$\pi$, K, and p selection and identification}
\label{sec:piKp}

Charged-particle tracks were reconstructed from clusters of energy depositions in the ITS and TPC using a Kalman filter algorithm.  Particles at midrapidity, with a pseudorapidity within \etarange{0.8}, were considered in this analysis.  Pions and kaons were selected in the transverse momentum range \ptrange{0.2}{3} and protons within \ptrange{0.4}{3}.  In order to ensure high-quality tracking, the tracks were required to be reconstructed from at least 70 clusters in the TPC (from a possible maximum of 159 for the longest tracks), with at least 80\% of the possible findable clusters, and the $\chi^2$ per cluster had to be less than 4.  The tracks were required to include at least one cluster from the SPD.  In the tracking procedure, refitting to both the ITS and TPC space points was required (see Ref.~\cite{ALICE:2014sbx} for more details on the tracking algorithm).  The analysis focuses on primary particles, defined in ALICE as those with mean proper lifetimes $\tau$ larger than 1 cm/$c$ which are either produced directly in the collision or from decays of particles with $\tau < 1$ cm/$c$~\cite{ALICE:2017hcy}.  To reduce contamination from secondary particles (those not produced in the original \pp{} collision), the distance of closest approach (DCA) to the primary vertex was required to be within \mbox{2 cm} in the longitudinal direction and $7\sigma$ in the transverse plane, corresponding to \mbox{DCA$_{xy}<0.0105+0.035p_\mathrm{T}^{-1.1}$ cm} (for \pt in \GeVc).  

Particles were identified from their specific energy loss in the TPC and flight time to the TOF detector.  For low-momentum tracks ($p^{\pi,\mathrm{e}} < 0.4$~\GeVc, $p^{\mathrm{K}} < 0.6$~\GeVc, $p^{\mathrm{p}} < 0.9$~\GeVc) only the TPC information was used for PID since the discriminating power of the TPC is high and many of the tracks do not reach the TOF detector.  For each track in the TPC, the measured \dEdx{} was compared to the expected signal for each particle type ($i = \pi$, K, p, e) and quantified in terms of the number of standard deviations ($n\sigma^i_{\mathrm{TPC}}$) away from a Bethe--Bloch parametrisation of the detector response.  At higher momentum, if a TPC track was matched to a signal in the TOF, then the velocity measured from the TOF detector was combined with the particle's momentum to obtain its mass, $m$.  The measured $m^2$ was compared with the expectations for each particle species and quantified by $n\sigma^i_{\mathrm{TOF}}$.  For these tracks with a TOF signal, the PID information was determined by using the combined TPC and TOF information to improve the purity, $n\sigma^i=\sqrt{(n\sigma^i_\mathrm{TPC})^2+(n\sigma^i_\mathrm{TOF})^2}$.  For tracks without a matching TOF signal, only the TPC information was used, $n\sigma^i= n\sigma^i_\mathrm{TPC}$.  Each track was then associated with the particle species with the smallest $|n\sigma^i|$.  Residual contamination from wrongly-identified tracks was corrected statistically using a procedure described in Section~\ref{sec:misid}.  Selected tracks were required to be within $|n\sigma^i|<4$ for at least one of the particle species hypotheses, and not used in the reconstruction of any $\Xi$ baryon candidates.  Note that, due to the abundance of pions compared to electrons, a pion veto was applied such that a track satisfying the $n\sigma^i < 4$ criterion for both electrons and pions was always classified as a pion, even if $n\sigma^{\mathrm{e}} < n\sigma^{\pi}$.  The pion sample also included a negligibly small ($<1\%$) fraction of misidentified muons, which were subtracted off the yields at a later step according to the $\mu/\pi$ ratio obtained in \pythia{} Monte Carlo simulations.  

With these selection criteria applied, the reconstruction efficiency of pions was around $\varepsilon_{\pi}\sim 71\%$ for the lowest momentum interval ($0.2 < p < 0.3$~\GeVc), rising to a maximum of $\varepsilon_{\pi}\sim 86\%$ at intermediate momentum ($p\sim 2$~\GeVc) and declining to $\varepsilon_{\pi}\sim 81\%$ above $p = 5$~\GeVc.  The kaon efficiency was strongly momentum-dependent in the momentum range considered in this analysis, rising monotonically from a minimum of $\varepsilon_{\mathrm{K}} \sim 0.14$ in the lowest momentum interval ($0.2 < p < 0.3$~\GeVc) to a maximum of $\varepsilon_{\mathrm{K}}\sim 0.82$ above $p = 2.5$~\GeVc.  The tracking efficiency for protons was approximately $\varepsilon_{\mathrm{p}}\sim 80\%$ in the lowest momentum interval ($0.4 < p < 0.5$~\GeVc), rising to around $\varepsilon_{\mathrm{p}} \sim 86\%$ around $p\sim 1$~\GeVc, and was about $\varepsilon_{\mathrm{p}}\sim 82\%$ at high momentum ($p > 3$~\GeVc).  Due to annihilations with the detector material, the efficiency for antiprotons was slightly lower, with a minimum around $\varepsilon_{\overline{\mathrm{p}}}\sim 72\%$ at low momentum and a maximum of $\varepsilon_{\overline{\mathrm{p}}}\sim 82\%$.  All the efficiencies displayed small variations of the order of 2\% reflecting the changing detector performance in the different run periods over time.  

\subsection{$\Lambda$ and $\Xi$ reconstruction}
\label{sec:lambdaxi}

Primary $\Lambda(\overline{\Lambda})$ baryons were reconstructed in the kinematic range \ptrange{0.6}{12}, \etarange{0.72}, through their decays to (anti)protons and charged pions, $\Lambda \rightarrow \pim + \mathrm{p}$ and $\overline{\Lambda} \rightarrow \pip + \overline{\mathrm{p}}$ (branching ratio 63.9\%~\cite{Workman:2022ynf}), which leave a characteristic $V^0$ signature in the detector.  The $\Lambda$ signal peak in the proton-pion invariant mass ($m_{\mathrm{p}\pi}$) distribution was fitted with a double-Gaussian functional form, and the extracted mean and standard deviations of the two Gaussians were parametrised as a function of \pt{}.  The signal region was defined as $|m_{\mathrm{p}\pi}-\mu_{\Lambda}| < 3\sigma_{\Lambda,\mathrm{ wide}}$, where $\mu_{\Lambda}$ was the centre of the peak and $\sigma_{\Lambda,\mathrm{ wide}}$ was the width of the wider Gaussian (except in the lowest \pt{} bins where the wider Gaussian was insignificant and thus the width of the narrow Gaussian was used). Topological selection criteria on the $V^0$ decay, including the radial distance from the centre of the detector to the secondary vertex, DCA between the proton and pion tracks, and cosine of the pointing angle of the $\Lambda(\overline{\Lambda})$ momentum back to the primary vertex were applied.  Furthermore, misidentified \kzero were rejected by requiring that $m_{\pi\pi}$ lies more than 10~\MeVmass from the \kzero mass.  The individual decay products, the proton and pion daughters, were required to fall within the kinematic range \ptrange{0.15}{20} and \etarange{0.8}.  Additional selections were applied on the DCA of the daughters to the primary vertex, and on their specific energy loss in the TPC ($|n\sigma_{\mathrm{TPC}}| < 4$ for the relevant particle species).  Finally, since the TPC readout time is on the order of 100 $\mu{\rm s}$, particles from out-of-bunch pileup events may be combined to create fake $V^0$s.  Consequently, at least one daughter track was required to satisfy the ITS refit or have a matched cluster in the TOF detector, to reject $\Lambda$ candidates from out-of-bunch pileup.  

$\Xi^{-}(\overline{\Xi}^{+})$ baryons, which serve as the trigger particles in all the correlation functions described below, were reconstructed through their decays to $\Lambda(\overline{\Lambda})$ baryons and charged pions, $\Xi^{-} \rightarrow \pim + \Lambda$ and $\overline{\Xi}^{+} \rightarrow \pip + \Lambda$ (branching ratio 99.9\%~\cite{Workman:2022ynf}).  The $\Lambda(\overline{\Lambda})$ then decays to p$\pim(\overline{\mathrm{p}}\pip)$, giving this decay its characteristic ``cascade'' topology in the detector.  The $\Xi$ candidates were identified by their invariant mass ($m_{\Lambda\pi}$) in the kinematic range \ptrange{0.8}{12} and \etarange{0.72}.  The same double-Gaussian fitting procedure which was used for the $\Lambda$ baryons was applied here to define the $\Xi$ signal region, $|m_{\Lambda\pi}-\mu_{\Xi}| < 3\sigma_{\Xi,\mathrm{ wide}}$.  The combinatorial background from random $\Lambda\pi$ pairs was reduced by applying selections on the topological properties of the cascade decay.  In particular, criteria were imposed on the following properties of the reconstructed $\Xi$ candidate: the DCA to the primary vertex, the radial distance between the secondary decay vertex and the centre of the detector, and the cosine of the pointing angle of the $\Xi$ momentum to the primary vertex.  Looser selection criteria were applied to the properties of the reconstructed secondary $\Lambda$ daughter than the primary $\Lambda$ baryons above, with selections on the invariant mass, the DCA between the proton and pion daughters, and the radial distance from the tertiary vertex to the centre of the detector.  All the stable daughters of the cascade decay (one proton and two pions) were required to have a pseudorapidity within \etarange{0.8} and a transverse momentum within \ptrange{0.15}{20}.  Selection criteria were imposed on the specific energy loss of the daughters ($|n\sigma_{\mathrm{TPC}}| < 4$ for the relevant particle species) as well as their DCA to the primary vertex.  Finally, at least one daughter track was required to have a matched cluster in the TOF or to satisfy the ITS refit in order to reject $\Xi$ candidates which may have come from out-of-bunch pileup events.  

A detailed Monte-Carlo-based study was carried out to optimise the purity of the selected $\Lambda$ and $\Xi$ candidates without significantly reducing the detection efficiency.  For all of the topological and kinematic properties listed above, the signal-to-background ratio as a function of the selection variable was fit with a \pt{}-dependent functional form to determine the optimal selection criteria.   For the $\Lambda$ baryons, with the exception of the $V^0$ radius, a \pt{}-dependent selection did not lead to a significant improvement, and thus standard \pt{}-independent criteria~\cite{ALICE:2019avo} were used.  However, for the $\Xi$ baryons, this procedure led to an improved set of selection criteria.  
The specific selections imposed on the $\Lambda$ and $\Xi$ candidates are listed in Appendix~\ref{sec:cuts}.  
After applying all selection criteria, the efficiency of $\Lambda$ reconstruction ranged from $\varepsilon_{\Lambda} \sim 0.06$ for low \pt{} particles near the edges of the pseudorapidity acceptance to $\sim 0.4$ at $\pt = 4$~GeV/$c$ at midrapidity. The efficiency of $\Xi$ reconstruction ranged from $\varepsilon_{\Xi} \sim 0.007$ at low \pt near the edges of the $\eta$ acceptance to $\sim 0.4$ around $\pt = 5$~GeV/$c$ around $\eta\sim 0$.

\section{Analysis methodology}
\label{sec:analysis}

This analysis measured the per-trigger yields of pions, kaons, protons, $\Lambda$, and $\Xi$ baryons associated with a $\Xi$ baryon trigger, as defined in Eq.~\ref{eq:yield}.

\subsection{Efficiency and acceptance correction}
\label{sec:corr}

The signal correlation function was formed by constructing the angular distribution of trigger--associated pairs within the events ($N_\mathrm{pairs}$), and dividing by the total number of trigger particles ($N_\mathrm{trig}$): 
\begin{equation}
S(\Delta y, \Delta\varphi)=\dfrac{1}{N_\mathrm{trig}}\dfrac{\mathrm{d}^2N^\mathrm{sig}_\mathrm{pairs}}{\mathrm{d}\Delta y\, \mathrm{d}\Delta\varphi}.
\end{equation}
To account for the tracking and reconstruction efficiencies of the detector, each trigger and associated particle was weighted by the inverse of the single-particle efficiency, $1/\varepsilon$.  In $\Xi\Xi$ correlations, where the trigger and associated particles are identical, an additional $\pt^\mathrm{trig} > \pt^\mathrm{assoc}$ requirement was imposed to avoid double counting of pairs.  

Furthermore, because of the limited and non-uniform acceptance of the detector, the $S(\Delta y,\Delta\varphi)$ distribution is convolved with the experimental two-particle acceptance. To measure and correct for this detector effect, an event mixing technique was used.  Each trigger particle was correlated with associated particles from different events to obtain
\begin{equation}
\label{eq:B}
B(\Delta y, \Delta\varphi)=\alpha\dfrac{\mathrm{d}^2N^\mathrm{mixed}_\mathrm{pairs}}{\mathrm{d}\Delta y\, \mathrm{d}\Delta\varphi}.
\end{equation}
Here, $\alpha$ is a normalisation constant chosen such that the value at $B(0,0)$ equals unity, defined such that the pair acceptance probability for reconstructed particles traveling in the same direction is 100\%.  The value of $\alpha$ was calculated from a fit to the projection $B(\Delta y)$ to improve statistical precision.  

In order to accurately model the detector acceptance, the mixed events should match the characteristics of the signal events as closely as possible. To ensure similarity between signal and mixed events, the mixed events were matched such that the difference in the number of tracks, $\Delta N_{\mathrm{tracks}}$, and the difference in $z$-vertex position, $\Delta z_\mathrm{vtx}$, were minimised, requiring that $\Delta N_{\mathrm{tracks}} \leq 5$ and $\Delta z_{\mathrm{vtx}} < 1$~cm. Each signal event (containing a $\Xi$ trigger) was correlated with up to 60 different events.  It was not required that the events used for mixing contained $\Xi$ triggers.  

Finally, dividing the same-event correlation function $S(\Delta y, \Delta\varphi)$ by the mixed-event distribution $B(\Delta y,\Delta\varphi)$ resulted in the efficiency- and acceptance-corrected per-trigger yield,
\begin{equation}
Y(\Delta y, \Delta\varphi)=\dfrac{S(\Delta y, \Delta\varphi)}{B(\Delta y,\Delta\varphi)}.
\label{eq:correlation measurement}
\end{equation}

\subsection{Misidentification correction\label{sec:misid}}

While the kinematic and topological selection criteria described in Section~\ref{sec:lambdaxi} significantly reduced the number of random p$\pi$ and $\Lambda\pi$ pairs which were misidentified as coming from $\Lambda$ and $\Xi$ decays, there still remained a small combinatorial background under the signal peaks in the $m_{\mathrm{p}\pi}$ and $m_{\Lambda\pi}$ distributions.  The contribution to the correlation function coming from these combinatorial pairs was removed through the sideband subtraction procedure.  The sideband regions were defined as $4\sigma_{\Xi,\mathrm{ wide}} < |m_{\Lambda\pi}-\mu_{\Xi}| < 7\sigma_{\Xi,\mathrm{ wide}}$ for $\Xi$ baryons and $3\sigma_{\Lambda,\mathrm{ wide}} < |m_{\mathrm{p}\pi}-\mu_{\Lambda}| < 6\sigma_{\Lambda,\mathrm{ wide}}$ for $\Lambda$ baryons.  Since the combinatorial background is basically linear as a function of invariant mass, the summed correlation functions with respect to associated p$\pi$ and $\Lambda\pi$ pairs in the left and right sideband regions can be directly subtracted from the correlation functions in the signal region to obtain the corrected associated yields.  For $\Lambda$ baryons at low \pt{}, however, the background is non-linear, and in that kinematic region the combinatorial background was interpolated before subtraction.  

The associated pions, kaons, and protons were identified as described in Section~\ref{sec:piKp}, with tracks assigned to the particle species with the smallest corresponding $n\sigma^i$.  However, some amount of misidentification of the associated particles was inevitable and led to contamination in the resulting correlation function.  Instead of applying strict $n\sigma$ selections which would have led to a worsening of the statistical uncertainties, a method was developed which made use of linear algebra techniques to remove contamination due to misidentified particles.  The procedure described below made it possible to relate the measured $\mathbf{C}^{\rm track}$ to the misidentification-corrected $\mathbf{C}^{\rm particle}$, where $\mathbf{C}^{\rm track}$ ($\mathbf{C}^{\rm particle}$) is a column vector with either the same-event or mixed-event correlations for all track (particle) types, normalised to the number of tracks (particles). 

First, the misidentification matrix $\bf A$ was constructed, consisting of the coefficients $a_{ij}$ which are the fractions of the tracks identified as type $i$ which correspond to true particles of species $j$ (when $i=j$, $a_{ij}$ is the probability that a track is correctly identified).  Tracks were divided into eight classes corresponding to those identified as pions, kaons, protons, and electrons, each reconstructed with a TOF signal or without. Therefore $\bf A$ is an $8\times 4$ matrix.  

The misidentification fractions, $a_{ij}^{\mathrm{MC}}$, were obtained from a Monte Carlo simulation using events generated with the \pythia{} generator and propagated through a \geant{}3~\cite{Brun:1994aa} model of the ALICE detector. When measuring the misidentification fractions, the events were weighted in order to effectively match the multiplicity distribution of events containing a $\Xi$ baryon, to ensure that multiplicity-dependent effects, such as the probability for a track to be matched to a signal in the TOF, were correctly modelled. 

To obtain the misidentification fractions in data, $a_{ij}^{\mathrm{data}}$, it was assumed that the \geant{}3 detector description was correct to first order, and the following iterative minimisation procedure was performed to account for the particle ratios being different between \pythia{} and the experimental data. If the fraction of particles of species $j$ in the Monte Carlo is $x_j^{\rm MC}$, and in data it is $x_j^\mathrm{data}$, the misidentification fraction in data simply scales by $x_j^\mathrm{data}/x_j^{\rm MC}$, apart from a normalisation factor. Accounting for the normalisation, the misidentification fraction for tracks of type $i$ corresponding to particles of species $j$ in data can be written as
\begin{equation}
a_{ij}^\mathrm{data}=\dfrac{a_{ij}^\mathrm{MC} x_j^\mathrm{data}/x_j^\mathrm{MC}}{\sum_k a_{ik}^\mathrm{MC}x_k^\mathrm{data}/x_k^\mathrm{MC}}.
\label{eq:a_ij}
\end{equation}
If $y_i$ was the fraction of tracks of type $i$, one can form the matrix equation
\begin{equation}
\mathbf{A^\mathrm{T}y=x},
\label{eq:Ay=x}
\end{equation}
where $\mathbf{y}$ is a column vector with all $y_i$ and $\mathbf{x}$ is a column vector with all $x_j$. This identity is valid both in Monte Carlo and in data, and Eq.~\ref{eq:a_ij} forms the link between the two. Hence, Eqs.~\ref{eq:a_ij} and~\ref{eq:Ay=x} can be combined into a single matrix equation. Since $\mathbf{A}^\mathrm{MC}$ and $\mathbf{x}^\mathrm{MC}$ are known, by setting $\mathbf{y}\equiv\mathbf{y}^\mathrm{data}$, this combined system can be solved iteratively to obtain $\mathbf{A}^\mathrm{data}$ and $\mathbf{x}^\mathrm{data}$, the misidentification fractions and particle ratios in data. This procedure was performed separately in each individual momentum interval, and only for tracks in events containing a $\Xi$ baryon since the correlation with a $\Xi$ trigger will also affect the total yield. 

The same- and mixed-event correlations were constructed in each momentum interval for each track type.  Once the misidentification matrix ${\bf A}^{\mathrm{data}}$ had been measured, it was applied to the correlation functions to unfold them for misidentification.  To obtain the correlations for each particle species, one needs to solve the system
\begin{equation}
\mathbf{AC}^{\rm particle}=\mathbf{C}^{\rm track}.
\end{equation}
Since $\bf A$ is an $8\times 4$ matrix, this system is overdetermined and a least-squares solution is required to maximise the statistical significance. This has the solution 
\begin{equation}
\mathbf{C}^{\rm particle}=(\mathbf{A^\mathrm{T}WA})^\mathrm{-1}\mathbf{A^\mathrm{T}W}\mathbf{C}^{\rm track},
\label{eq:inverse}
\end{equation}
where the weight matrix $\mathbf{W}=\mathrm{diag}(\mathbf{N}^{\rm track})$ and the vector $\mathbf{N}^{\rm track}$ contain the number of tracks of each type. These were not corrected for efficiency, since $\mathbf{W}$ is used to optimise the statistical precision and cancels exactly in Eq.~\ref{eq:inverse}. 

Finally, to obtain the unnormalised correlation functions $S(\Delta y,\Delta\varphi)$ and $B(\Delta y,\Delta\varphi)$ used in Eq.~\ref{eq:correlation measurement}, $\mathbf{C}^{\rm particle}$ was multiplied by 
\begin{equation}
\mathbf{N}^{\rm particle}=\mathcal{E}(p)\mathbf{A^\mathrm{T}N}^{\rm track},
\label{eq:correlation efficiency corr}
\end{equation}
where $\mathcal{E}(p)$ is a diagonal matrix with all particle efficiencies, $\varepsilon_i$. The results from the different momentum intervals were merged before dividing the same-event and mixed-event correlation functions.

\subsection{Feeddown corrections}
\label{sec:feeddown}

The selected p($\overline{\mathrm{p}}$) and $\Lambda(\overline{\Lambda})$ baryons include not only particles produced from the primary collision vertex, but also secondary hadrons which come from the weak decays of heavier baryons.  In particular, non-negligible fractions of the identified (anti)protons come from $\Lambda(\overline{\Lambda})$ decays, and of the reconstructed $\Lambda(\overline{\Lambda})$ baryons come from $\Xi^{-}(\overline{\Xi}^{+})$ and $\Xi^0(\overline{\Xi}^{0})$ decays.  In order to report the correlations of $\Xi^{-}$ and $\overline{\Xi}^{+}$ hadrons with primary (anti)protons and $\Lambda(\overline{\Lambda})$ baryons, the contribution from these secondary particles was removed using the same procedure as in Ref.~\cite{ALICE:2019avo}.  

\pythia{} Monte Carlo simulations were used to construct feeddown matrices, containing the (efficiency-corrected) average number of detected daughter particles in the target \pt{} interval per generated mother particle, as a function of the \pt{} and $\eta$ of the mother. For the feeddown correction, these coefficients were then used as weights when constructing the correlation functions.

The contribution of the feeddown from $\Xi^0(\overline{\Xi}^0)$ baryons, which decay to $\Lambda\pi^0(\overline{\Lambda}\pi^0)$ and cannot be easily reconstructed in the ALICE detector, was estimated using the \pythia{} and \eposlhc{} Monte Carlo generators.  It was observed in both generators that the $\Xi^0(\overline{\Xi}^0)$ and $\Xi^{-}(\overline{\Xi}^{+})$ production rates were similar, as well as the shapes of the $\Xi^{-}\Xi^0$ and $\Xi^{-}\Xi^{-}$ same-baryon-number correlations (and their charge conjugates, $\overline{\Xi}^{+}\overline{\Xi}^0$ and $\overline{\Xi}^{+}\overline{\Xi}^{+}$).  However, \pythia{} showed a difference between the opposite-baryon-number correlation functions, where the correlation of the associated $\Xi^0(\overline{\Xi}^{0})$ on the near-side is weaker by 70\% than for the $\Xi^{-}(\overline{\Xi}^{+})$.  Therefore, for the purposes of the feeddown correction, the suppression of opposite-baryon-number $\Xi^{-}\overline{\Xi}^0$ (and $\overline{\Xi}^{+}\Xi^{0}$) correlations was assumed to be half of this difference, i.e.\ 35\%, with respect to $\Xi^{-}\overline{\Xi}^{+}$ (and $\overline{\Xi}^{+}\Xi^{-}$), and the two extremes (no suppression or 70\% suppression) covered within the systematic uncertainties. In practice, the feeddown contribution from $\Xi$ baryons to $\Lambda$ was measured as
\begin{equation}
N_\Xi^\mathrm{feeddown}=\left\{\begin{array}{lr}
(f_{\Xi^-}+f_{\Xi^0})N_{\Xi}^{\rm SB}&\hspace{0.5cm}\textrm{same baryon number,}\\
(f_{\Xi^-}+(1-r)f_{\Xi^0})N_{\Xi}^{\rm OB}+rf_{\Xi^0}N_{\Xi}^{\rm SB} &\hspace{0.5cm}\textrm{opposite baryon number,}
\end{array}\right.
\label{eq:xi_feeddown}
\end{equation}
where $N_{\Xi}^{\rm SB}$ and $N_{\Xi}^{\rm OB}$ are the measured numbers of same- and opposite-baryon number $\Xi\Xi$ pairs, $f_{\Xi^-}$ and $f_{\Xi^0}$ are the probabilities for the $\Lambda$ daughter particle of a charged and neutral $\Xi$, respectively, to appear within the $\Lambda$ sample, and $r$ is the reduction factor for $\Xi^0(\overline{\Xi}^0)$.

The final correlation function was then convolved by the autocorrelation function between mother and daughter, i.e.\ the expected smearing from the decay, which was simulated in Monte Carlo but weighted with the differences in the \pt{} spectra between Monte Carlo and data. Since this could only be done (at least without introducing any biases) to the final correlation function, the contribution from feeddown was not subtracted from the correlation function until the very end.

Monte Carlo studies demonstrated that contributions to the correlation function from other sources, such as $\mathrm{K^0_s}$ decays into pions and knock-out protons from the detector material, are small or negligible, and their minor effects were taken into account in the efficiency corrections.  

\subsection{Systematic uncertainties}
\label{sec:systuncert}
In order to assess the experimental systematic uncertainty on the correlation functions (using the procedure described in Section~\ref{sec:corruncert}), the event selection, particle reconstruction and identification, and efficiency and feeddown determination were varied.  The analysis methodology was verified with a Monte Carlo closure test.  

\subsubsection{Event selection}

The $z_\mathrm{vtx}$ range of the selected events was reduced from \mbox{$|z_\mathrm{vtx}|<10$ cm} to \mbox{$|z_\mathrm{vtx}|<8$ cm}, which reduced the statistical precision of the measurement but made the detector acceptance more uniform by avoiding edge effects at large $\eta$.  

\subsubsection{Particle reconstruction, selection, and identification}

The systematic uncertainty on the reconstruction of the primary pions, kaons, and protons, described in Section~\ref{sec:piKp}, was estimated by loosening the selection criteria on the minimum number of TPC clusters per track (to 60), maximum $\chi^2$ per cluster (from 4 to 5), and maximum DCA of the track to the primary vertex in the longitudinal direction (to 3 cm).  The PID selection criteria were also tightened to $|n\sigma^i|<2$ (for $i=\pi\mathrm{,K,p,e}$).  The analysis was repeated using a fixed momentum value, $p = 0.6\,\mathrm{GeV}/c$ for all particle types, as a starting point for using TOF information for PID.

The \pt{}-dependent selection criteria applied to the properties of the reconstructed $\Lambda(\overline{\Lambda})$ and $\Xi$ candidates, described in Section~\ref{sec:lambdaxi} and listed explicitly in Appendix~\ref{sec:cuts}, were also loosened and tightened.  Specifically, the criteria varied in the reconstruction of the $\Lambda(\overline{\Lambda})$ candidates were: the DCA of the pion and proton daughters to the primary vertex, the DCA of the daughters to each other, the radial location of the secondary vertex in the detector, and the cosine of the pointing angle of the $V^0$ momentum back to the primary vertex.  In the cascade reconstruction, the following selection criteria were varied: the DCAs of the three daughter tracks to the primary vertex, the invariant mass of the $\Lambda(\overline{\Lambda})$ daughter, the DCA of the $V^0$ daughters, the radial distances of the $V^0$ and cascade vertices, the DCA of the cascade to the primary vertex, and the cosine of the pointing angle.  A tighter pileup selection was used for the $\Xi$ reconstruction, requiring two daughter tracks (instead of one) with either a successful ITS refit or a TOF hit.

\subsubsection{Acceptance, efficiency, and feeddown corrections}

The procedures for calculating the efficiency and feeddown correction factors were also varied, by removing the event multiplicity weighting, and obtaining the coefficients as a function of \pt{} (instead of \pt{} and $\eta$).  Since the ALICE detector model in \geant{}3 is not perfect, an additional uncertainty on the tracking efficiencies due to the material budget has been estimated as 0.7\% for pions, 0.5\% for kaons, and 1.5\% for protons~\cite{ALICE:2020nkc}. The suppression factor of the opposite-baryon-number $\Xi^{-}\overline{\Xi}^0$ (and $\overline{\Xi}^{+}\Xi^{0}$) correlations with respect to $\Xi^{-}\overline{\Xi}^{+}$ (and $\overline{\Xi}^{+}\Xi^{-}$), used in the estimation of the feeddown from $\Xi^0$ to $\Lambda$, was changed to $r = 0$ and $r = 0.7$, in accordance with the \pythia{} indications (see Section~\ref{sec:feeddown}).  

In the sideband subtraction procedure, which corrects for the combinatorial p$\pi$ and $\Lambda\pi$ pairs misidentified as $\Lambda$ and $\Xi$ baryons, the sideband definitions were changed to $4-7\sigma$ for $\Xi$ and $5-8\sigma$ for $\Lambda$. 

Finally, the uncertainty on the fit value of the event mixing normalisation, $\alpha$ (see Eq.~\ref{eq:B}), was assigned as a systematic uncertainty. 

\subsubsection{Systematic effects of the analysis procedure}

The full analysis procedure was verified utilising a Monte Carlo closure test, both at the level of the single-particle spectra and the correlation functions.  The analysis, including all correction procedures, was performed on reconstructed Monte Carlo particles from \pythia{} which had been propagated through a \geant{}3 model of the ALICE detector, and the results were compared to $\Xi$--hadron correlation functions obtained from the generator-level Monte Carlo particles. The disagreement between reconstructed- and generated-level results should quantify the systematic uncertainty related to the analysis method itself, and in particular to the misidentification correction procedure (Sec.~\ref{sec:misid}), and was assigned as a systematic source.  Due to the limited statistics of the $\Xi\Xi$ correlations in the reconstruction-level Monte Carlo, the uncertainties from the $\Xi\Lambda$ closure test are used instead.  

Since the closure test is a comparison of Monte Carlo to itself, it is not sensitive to systematic effects which may arise from differences between Monte Carlo and data. Therefore, to further test the particle reconstruction and identification methods, and in particular the misidentification and feeddown corrections, the \pt{} spectra of pions, kaons, protons, $\Lambda$, and $\Xi$ baryons are compared to the spectra published in Ref.~\cite{ALICE:2020jsh}, in which they were measured using different techniques.  The small observed differences were included as additional uncertainties in this analysis. 

\subsubsection{Uncertainties on the correlation function\label{sec:corruncert}}

To quantify the effect of each systematic variation, the ratio of the correlation function obtained after the variation with respect to the nominal one was calculated.  The statistical uncertainties on the ratios were calculated under the assumption that the variation and the default results were maximally correlated.  The ratio for each variation was fit with a simple function to reduce the impact of point-to-point statistical fluctuations in the systematic uncertainty estimation.  It was observed that the ratio was always consistent with a constant shift along the $\Delta y$ axis, and thus fit with a flat line to obtain the relative uncertainty.  Some systematic variations, however, showed a modulation in the ratio as a function of $\Delta\varphi$. Consequently, the ratios of the $\Delta\varphi$ projections were fit with a periodic function of the form $a+b\cos(\Delta\varphi)$.  For the differences between opposite- and same-sign correlations, the fit functions were applied directly to the difference between the systematic variation and the default, instead of to the ratio, to obtain the absolute uncertainty.  The values of the fit functions defined the estimated relative uncertainty at each point in $\Delta y$ or $\Delta\varphi$.  The systematic variation was deemed significant if the values of the fit parameters were larger than the uncertainty on the fit parameters.  This was equivalent to a Barlow check with a Barlow criterion of one~\cite{Barlow:2002yb}.  Systematic uncertainties were calculated directly on the integrated yields and near-side widths shown in Sections~\ref{sec:yields} and~\ref{sec:multdep}.  

For the OS and SS (and OB and SB) $\Xi\pi$, $\Xi\mathrm{K}$, and $\Xi\mathrm{p}$ correlation functions, the systematic uncertainties from the individual sources are generally less than 1\%. The largest uncertainties, which reach the few-percent level, are generally related to the PID selection criteria ($\sim 1.5\%$ for kaons and protons), the multiplicity weighting ($\sim 1\%$ for kaons), the Monte Carlo closure and the tracking efficiency (up to $2.5\%$ for protons). The uncertainties on the OB and SB $\Xi\Lambda$ and $\Xi\Xi$ correlations are slightly larger, on the level of a few percent, and the most significant sources are the topological selection criteria variations (up to $9\%$ for $\Xi$ baryons), pileup rejection criteria (up to $5.2\%$ for $\Lambda$ baryons and $7.5\%$ for $\Xi$), and sideband definitions (up to $\sim 30\%$ for $\Xi$ baryons).  The uncertainties are also calculated directly for the OS--SS and OB--SB differences, where the only uncertainties which exceed the percent level are from the $\Lambda$ and $\Xi$ topological selections (up to $2\%$ for $\Xi\Lambda$), the pileup rejection (up to $\sim 5\%$ for $\Xi\Lambda$), the $\Xi^0$ contribution (up to $3.2\%$ for $\Xi\Lambda$), the Monte Carlo closure ($15\%$ for $\Xi\pi$), and the spectra closure (up to $2\%$ for $\Xi\Lambda$).  The individual uncertainties are added in quadrature to obtain the total uncertainties, which are generally in the range of a few percent for minimum bias events.  

\section{Results and discussion}
\label{sec:results}

\subsection{Multiplicity-integrated correlation functions}

\begin{figure}[t]
    \begin{center}
    \includegraphics[width = 0.45\textwidth]{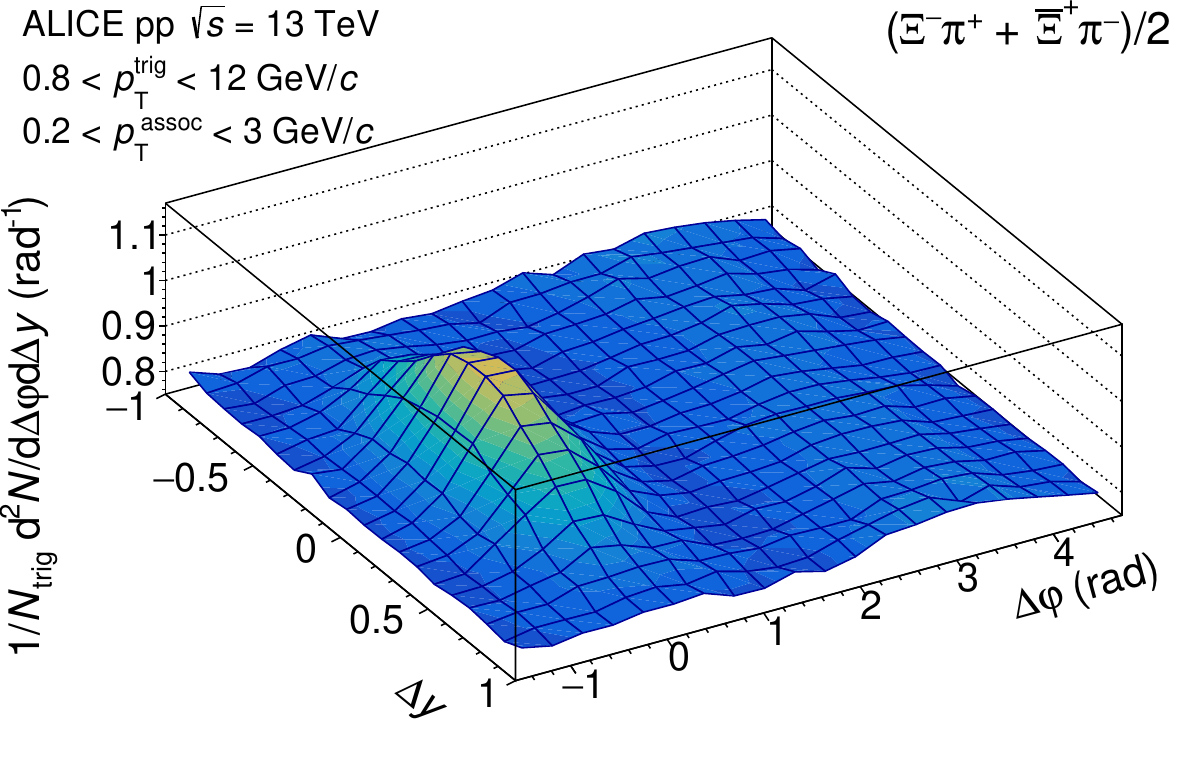}~\includegraphics[width = 0.45\textwidth]{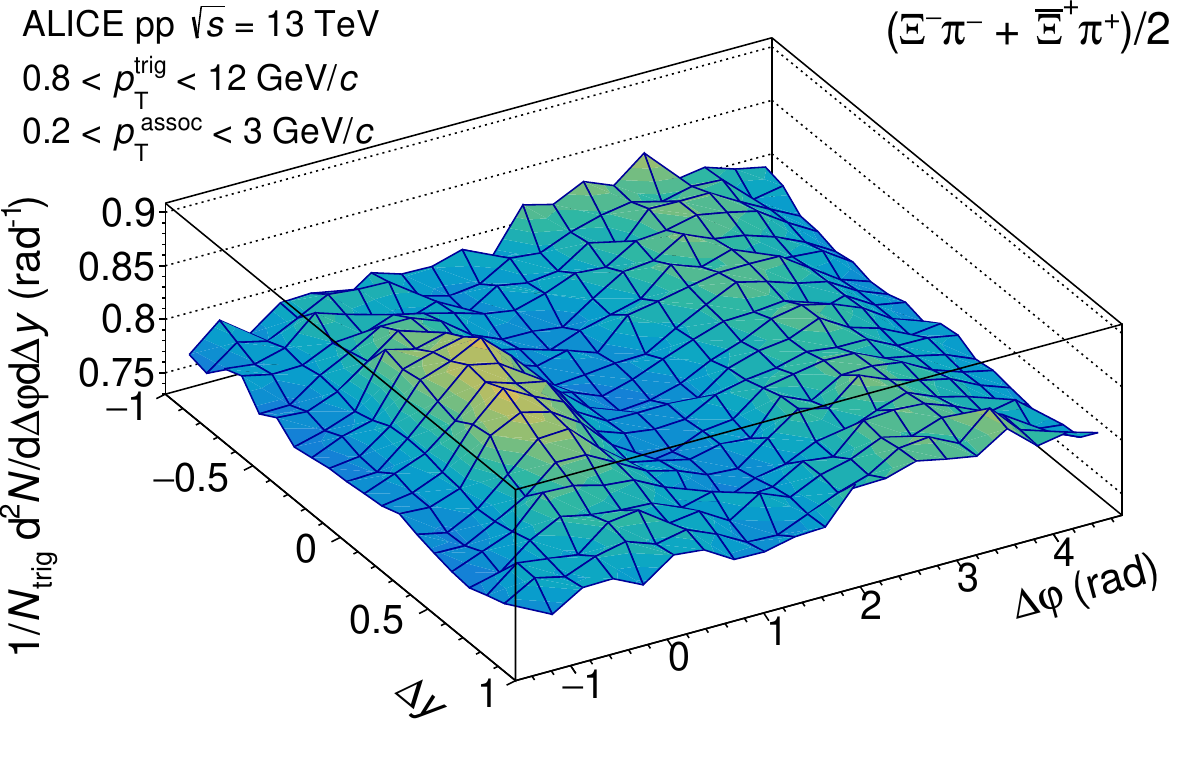} 
    \includegraphics[width = 0.45\textwidth]{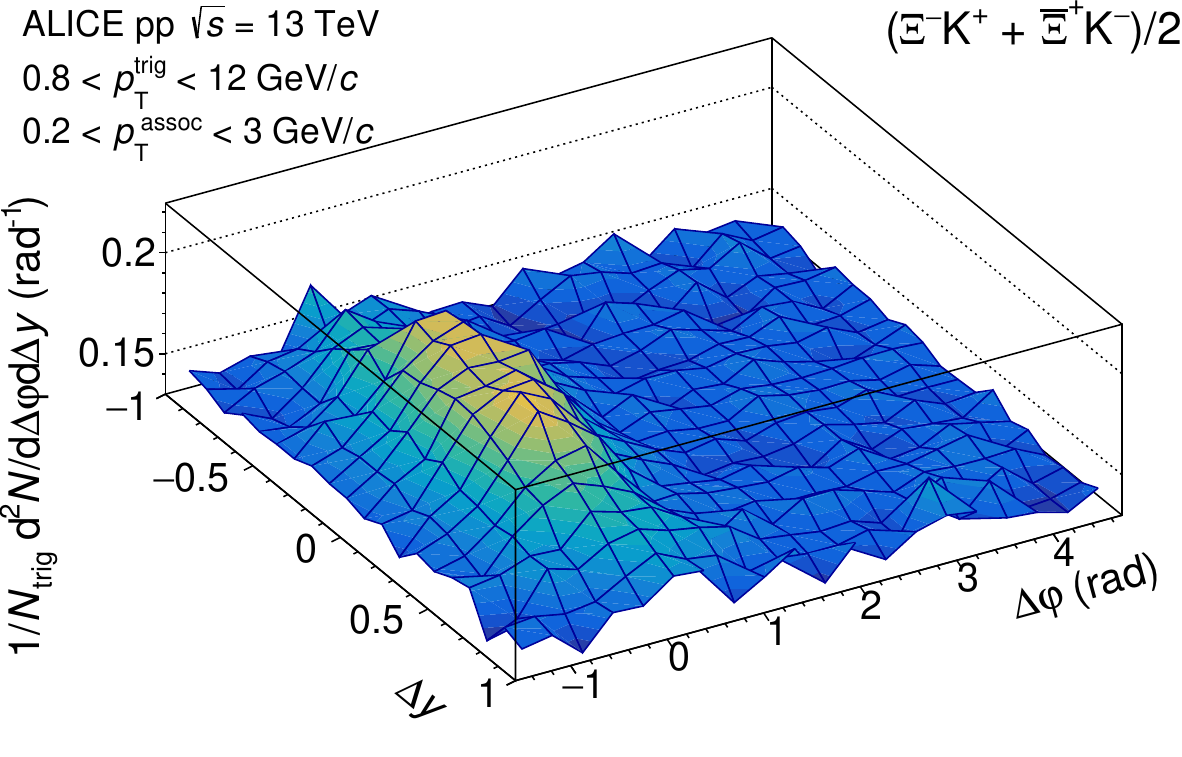}~\includegraphics[width = 0.45\textwidth]{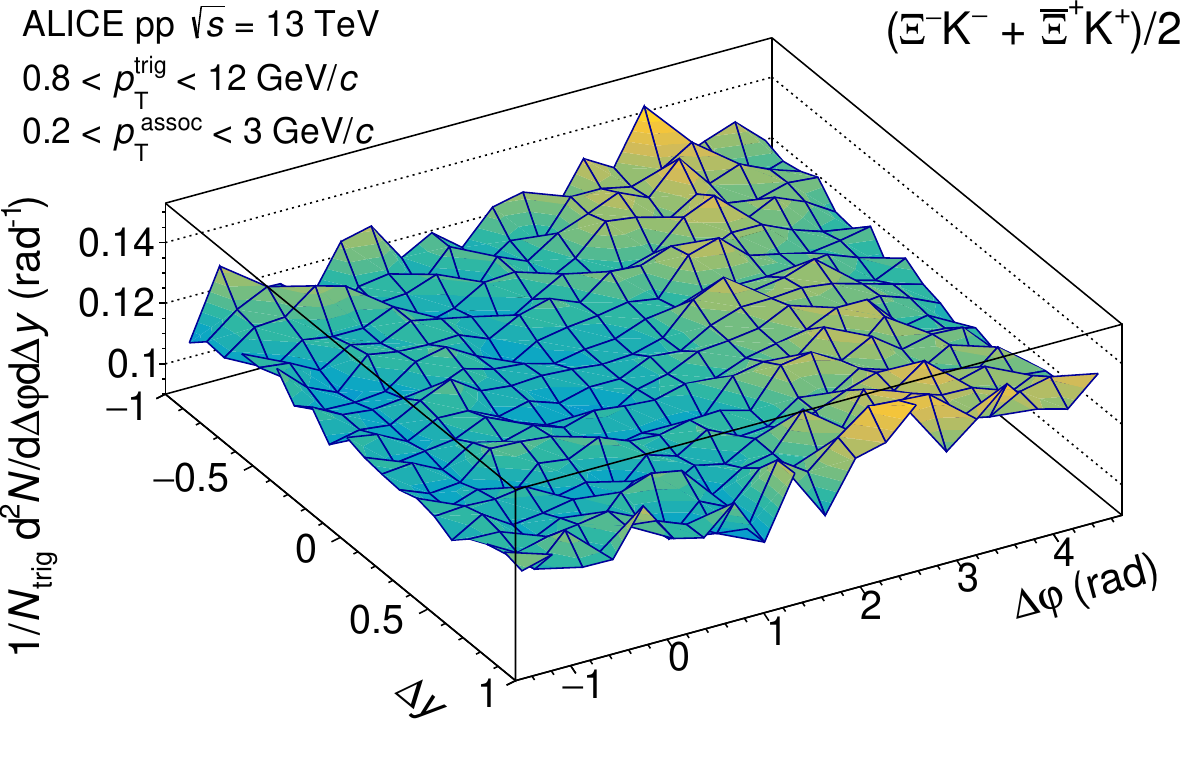}
    \end{center}
    \caption{$\Xi\pi$ (top row) and $\Xi\mathrm{K}$ (bottom row) per-trigger yields in $(\Delta y,\Delta\varphi)$ for particle pair combinations with the opposite (left column) and same (right column) electric charge, measured in \pp{} collisions at $\sqrt{s} = 13$ TeV. }
    \label{fig:corr_mesons}
\end{figure}

\begin{figure}[tbp]
    \begin{center}
    \includegraphics[width = 0.33\textwidth]{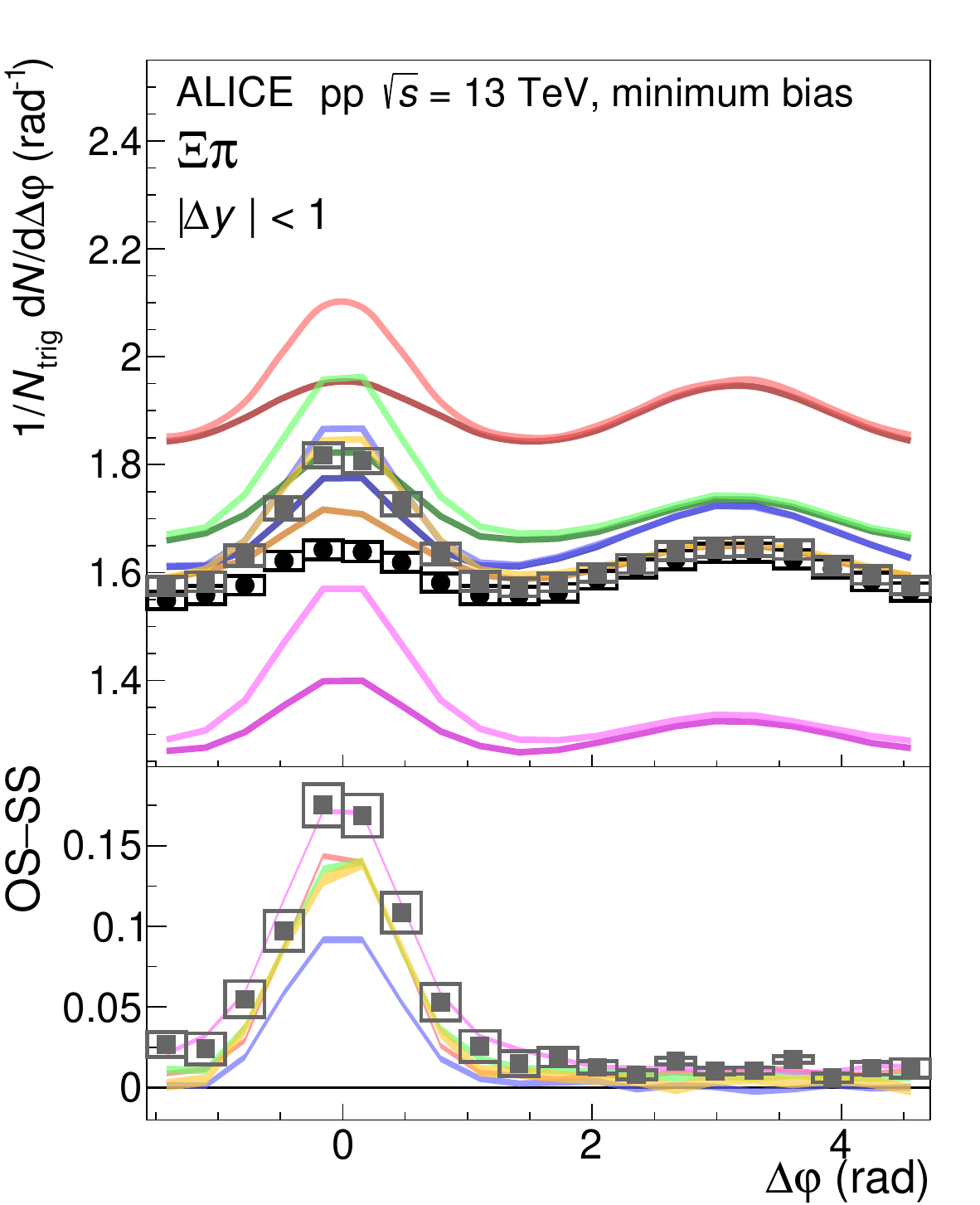}~\includegraphics[width = 0.33\textwidth]{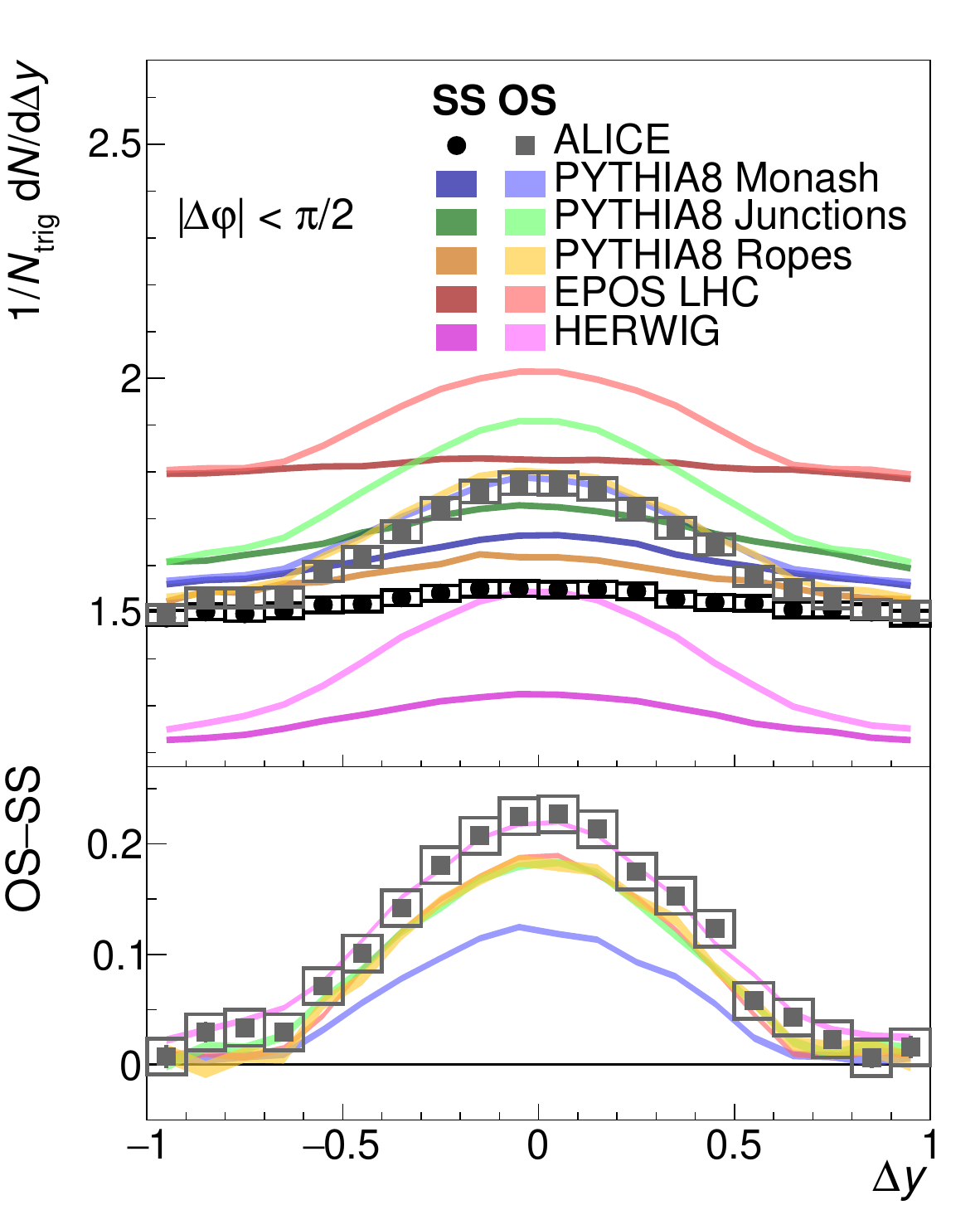}~\includegraphics[width = 0.33\textwidth]{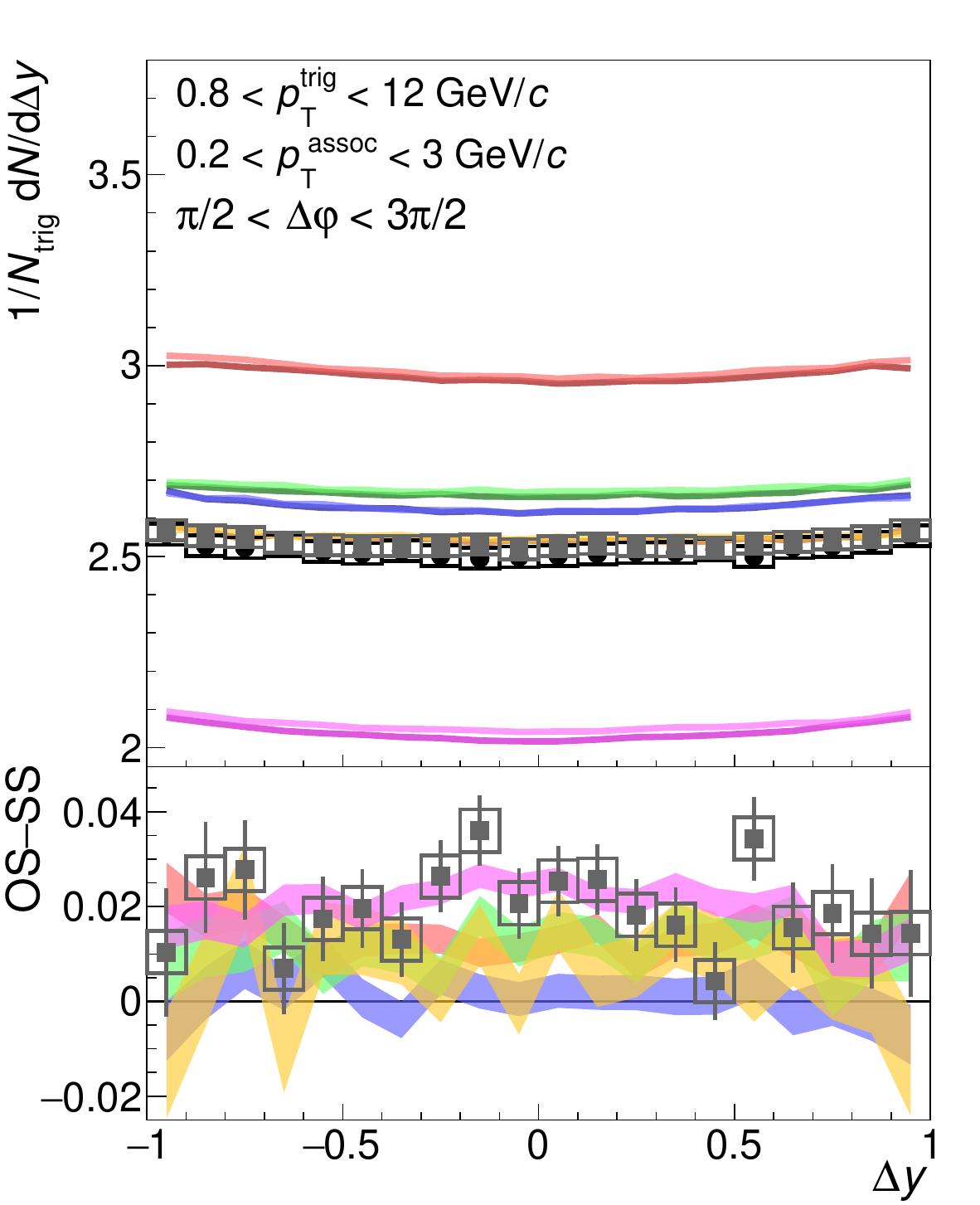}
    \end{center}
    \caption{$\Xi^-\pi^+$ and $\Xi^-\pi^-$ (and charge conjugate) correlation functions projected onto $\Delta\varphi$ ($|\Delta y| < 1$, left), the near-side on $\Delta y$ ($|\Delta\varphi| < \pi/2$, middle), and the away-side on $\Delta y$ ($\pi/2 < \Delta\varphi < 3\pi/2$, right). Opposite-sign ($\Xi^{-}\pi^{+}+\overline{\Xi}^{+}\pi^{-}$) correlations are shown in grey squares, the same-sign ($\Xi^{-}\pi^{-}+\overline{\Xi}^{+}\pi^{+}$) correlations are black circles; the OS--SS difference is displayed in the bottom panels.  Statistical and systematic uncertainties are represented by bars and boxes, respectively.  The ALICE data are compared with the following models: \pythia{} Monash tune (blue), \pythia{} with junctions enabled (green), \pythia{} with junctions and ropes (yellow), \eposlhc{} (red), and \herwig{} (pink).}
    \label{fig:xipi}
\end{figure}

\begin{figure}[tbp]
    \begin{center}
    \includegraphics[width = 0.33\textwidth]{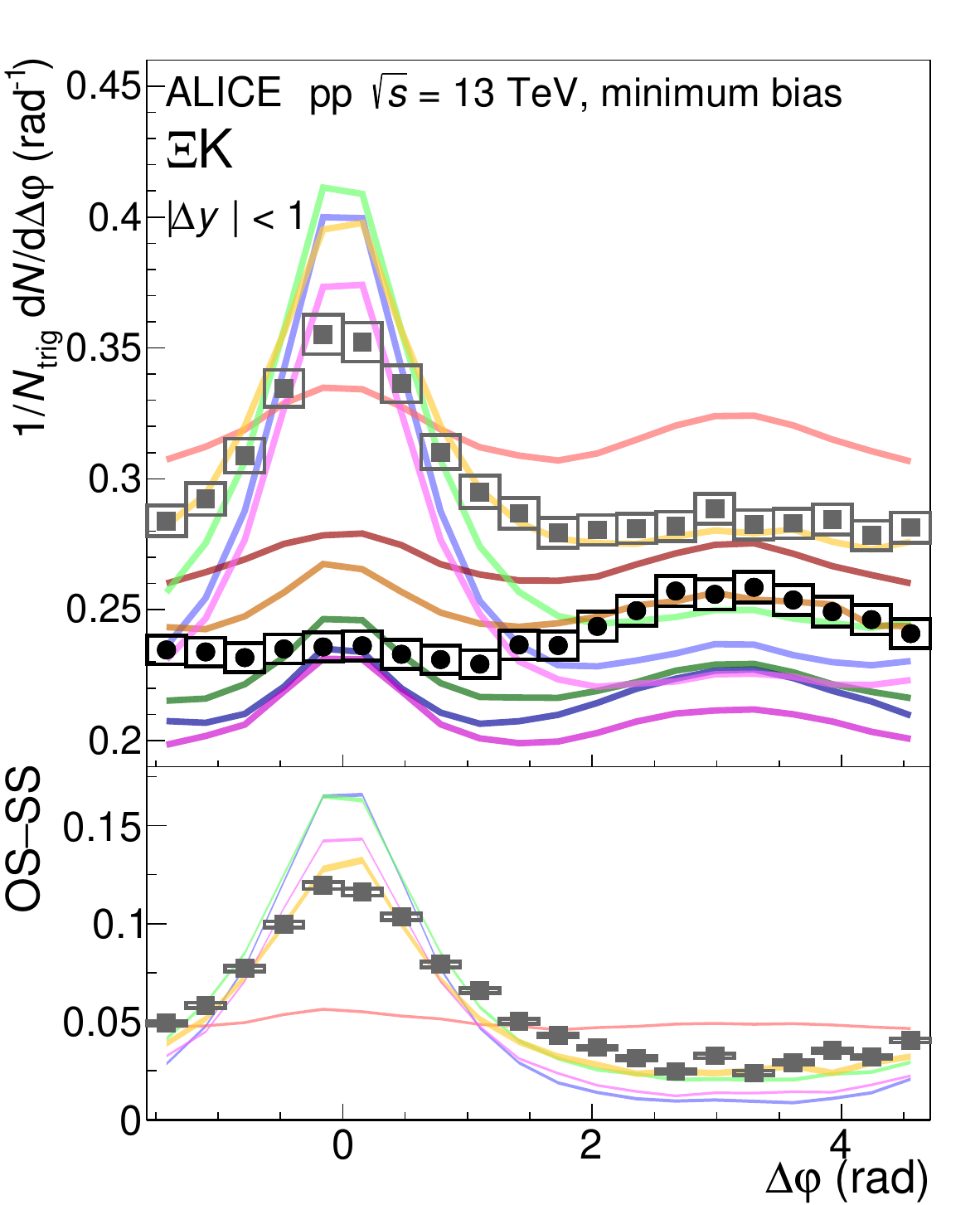}~\includegraphics[width = 0.33\textwidth]{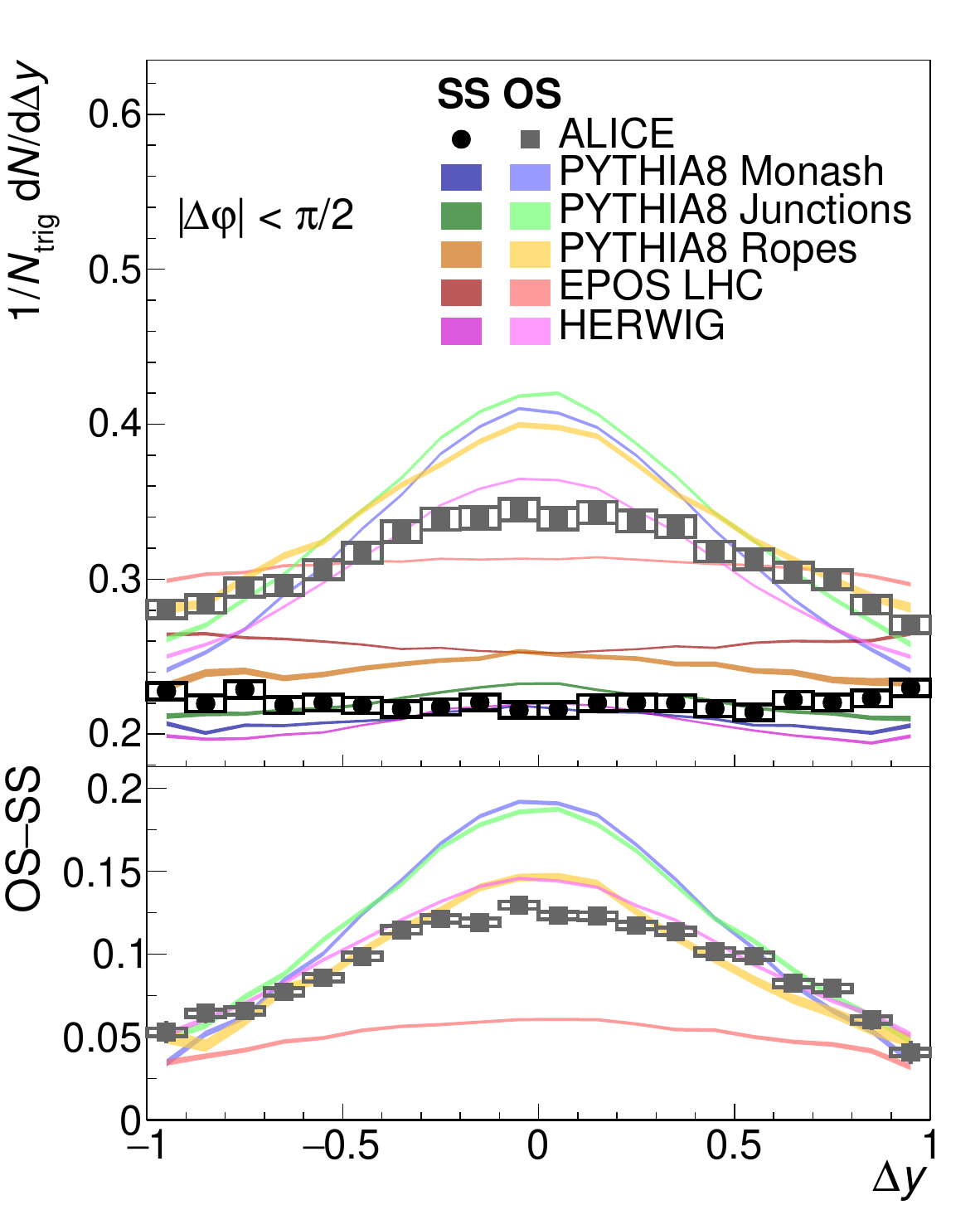}~\includegraphics[width = 0.33\textwidth]{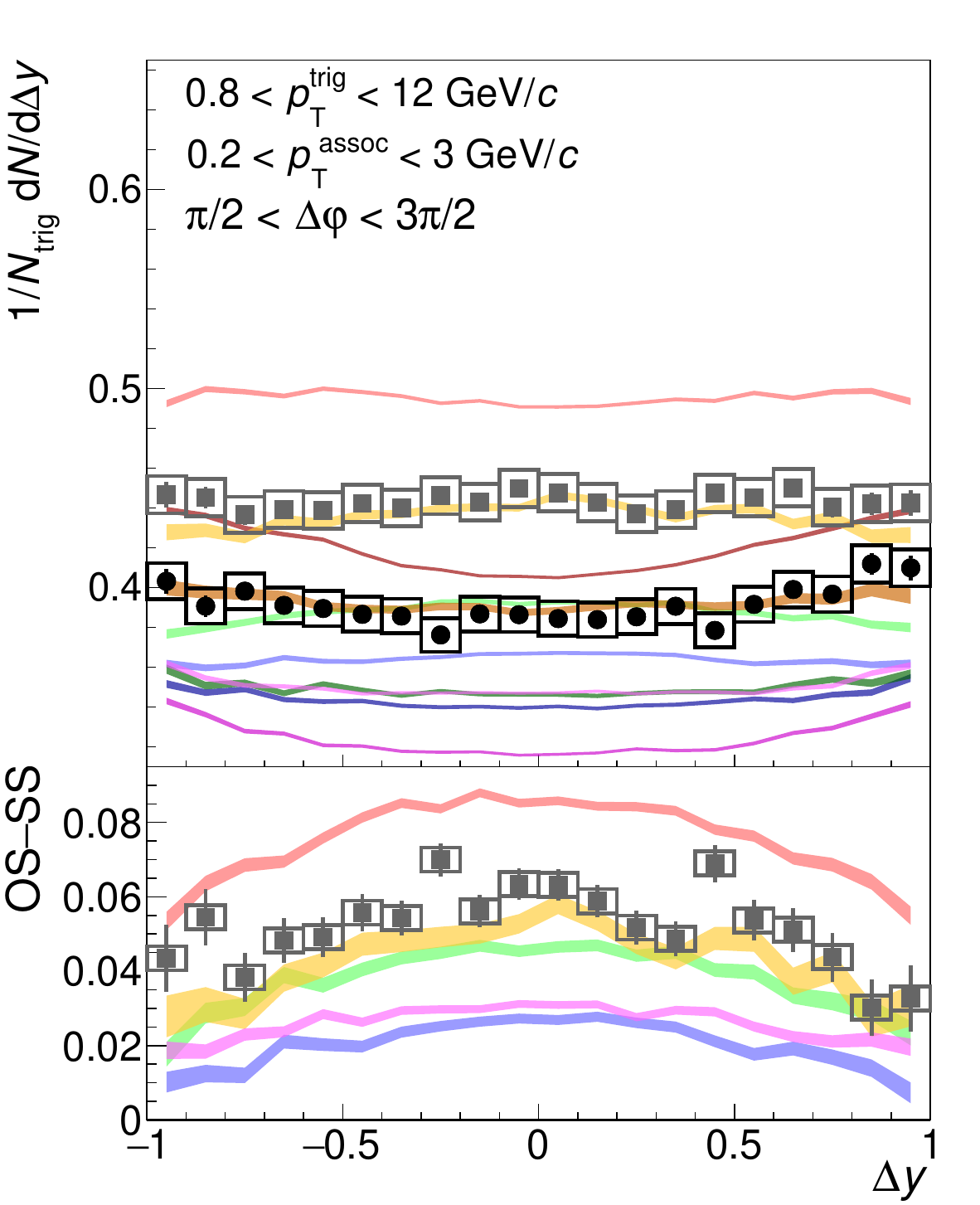}
    \end{center}
    \caption{$\Xi^-\mathrm{K}^+$ and $\Xi^-\mathrm{K}^-$ (and charge conjugate) correlation functions projected onto $\Delta\varphi$ ($|\Delta y| < 1$, left), the near-side on $\Delta y$ ($|\Delta\varphi| < \pi/2$, middle), and the away-side on $\Delta y$ ($\pi/2 < \Delta\varphi < 3\pi/2$, right). Opposite-sign ($\Xi^{-}\mathrm{K}^{+}+\overline{\Xi}^{+}\mathrm{K}^{-}$) correlations are shown in grey squares, the same-sign ($\Xi^{-}\mathrm{K}^{-}+\overline{\Xi}^{+}\mathrm{K}^{+}$) correlations are black circles; the OS--SS difference is displayed in the bottom panels.  Statistical and systematic uncertainties are represented by bars and boxes, respectively.  The ALICE data are compared with the following models: \pythia{} Monash tune (blue), \pythia{} with junctions enabled (green), \pythia{} with junctions and ropes (yellow), \eposlhc{} (red), and \herwig{} (pink).}
    \label{fig:xiK}
\end{figure}

\begin{figure}[t]
    \begin{center}
    \includegraphics[width = 0.45\textwidth]{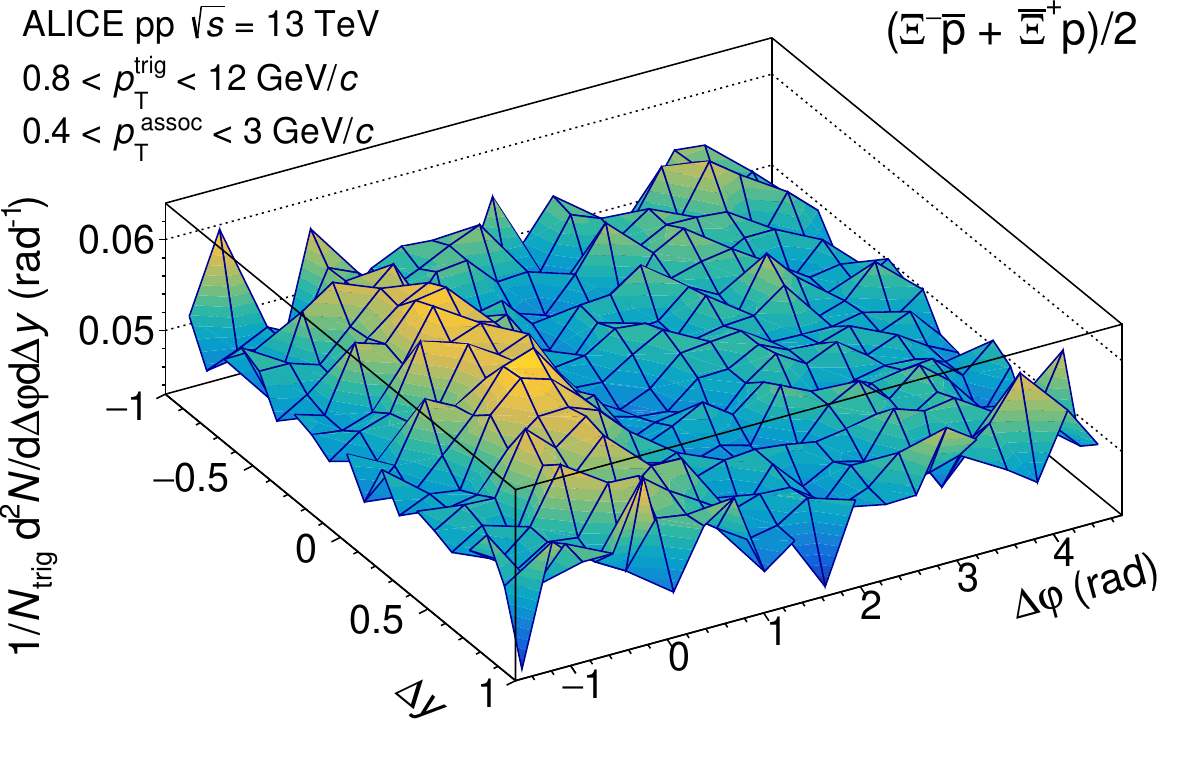}~\includegraphics[width = 0.45\textwidth]{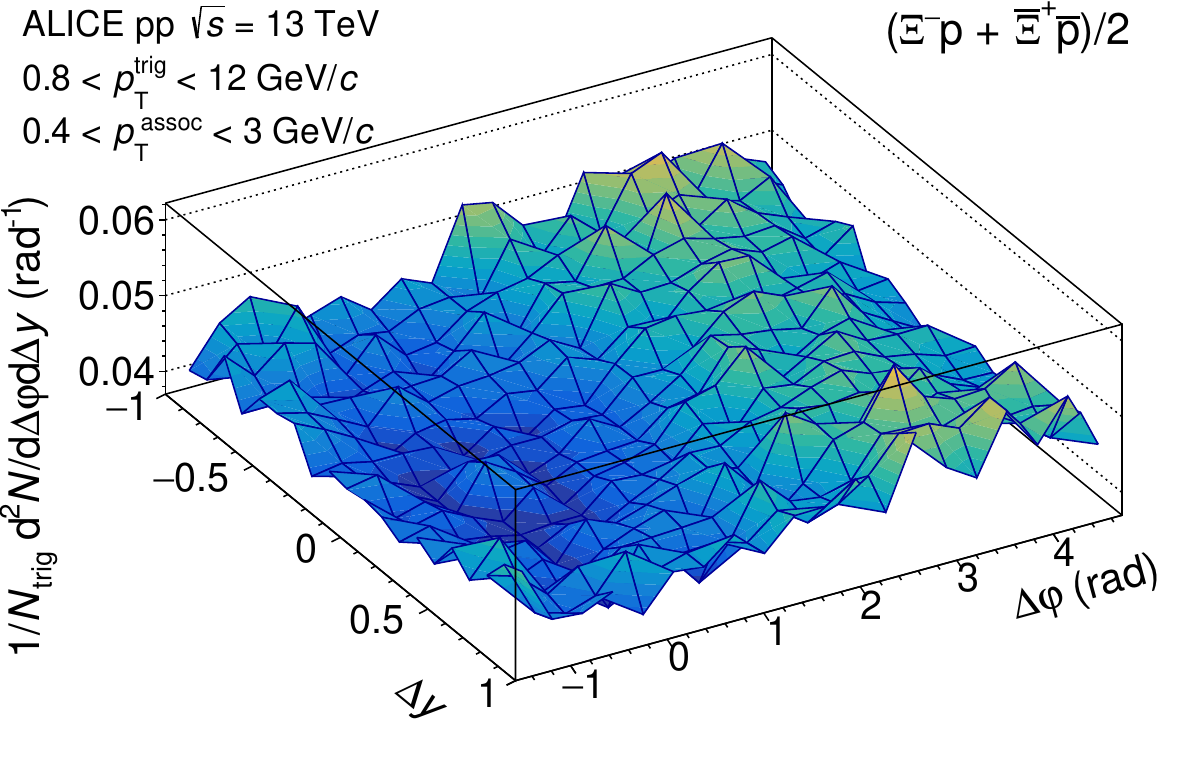}
        \includegraphics[width = 0.45\textwidth]{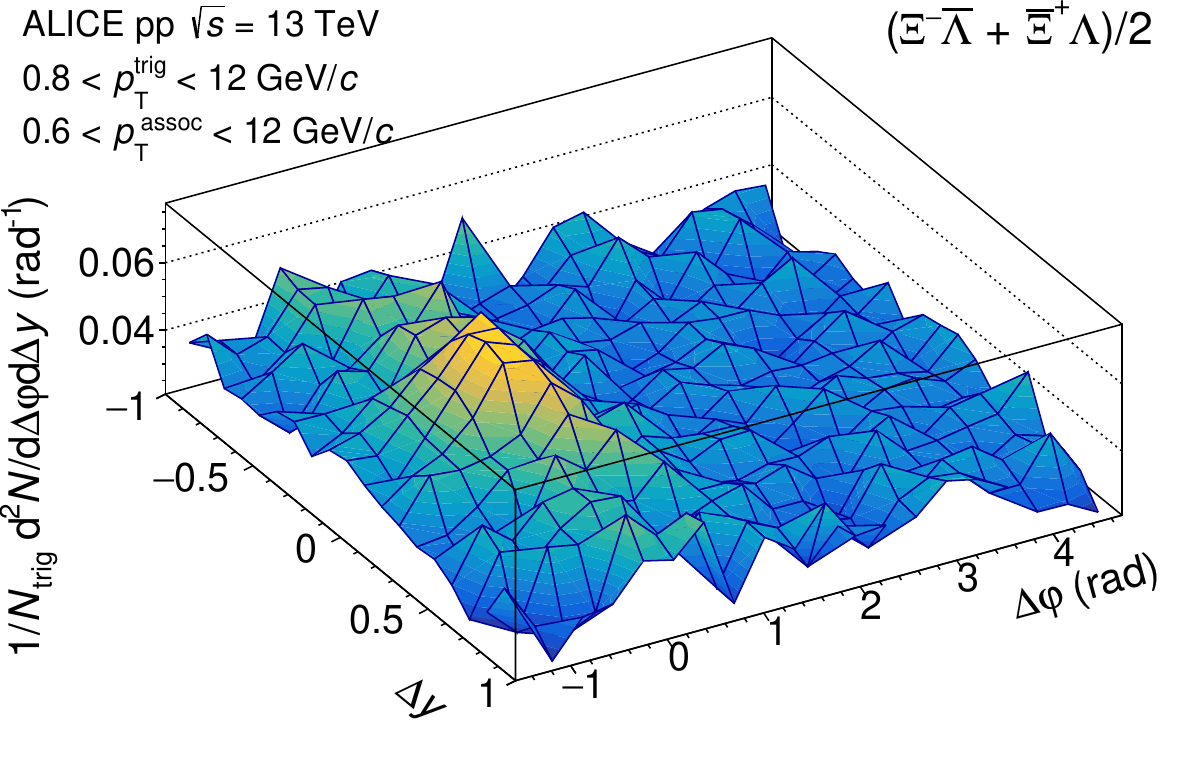}~ \includegraphics[width = 0.45\textwidth]{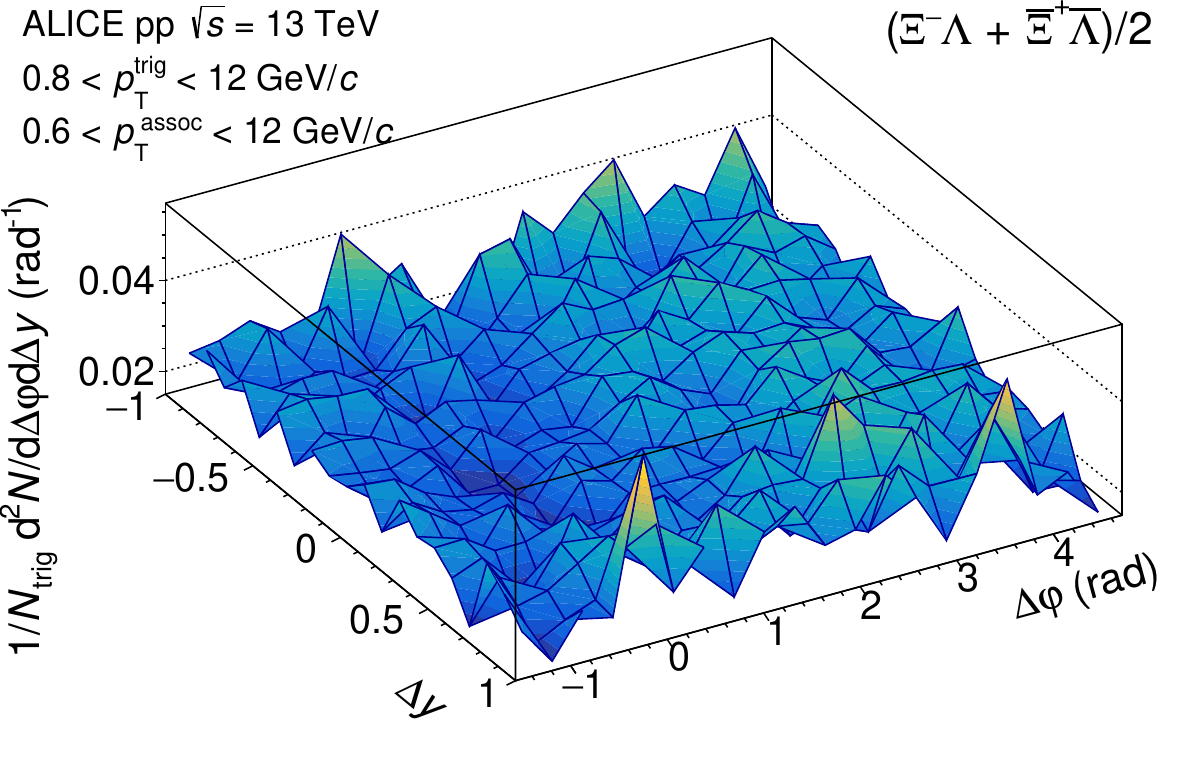}
            \includegraphics[width = 0.45\textwidth]{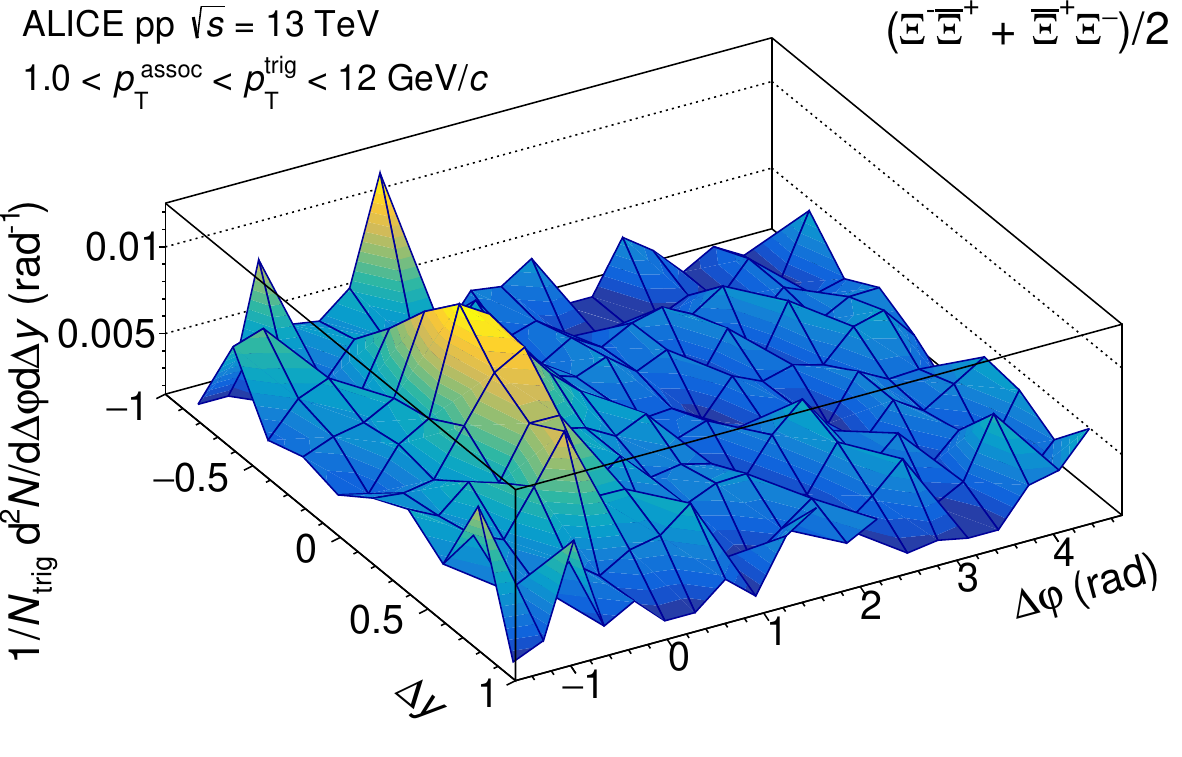}~\includegraphics[width = 0.45\textwidth]{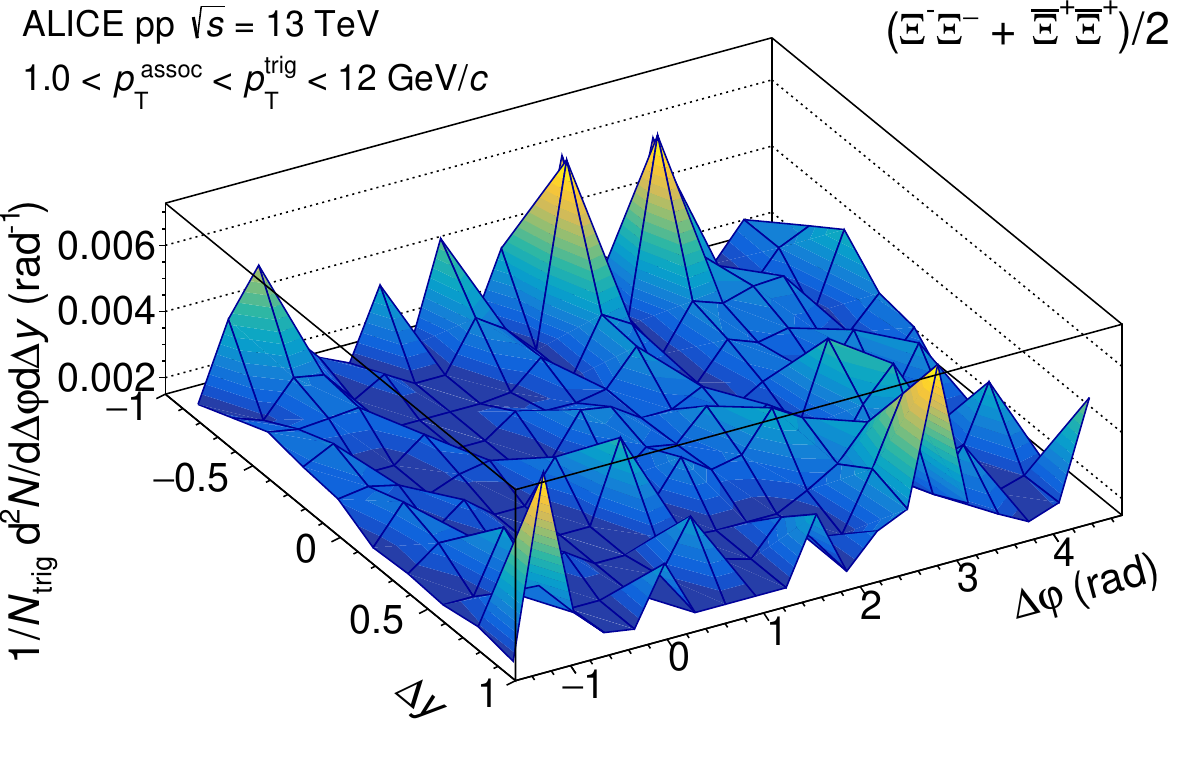}
    \end{center}
    \caption{$\Xi\mathrm{p}$ (top row), $\Xi\Lambda$ (middle row), and $\Xi\Xi$ (bottom row) per-trigger yields in $(\Delta y,\Delta\varphi)$ for particle pair combinations with the opposite (left column) and same (right column) baryon number, measured in \pp{} collisions at $\sqrt{s} = 13$ TeV. }
    \label{fig:corr_baryons}
\end{figure}

\begin{figure}[tbp]
    \begin{center}
    \includegraphics[width = 0.33\textwidth]{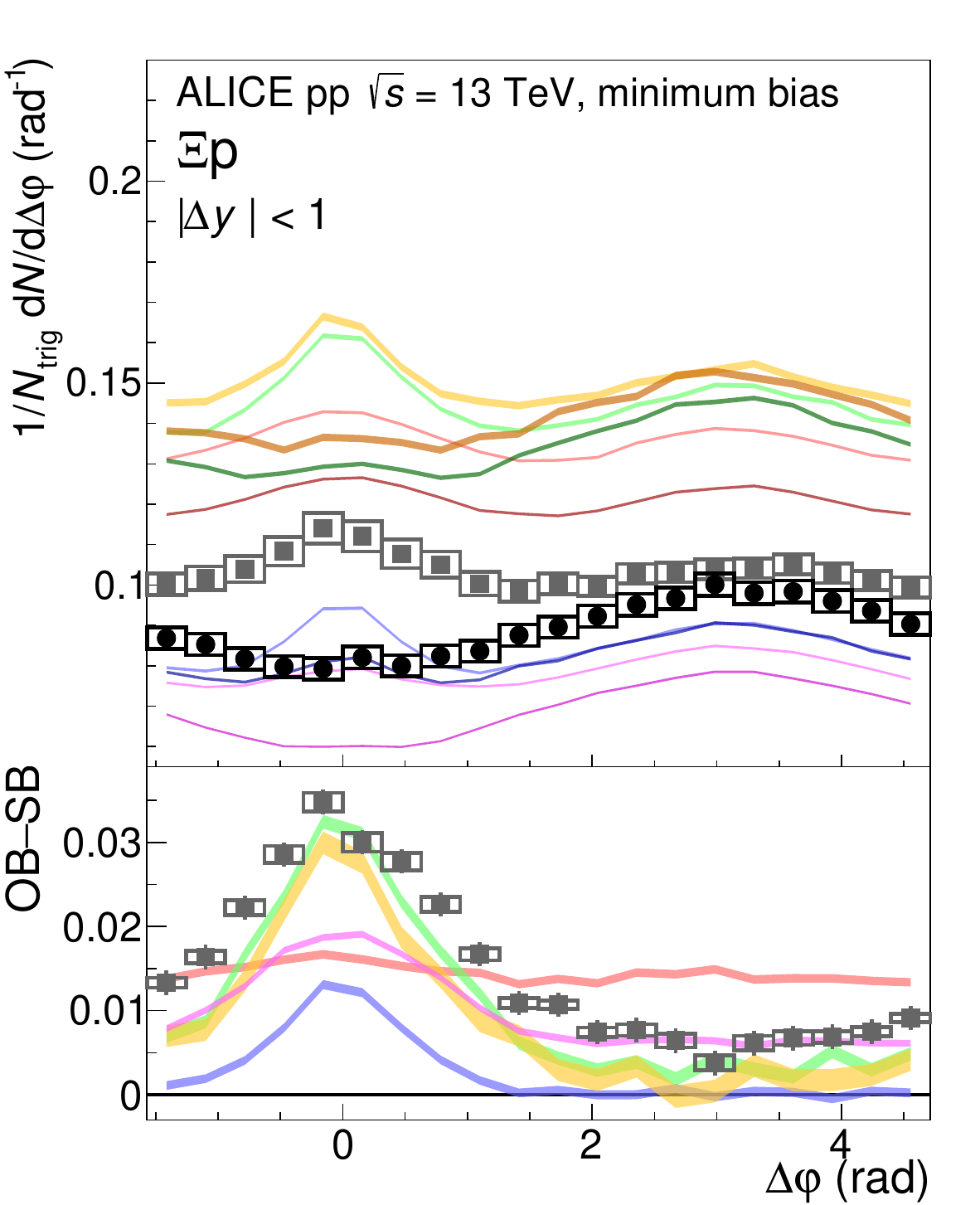}~\includegraphics[width = 0.33\textwidth]{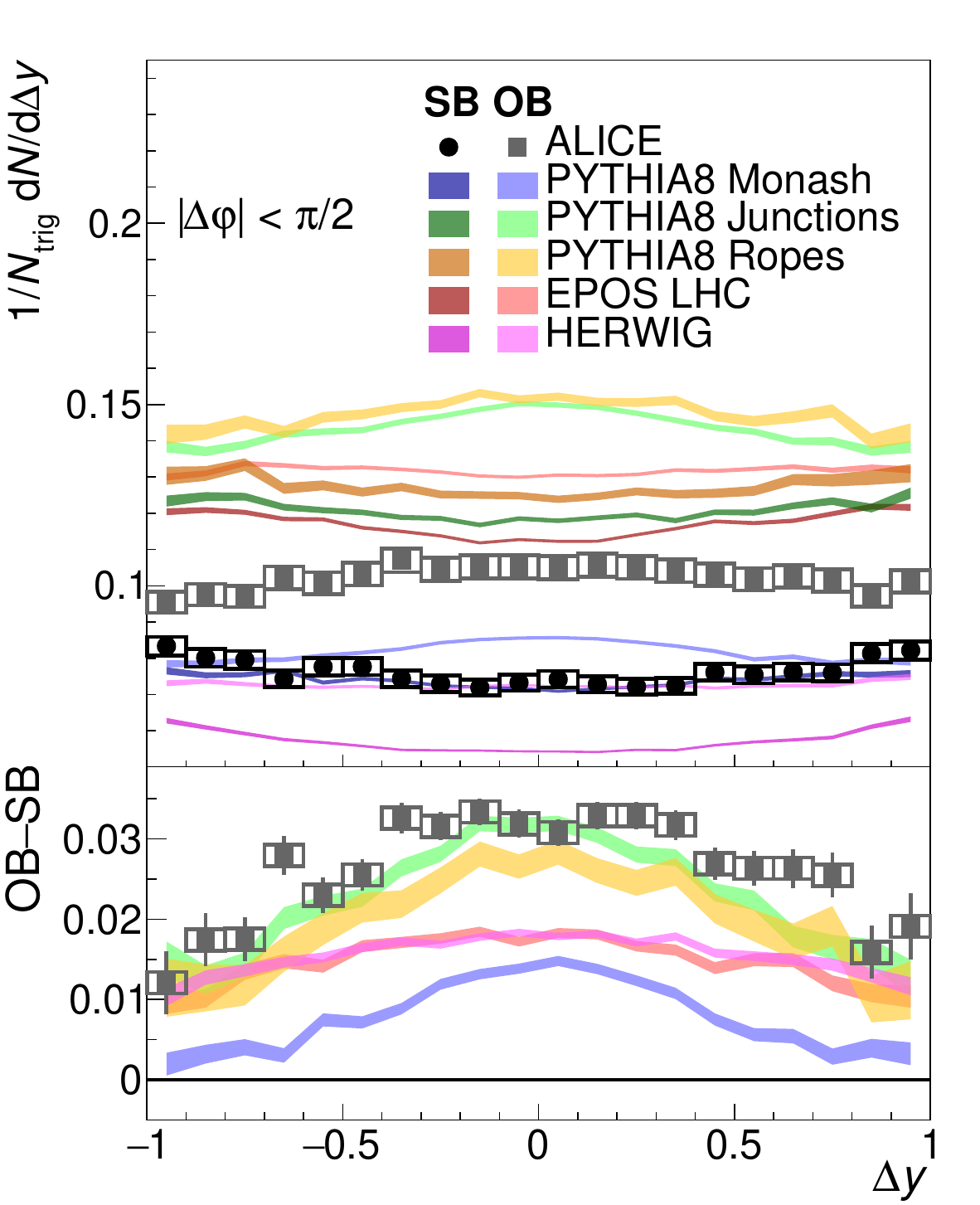}~\includegraphics[width = 0.33\textwidth]{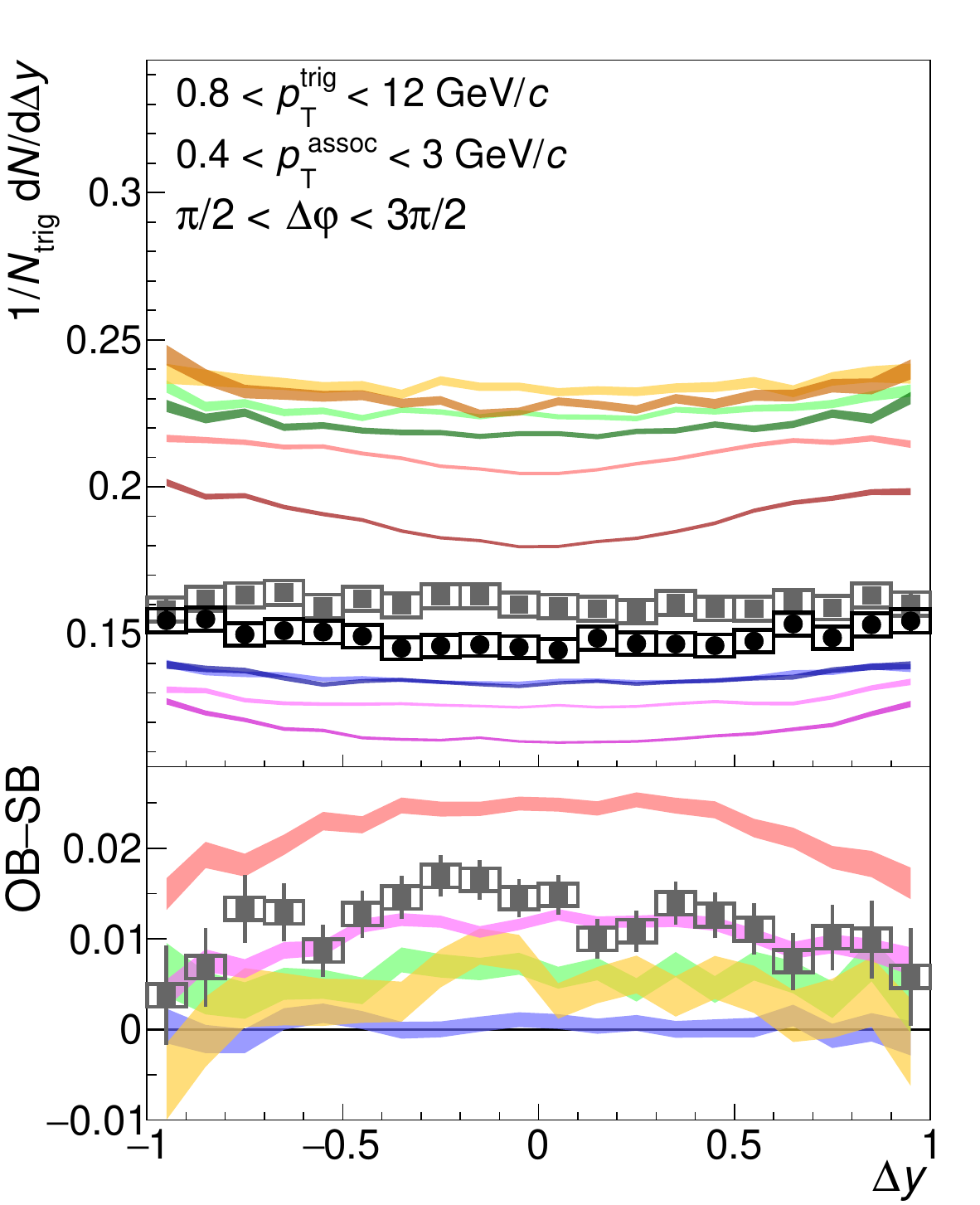}
    \end{center}
    \caption{$\Xi^-\overline{\mathrm{p}}$ and $\Xi^-\mathrm{p}$ (and charge conjugate) correlation functions projected onto $\Delta\varphi$ ($|\Delta y| < 1$, left), the near-side on $\Delta y$ ($|\Delta\varphi| < \pi/2$, middle), and the away-side on $\Delta y$ ($\pi/2 < \Delta\varphi < 3\pi/2$, right). Opposite-baryon-number ($\Xi^{-}\overline{\mathrm{p}}+\overline{\Xi}^{+}\mathrm{p}$) correlations are shown in grey squares, the same-baryon-number ($\Xi^{-}\mathrm{p}+\overline{\Xi}^{+}\overline{\mathrm{p}}$) correlations are black circles; the OB--SB difference is displayed in the bottom panels.  Statistical and systematic uncertainties are represented by bars and boxes, respectively.  The ALICE data are compared with the following models: \pythia{} Monash tune (blue), \pythia{} with junctions enabled (green), \pythia{} with junctions and ropes (yellow), \eposlhc{} (red), and \herwig{} (pink).}
    \label{fig:xip}
\end{figure}

\begin{figure}[tbp]
    \begin{center}
        \includegraphics[width = 0.33\textwidth]{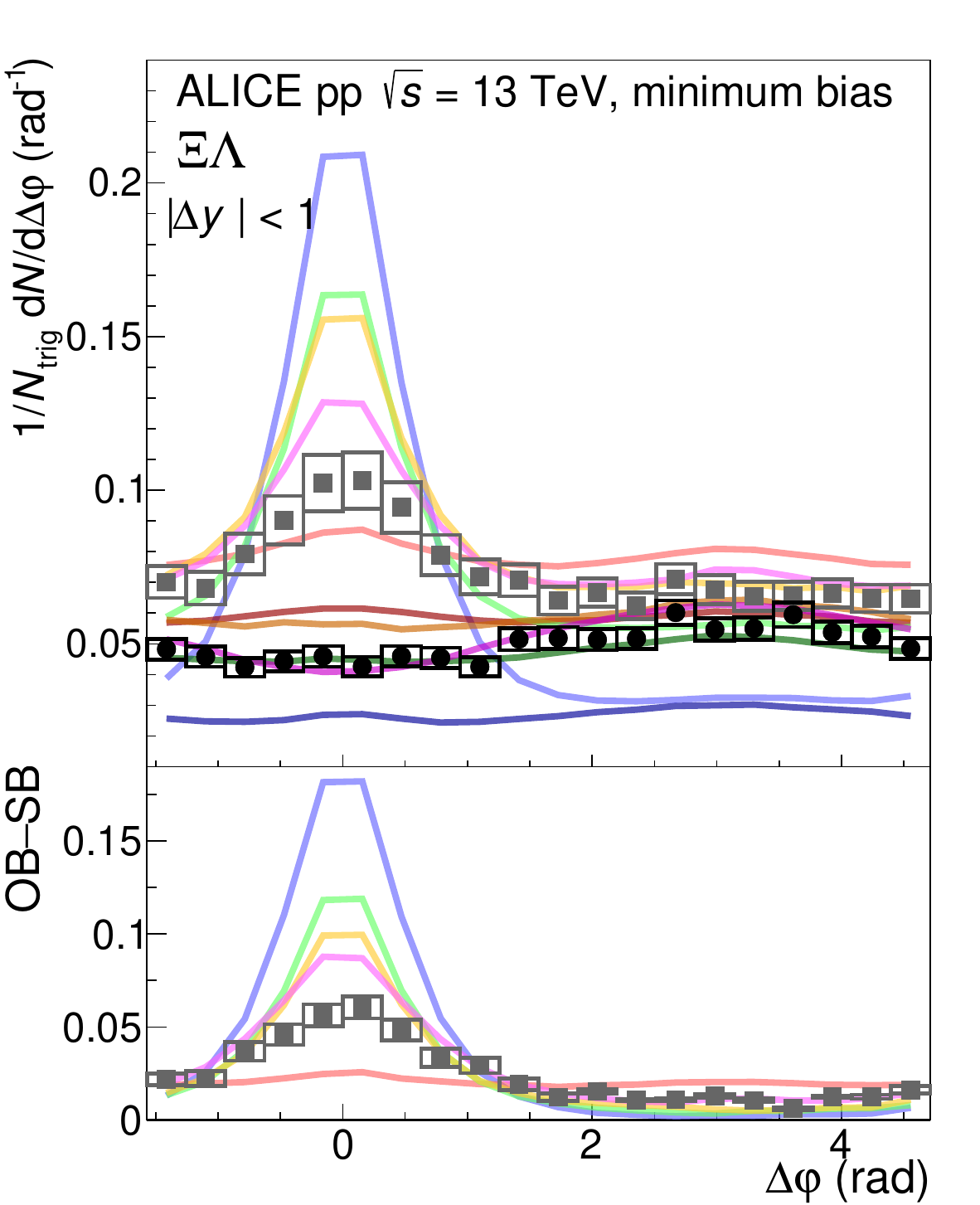}~\includegraphics[width = 0.33\textwidth]{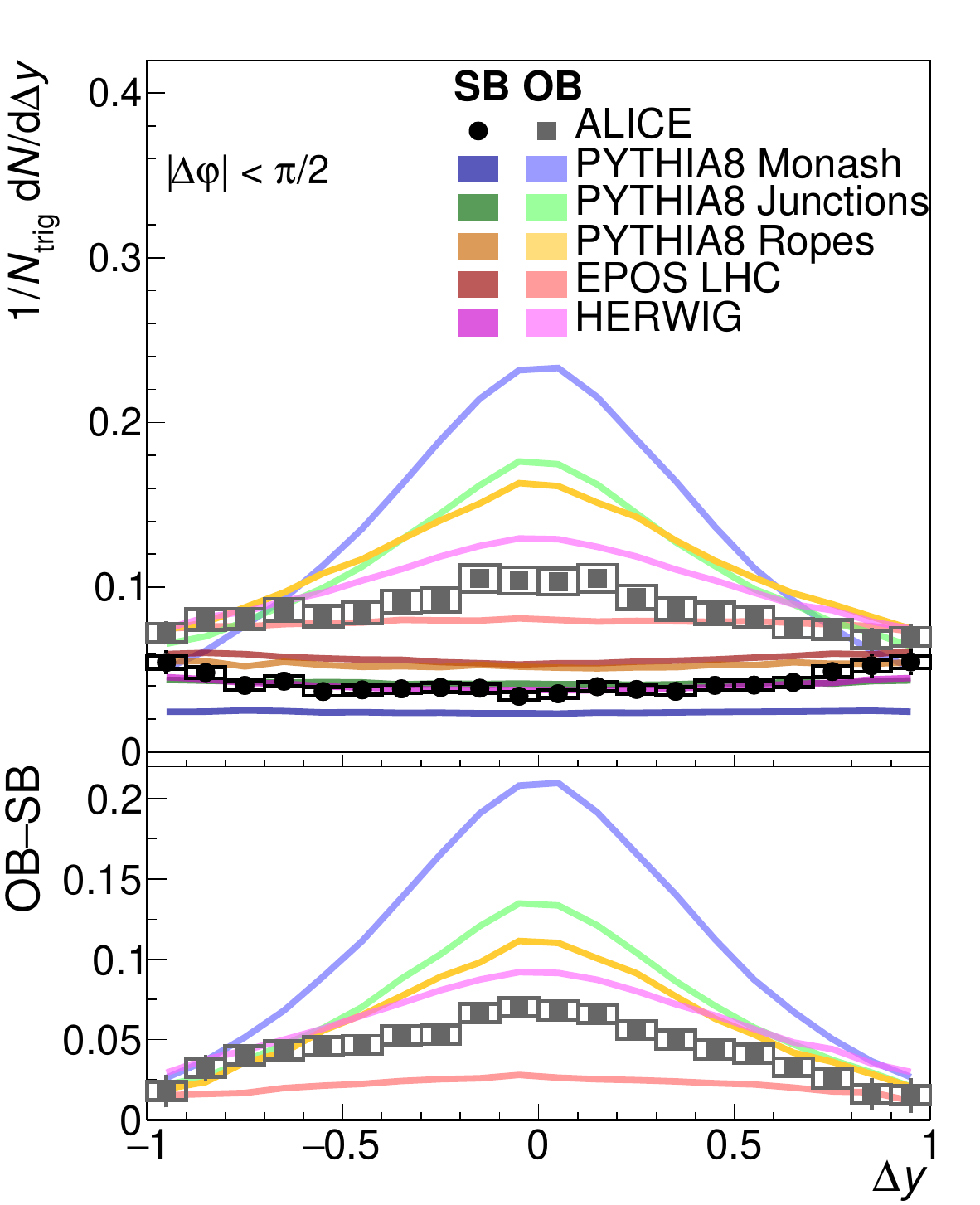}~\includegraphics[width = 0.33\textwidth]{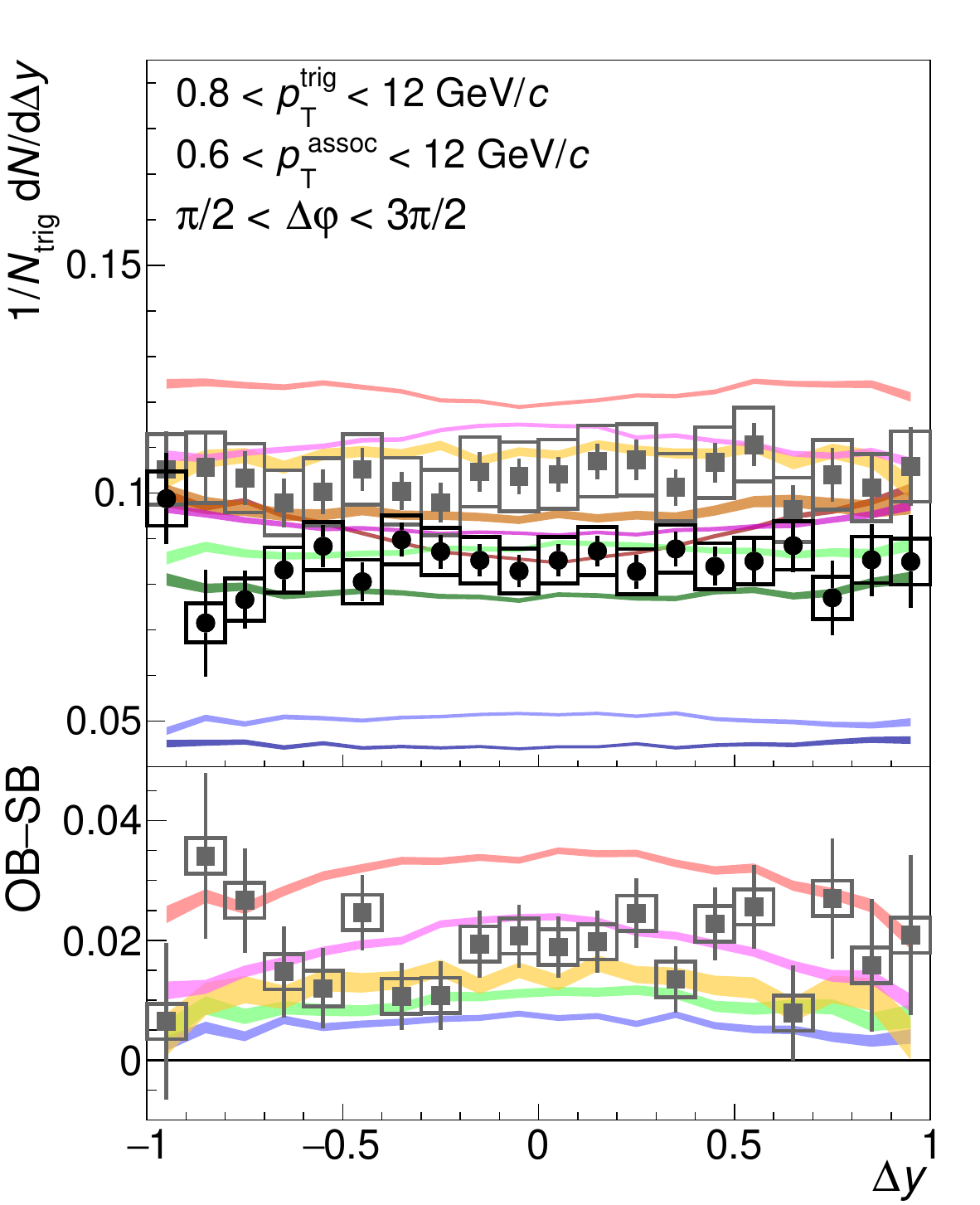}
    \end{center}
    \caption{$\Xi^-\overline{\Lambda}$ and $\Xi^-\Lambda$ (and charge conjugate) correlation functions projected onto $\Delta\varphi$ ($|\Delta y| < 1$, left), the near-side on $\Delta y$ ($|\Delta\varphi| < \pi/2$, middle), and the away-side on $\Delta y$ ($\pi/2 < \Delta\varphi < 3\pi/2$, right). Opposite-baryon-number ($\Xi^{-}\overline{\Lambda}+\overline{\Xi}^{+}\Lambda$) correlations are shown in grey squares, the same-baryon-number ($\Xi^{-}\Lambda+\overline{\Xi}^{+}\overline{\Lambda}$) correlations are black circles; the OB--SB difference is displayed in the bottom panels.  Statistical and systematic uncertainties are represented by bars and boxes, respectively.  The ALICE data are compared with the following models: \pythia{} Monash tune (blue), \pythia{} with junctions enabled (green), \pythia{} with junctions and ropes (yellow), \eposlhc{} (red), and \herwig{} (pink).}
    \label{fig:xilambda}
\end{figure}

\begin{figure}[tbp]
    \begin{center}
        \includegraphics[width = 0.33\textwidth]{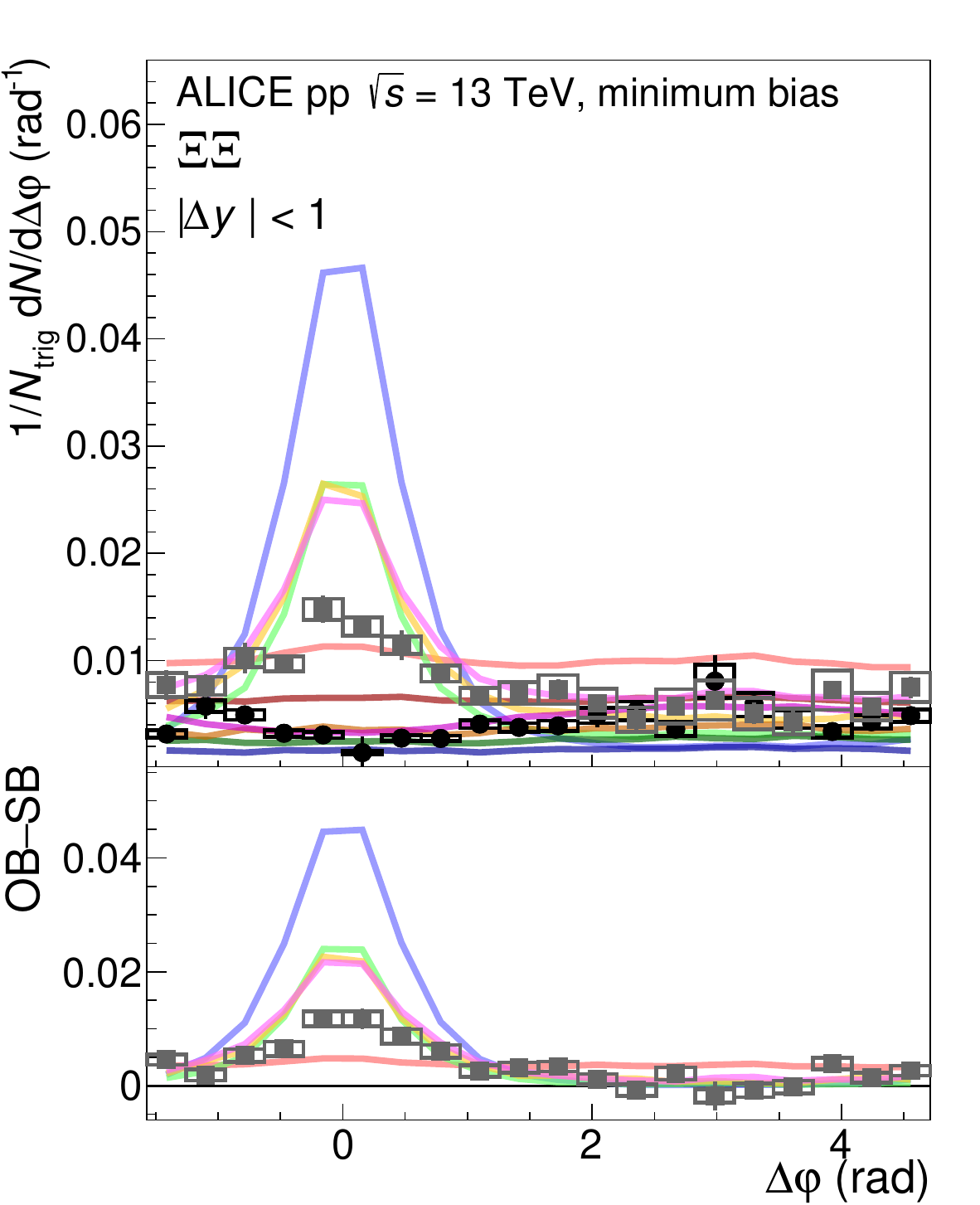}~\includegraphics[width = 0.33\textwidth]{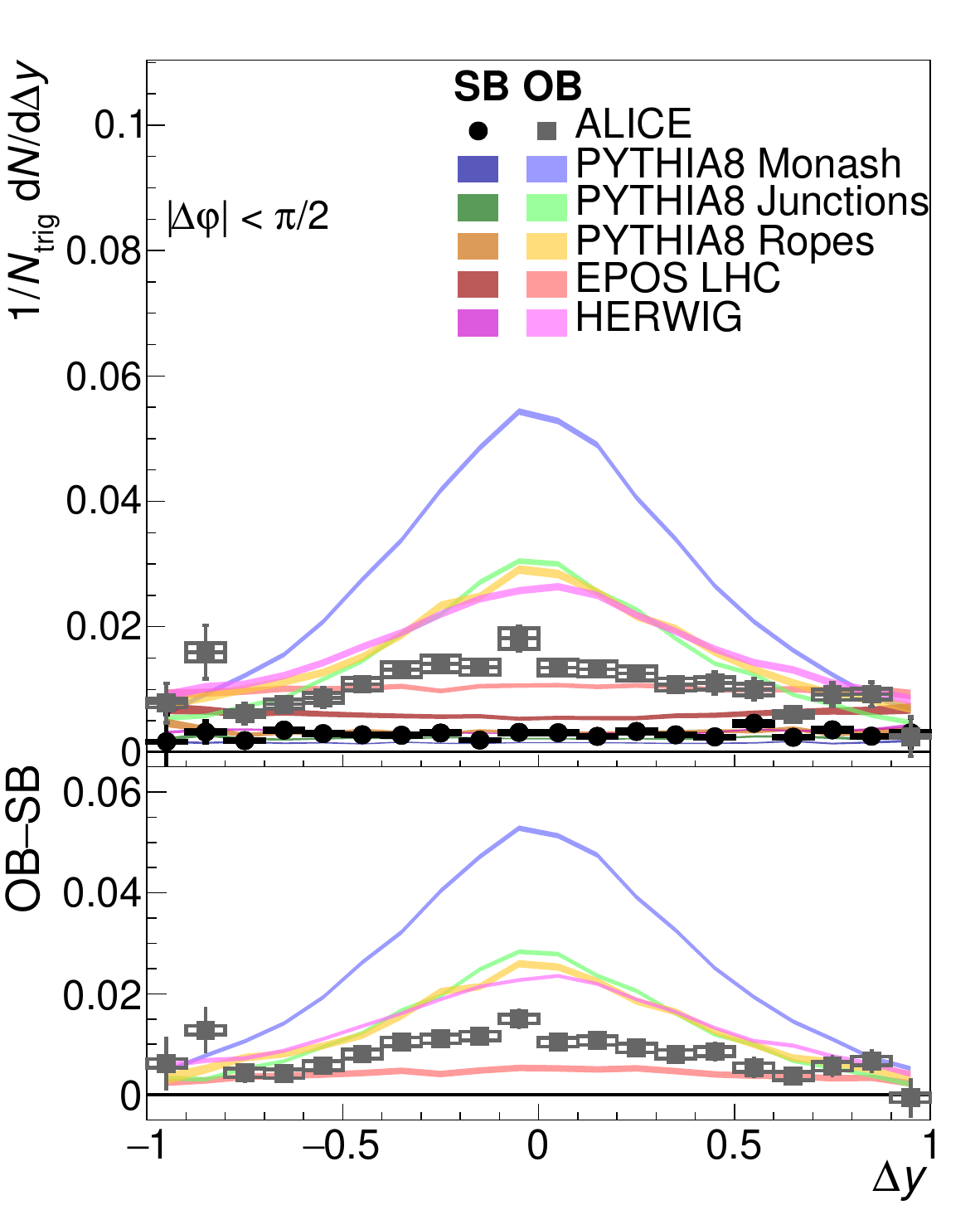}~\includegraphics[width = 0.33\textwidth]{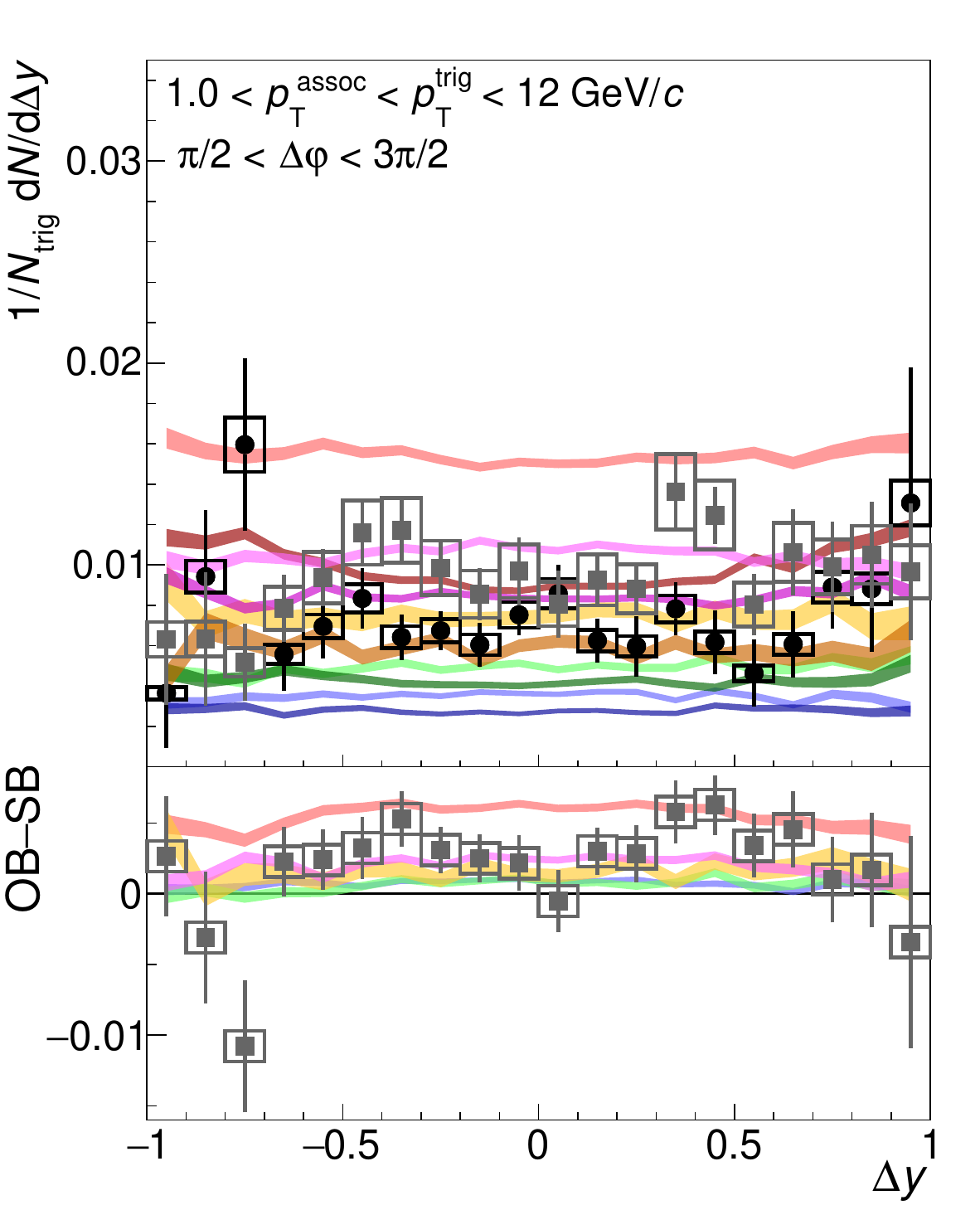}
    \end{center}
    \caption{$\Xi^-\overline{\Xi}^+$ and $\Xi^-\Xi^-$ (and charge conjugate) correlation functions projected onto $\Delta\varphi$ ($|\Delta y| < 1$, left), the near-side on $\Delta y$ ($|\Delta\varphi| < \pi/2$, middle), and the away-side on $\Delta y$ ($\pi/2 < \Delta\varphi < 3\pi/2$, right). Opposite-baryon-number ($\Xi^{-}\overline{\Xi}^{+}+\overline{\Xi}^{+}\Xi^{-}$) correlations are shown in grey squares, the same-baryon-number ($\Xi^{-}\Xi^{-}+\overline{\Xi}^{+}\overline{\Xi}^{+}$) correlations are black circles; the OB--SB difference is displayed in the bottom panels.  Statistical and systematic uncertainties are represented by bars and boxes, respectively.  The ALICE data are compared with the following models: \pythia{} Monash tune (blue), \pythia{} with junctions enabled (green), \pythia{} with junctions and ropes (yellow), \eposlhc{} (red), and \herwig{} (pink).}
    \label{fig:xixi}
\end{figure}

The per-trigger yields of pions and kaons associated with $\Xi$ baryons are shown in Figure~\ref{fig:corr_mesons}.  In Figures~\ref{fig:xipi} and~\ref{fig:xiK} the $\Xi\pi$ and $\Xi\mathrm{K}$ correlations are projected on $\Delta\varphi$ (for $|\Delta y| < 1$) and $\Delta y$ on the near-side ($|\Delta\varphi| < \pi/2$) and away-side ($\pi/2 < \Delta\varphi < 3\pi/2$). Furthermore, the differences between the opposite-sign and same-sign correlations, related to the balance function as described above, are shown.  Figure~\ref{fig:corr_baryons} reports the associated yields in $(\Delta y, \Delta\varphi)$ of protons, $\Lambda$, and $\Xi$ baryons, and the projections onto the $\Delta\varphi$ and $\Delta y$ axes can be found in Figures~\ref{fig:xip}--\ref{fig:xixi}. 

The (unsubtracted) per-trigger yields, $Y(\Delta y,\Delta\varphi)$, contain correlations not only related to the production of the quantum numbers of the $\Xi$ baryon but also other aspects of the events which produce correlations in momentum space.  These are especially evident in the $\Xi\pi$ correlations (Figures~\ref{fig:corr_mesons} and~\ref{fig:xipi}), since pions make up the majority of the particles produced in \pp{} collisions, and particularly in $\Xi^{-}\pi^{-}$ correlations where the hadrons do not share any quark--antiquark pairs.  A significant flat pedestal, known as the underlying event, is apparent in these correlations and represents pion production uncorrelated with the $\Xi$ baryon.  However, there are also peaks on the near-side and away-side, localised around $(\Delta y,\Delta\varphi) = (0,0)$ and $\Delta\varphi=\pi$.  These peaks can be attributed to $\Xi$ and $\pi$ production within the same and back-to-back (mini)jets, respectively.  The differences between the $\Xi^{-}\pi^{+}$ and $\Xi^{-}\pi^{-}$ correlations, most easily visible in Figure~\ref{fig:xipi}, can be attributed to the presence of a d$\overline{\rm d}$ pair in the former combination and a dd pair in the latter, as well as to the effects of electric charge balancing in the $\Xi^{-}\pi^{+}$ correlation function. Note that the same argument holds for the charge conjugate pairs, $\overline{\Xi}^{+}\pi^{-}$ and $\overline{\Xi}^{+}\pi^{+}$, which are included in the reported correlation functions.  

In contrast with the pions, the $\Xi\mathrm{K}$ correlations (Figures~\ref{fig:corr_mesons} and~\ref{fig:xiK}) show rather different behaviour.  There is little or no near-side peak observed in the same-charge $\Xi^{-}\kam$ correlations, which share no quark--antiquark pairs, demonstrating the difficulty of producing three strange quarks in separate processes within the same (mini)jet.  Meanwhile, a significant away-side peak is observed, indicative of the overall production rate of kaons in (mini)jet fragmentation.  The opposite-sign $\Xi^{-}\kap$ correlations, on the other hand, show a strong near-side peak, which can be attributed to the correlation of the s$\overline{\rm s}$ pair created in a single process.  The near-side peak in $\Xi^{-}\kap$ correlations is wider than the corresponding $\Xi^{-}\pip$ peak, which may be attributed to early-stage diffusion of strange quarks prior to hadronisation. 

The effects of baryon number production, conservation, and dispersion can be observed in the $\Xi^{-}\mathrm{p}$ correlations (Figures~\ref{fig:corr_baryons} and~\ref{fig:xip}).  As has been demonstrated in previous analyses~\cite{ALICE:2016jjg}, the production of multiple baryons (or multiple antibaryons) within the same (mini)jet is highly disfavoured, leading to a depletion of the correlation function on the near-side.  No such dip is seen in the baryon--antibaryon correlations.  Similarly, the near-side dip in same-baryon-number correlations and near-side peak in opposite-baryon-number correlations can be seen in the $\Xi\Lambda$ (Figures~\ref{fig:corr_baryons} and~\ref{fig:xilambda}) and $\Xi\Xi$ (Figures~\ref{fig:corr_baryons} and~\ref{fig:xixi}) correlations.  The near-side peak is also observed to be broader in $\Xi-$baryon correlations than $\Xi-$meson, which may indicate the early decoupling and diffusion of baryon number, as was observed for strangeness above.  When going from $\Xi^{-}\overline{\mathrm{p}}$ to $\Xi^{-}\overline{\Lambda}$ to $\Xi^{-}\overline{\Xi}^{+}$ correlations, a sequential enhancement in the amount of correlated particle production (i.e.\ the magnitude of the near-side peak relative to the level of the underlying event) is observed.  This is likely related to the production of zero, one, or two s$\overline{\rm s}$ pairs, respectively.  

\subsection{Comparison to Monte Carlo models}

The $\Delta y$ and $\Delta\varphi$ projections in Figures~\ref{fig:xipi}--\ref{fig:xiK} and~\ref{fig:xip}--\ref{fig:xixi} are compared to calculations from Monte Carlo event generators.  Three configurations of \pythia{} are shown: (1) the default Monash tune, (2) when junctions are turned on, and (3) when both ropes and junctions are enabled.  Predictions from (4) \herwig{} and (5) \eposlhc{} are also included.  

\begin{figure}[tb]
    \begin{center}
    \includegraphics[width = 0.9\textwidth]{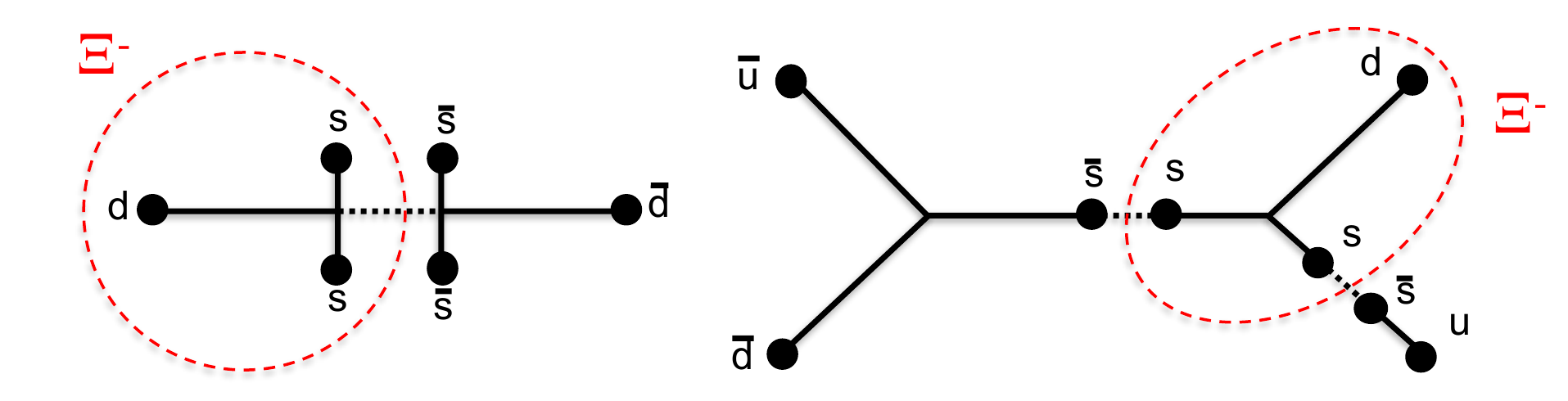}
    \end{center}
    \caption{A schematic representation of the production of a $\Xi^-$ baryon via the mechanisms of diquark string breaking (left) and baryon junction formation (right).  Figure taken from Ref.~\cite{Adolfsson:2020dhm}.}
    \label{fig:strings}
\end{figure}

The \pythia{} Monash tune~\cite{Skands:2014pea} includes the colour reconnection mechanism~\cite{Christiansen:2015yqa}, in which individual MPI systems may be colour-connected.  \pythia{} does not include the formation of a QGP-like medium, but colour reconnection allows it to capture some of the multiplicity-dependent and flow-like signals observed in data, like the enhancement of the p/$\pi$ ratio at intermediate \pt{}~\cite{OrtizVelasquez:2013ofg}.  In the default tune of \pythia{}, baryons are produced through diquark string breaking, which generally leads to an underestimation of the baryon yields.  The junction mechanism~\cite{Bierlich:2015rha} introduces another process for producing baryons that may also lead to significantly different correlation structures.  As shown in the illustration in Figure~\ref{fig:strings}, the diquark breaking mechanism implies that the production of a $\Xi^{-}$ baryon normally leads to the production of a $\overline{\Xi}^{+}$, while junctions allow more possibilities for the strangeness and baryon number to be balanced with singly-strange baryons ($\Lambda$) and strange mesons (K).  The introduction of colour ropes~\cite{colourrope,Bierlich:2014xba} allows strings to fuse together in high-density regions, thus increasing the string tension, leading to the enhanced production of strange mesons (and baryons, via junctions). For the $\Xi^{-}(\overline{\Xi}^{+})$ baryons studied here, ropes are expected to enhance the overall strange quark production rates but have similar correlation patterns as strings since the hadronisation mechanism still proceeds via diquark breaking.  

\herwig{} includes its own colour reconnection model, as well as a mechanism for non-perturbative gluon splittings which is necessary to describe strangeness enhancement through the production of s$\overline{\rm s}$ pairs.  The core--corona model, \eposlhc{}~\cite{Pierog:2013ria}, also includes parton showers from string-breaking processes in the corona, but the core is a dense, thermalised, many-particle system in the grand-canonical limit.  In particular, quantum numbers are only conserved globally, not locally, in the core.  The multiplicity dependence of various observables is described in \eposlhc{} by the transition from low-multiplicity corona-dominated collisions to high-multiplicity systems with a large core component.  

\textbf{Magnitude of the underlying event}\\
The overall magnitude of the underlying event in the $\Xi\pi$ correlations (Figure~\ref{fig:xipi}) is described well by the default \pythia{} Monash tune, likely because this model has been tuned to underlying event measurements at the LHC~\cite{Skands:2014pea}.  \pythia{} Monash also shows decent agreement with the level of the underlying event for the other associated hadron species, which is expected because this tune can describe the single-particle spectra~\cite{ALICE:2015ial}.  The rope and junction tunes of \pythia{} also show reasonable agreement with the level of the underlying event, with the exception of the $\Xi\mathrm{p}$ correlations, where the models significantly overpredict the number of protons per trigger $\Xi$ baryon.  On the other hand, while \eposlhc{} has also been tuned to match the inclusive spectra of identified hadrons~\cite{ALICE:2015ial}, it surprisingly does not capture the magnitude of the underlying event in $\Xi-$hadron correlations.  It is hypothesised that in \eposlhc{}, $\Xi$ production occurs mainly in events with higher-than-average multiplicity, and the multiplicity dependence of the single-particle spectra may not be fully captured by the model. \herwig{} also appears not to accurately describe the level of the underlying event, particularly in $\Xi\pi$ correlations.

\textbf{Magnitude and shape of the near-side peak in unsubtracted correlations}\\
The shape of the near- and away-side jet peaks in the $\Xi\pi$ correlation functions is also described well by \pythia{} and \eposlhc{}, indicating that the fragmentation of (mini)jets is relatively well modelled.  However, the Monte Carlo descriptions of the near-side peak shapes for the other particle species are less accurate.  In particular, \pythia{} and \herwig{} tend to predict much more significant near-side peaks in $\Xi\mathrm{K}$, $\Xi\Lambda$, and $\Xi\Xi$ correlations than are observed in the data, indicating that strangeness is overproduced in (mini)jet fragmentation in the models.  The near-side peaks in these correlations in \pythia{} and \herwig{} are also narrower than those observed in the data, indicating that strange quark diffusion is more significant than anticipated by the model.  The near-side dip observed in same-baryon-number correlations (e.g.\ $\Xi^{-}\mathrm{p}$) is also challenging to reproduce in models, although can be captured by the junction tune of \pythia{} (as well as the rope tune, which includes the baryon junction mechanism).  Like \pythia{}, \herwig{} also tends to predict stronger correlations on the near-side than what is observed in data.  While \eposlhc{} can describe the shape of the near-side peak in $\Xi\pi$ correlations well, it predicts broader near-side structures for all other particle pairs, likely because local strangeness conservation is only implemented in the corona but not the core.  Similar evidence for longer-range correlations than predicted by string-breaking models has also been observed in measurements of the event-by-event fluctuations in net-proton production in central \PbPb{} collisions, which are in fact consistent with expectations from global baryon number conservation~\cite{ALICE:2019nbs}.  Finally, in \eposlhc{} and \herwig{} an unusual ``wing'' structure in $\Delta y$ is visible in the same-sign correlations (seen most clearly in the away-side $\Delta y$ projection in e.g. Figure~\ref{fig:xiK}), where a significant number of particle pairs are produced at larger $\Delta y$, indicating that the momentum distribution and conservation in the event are not precisely modelled.  

\textbf{Opposite-sign minus same-sign correlations}\\
\herwig{}, and to a lesser extent \pythia{} (except the Monash tune) and \eposlhc{}, are able to describe the OS--SS $\Xi\pi$ correlations, which again demonstrates that charge balancing and (mini)jet fragmentation are well modelled.  \pythia{} and \herwig{} are also able to describe most aspects of the $\Xi\mathrm{K}$ correlations (Figure~\ref{fig:xiK}); in particular the configuration with ropes shows good agreement with the OS--SS correlations, indicating that the electric charge and strangeness balancing is accurately modelled.  For all other particle species pairs between strange baryons and mesons, \eposlhc{} tends to produce a flat balance function in disagreement with the data, because in the core component strangeness is only conserved globally and local conservation of quantum numbers is not implemented.  To describe the $\Xi\mathrm{p}$ OB--SB correlations, the baryon junction mechanism is necessary, and the junction and rope tunes of \pythia{} describe the experimentally-measured $\Xi\mathrm{p}$ balance well.  However, for the other baryon--baryon pairs, $\Xi\Lambda$ and $\Xi\Xi$, all the \pythia{} tunes vastly overpredict the strength of the OB--SB correlation.  

\subsection{Balance function integrated yields\label{sec:yields}}

\begin{figure}[tbp]
    \includegraphics[width = 0.49\textwidth]{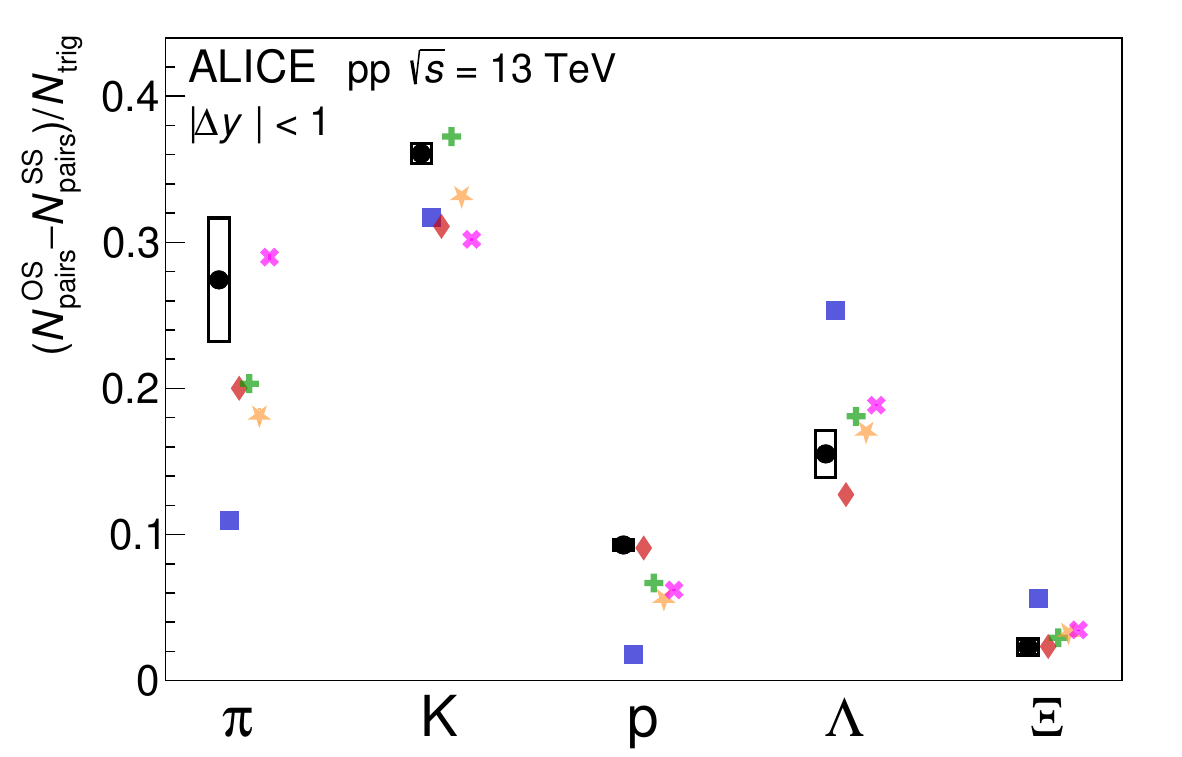}\\
    \includegraphics[width = 0.49\textwidth]{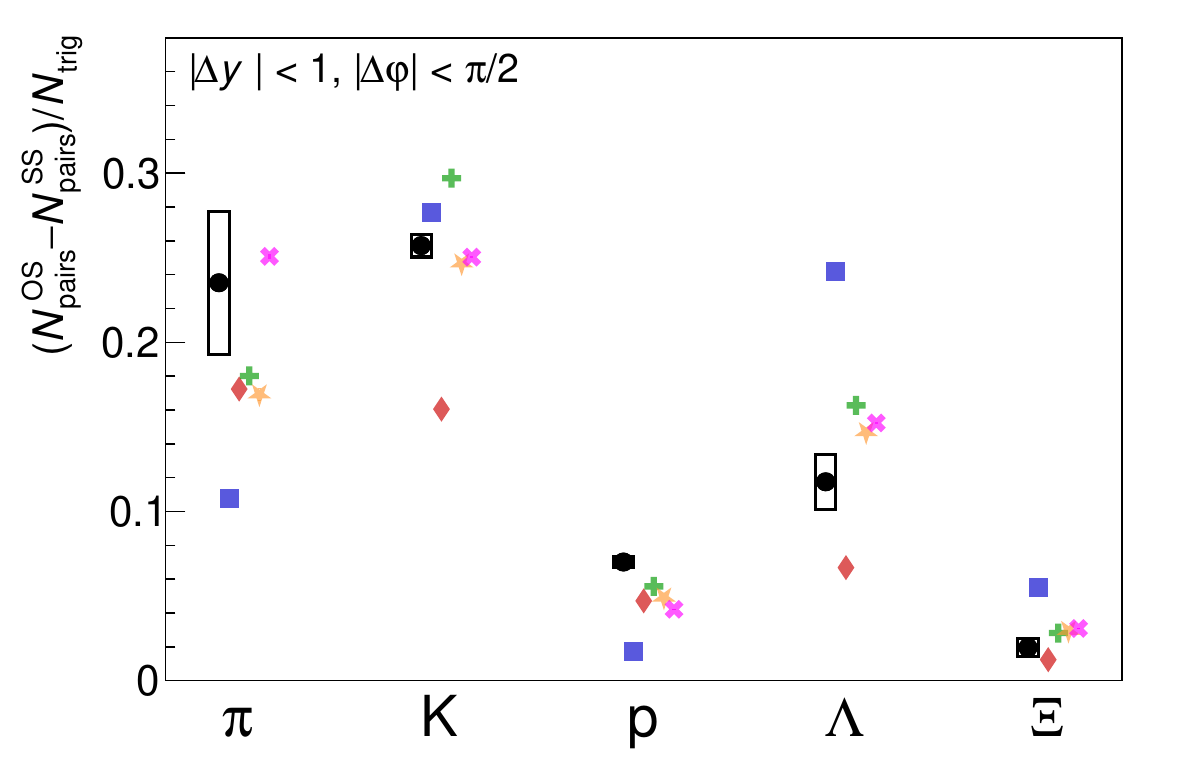}~
    \includegraphics[width = 0.49\textwidth]{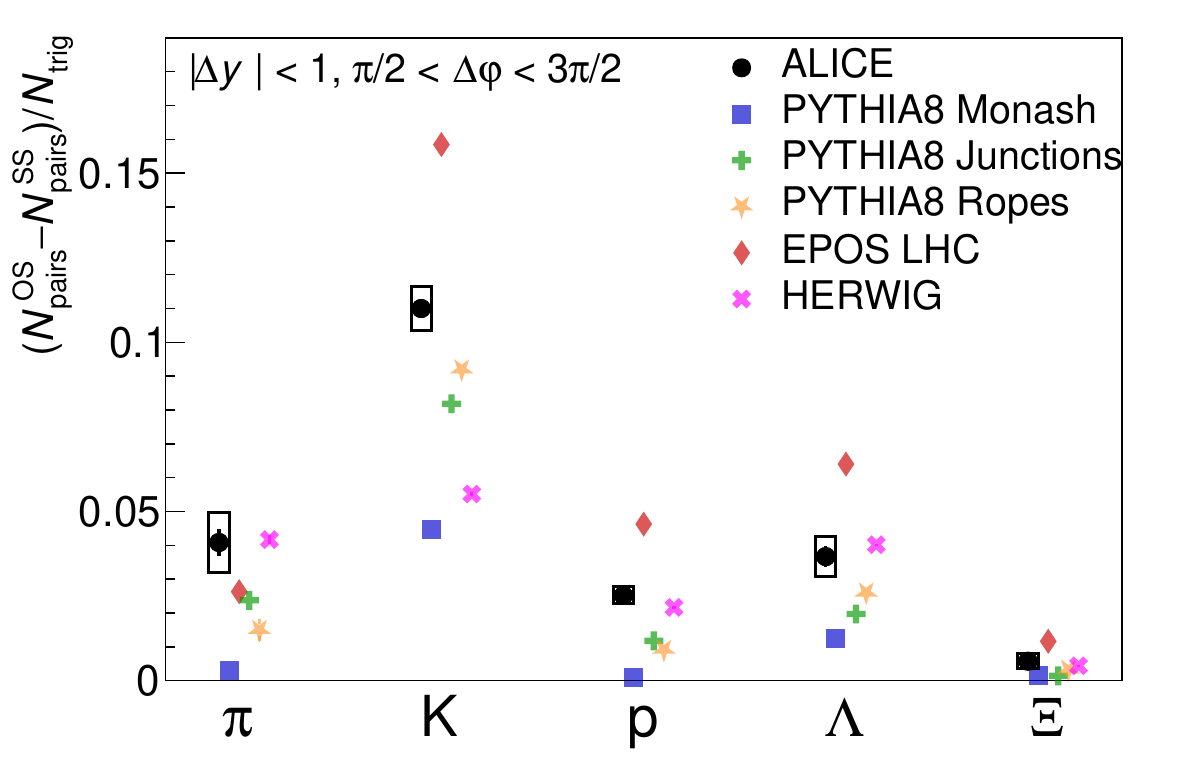}
    \caption{The OS--SS and OB--SB per-trigger yields for $\Xi\pi$, $\Xi\mathrm{K}$, $\Xi\mathrm{p}$, $\Xi\Lambda$, and $\Xi\Xi$ correlations are shown when integrated over all phase space (top), on the near-side ($|\Delta\varphi| < \pi/2$, bottom left), and on the away-side ($\pi/2 < \Delta\varphi < 3\pi/2$, bottom right).  Statistical and systematic uncertainties are represented by bars and boxes, respectively.  The ALICE data are compared with the following models: \pythia{} Monash tune (blue), \pythia{} with junctions enabled (green), \pythia{} with junctions and ropes (yellow), \eposlhc{} (red), and \herwig{} (pink). The statistical uncertainties on the model predictions are smaller than the marker sizes.}
    \label{fig:yields}
\end{figure}

The OS--SS and OB--SB yields were integrated over $\Delta\varphi$ and $\Delta y$ to probe the overall balancing of charges, both over the full correlation function ($-\pi/2 < \Delta\varphi < 3\pi/2$, $|\Delta y| < 1$) and over the near- ($-\pi/2 < \Delta\varphi < \pi/2$) and away-sides ($\pi/2 < \Delta\varphi < 3\pi/2$).  The results are compared to predictions from \pythia{}, \eposlhc{}, and \herwig{} in Figure~\ref{fig:yields}.  The large systematic uncertainty on the OS--SS pion yield is due to small relative uncertainties on the large number of pions produced per event, leading to larger absolute uncertainties compared with the other associated particle species.  

Furthermore, the balancing of the quantum numbers associated to the $\Xi$ baryon can be estimated through the following sums: 
\begin{itemize}
\item[$\bullet$] electric charge: net-$Q$ = net-$\pi$ + net-K + net-$\Xi$ \textminus~net-p,
\item[$\bullet$] strangeness: net-$S$ = net-K + net-$\Lambda$ + $2\times$net-$\Xi$,
\item[$\bullet$] baryon number: net-$B$ = net-p + net-$\Lambda$ + net-$\Xi$.
\end{itemize}
where the net-$\pi$ yield is obtained from the OS--SS ($\Xi^{-}\pi^{+} - \Xi^{-}\pi^{-}$) integral, and similarly for the other associated particle species.  Note that the net-p goes into the net-$Q$ calculation with a relative minus sign, since the quantity measured in Figure~\ref{fig:xip} is the opposite-baryon-number minus same-baryon-number difference ($\Xi^{-}\overline{\mathrm{p}} - \Xi^{-}\mathrm{p}$), which is the reverse of the opposite-charge minus same-charge difference ($\Xi^{-}\mathrm{p} - \Xi^{-}\overline{\mathrm{p}}$).  Recall that all correlations with respect to the $\Xi^{-}$ baryon, and their corresponding integrals, include the charge conjugate pairs (correlations with trigger $\overline{\Xi}^{+}$ baryons).  

If all the charges balancing the trigger $\Xi^{-}$ baryon ($Q = -1$, $S = -2$, $B = +1$) were captured in the phase space of this measurement, then the resulting associated net quantum numbers would be net-$Q = +1$, net-$S = +2$, net-$B = -1$.  In Table~\ref{tab:quantumnumbers} the balancing charges are expressed as percentages of these expected values (e.g.\ the associated net strangeness has been divided by 2, and the associated net baryon number is written as a positive number).  

\begin{table}[tbp]
\caption{The balancing of the quantum numbers opposite a trigger $\Xi$ baryon are calculated from the integral of the $\Xi$--hadron OS--SS and OB--SB correlation functions.  The experimentally-measured electric charge, strangeness, and baryon number balance are displayed as percentages of the expected values (net-$Q = +1$, net-$S = +2$, net-$B = -1$), and compared with the following models: \pythia{} Monash tune, \pythia{} with junctions enabled, \pythia{} with junctions and ropes, \eposlhc{}, and \herwig{}.  The uncertainties on the experimental data are statistical and systematic, respectively.  } \centering
\begin{tabular}{ l | c c c }\label{tab:quantumnumbers}
   & Electric charge $Q$ & Strangeness $S$ & Baryon number $B$ \\ \hline
  Data & $(56.5\pm 0.9\pm0.8)\%$ & $(28.1\pm 0.4\pm0.4)\%$ & $(27.1\pm 0.7\pm0.6)\%$ \\
  \pythia{} Monash & 47\% & 34\% & 33\% \\
  \pythia{} junctions & 54\% & 31\% & 28\% \\
  \pythia{} junctions \& ropes & 49\% & 28\% & 26\% \\
  \eposlhc{} & 44\% & 24\% & 24\% \\
  \herwig{} & 56\% & 28\% & 28\% \\
\end{tabular}
\end{table}

None of the values reported in Table~\ref{tab:quantumnumbers} reach 100\%, indicating that approximately 40\% of the balancing charges (estimated from the net-$Q$ calculation) are not captured within the kinematic phase space of the measurement, mainly due to the restricted $\pt{}$ range (and less so due to the $y$ and $\eta$ constraints).  Additionally, significant fractions of the strangeness are contained within the un-measured $\mathrm{K}^{0}$ and $\overline{\rm K}^{0}$ mesons, while some of the baryon number balance is hidden in the neutrons and $\Sigma^{\pm}$, $\Xi^{0}$, and $\overline{\Xi}^{0}$ baryons.  

As is expected for a model which produces $\Xi$ baryons through diquark breaking, \pythia{} predicts that most of the balancing (anti)baryon number is carried by the $\Lambda$ baryons (see Figure~\ref{fig:strings}) and very little is balanced by the protons.  With the introduction of the junction mechanism (also included in the rope configuration) in \pythia{}, the model predictions move closer to the experimental results.  The string breaking and cluster hadronisation mechanisms in \herwig{} are also able to capture the OS--SS yields reasonably well.  It is particularly interesting to note that, while \eposlhc{} does not capture the strangeness balance on the near-side or away-side individually (seen most prominently in the net-kaon yield), it does agree with the data over the full $\Delta\varphi$ range.  This is consistent with expectations since \eposlhc{} respects global, but not local, strangeness conservation in the core.  

\begin{figure}[tbp]
    \includegraphics[width = 0.33\textwidth]{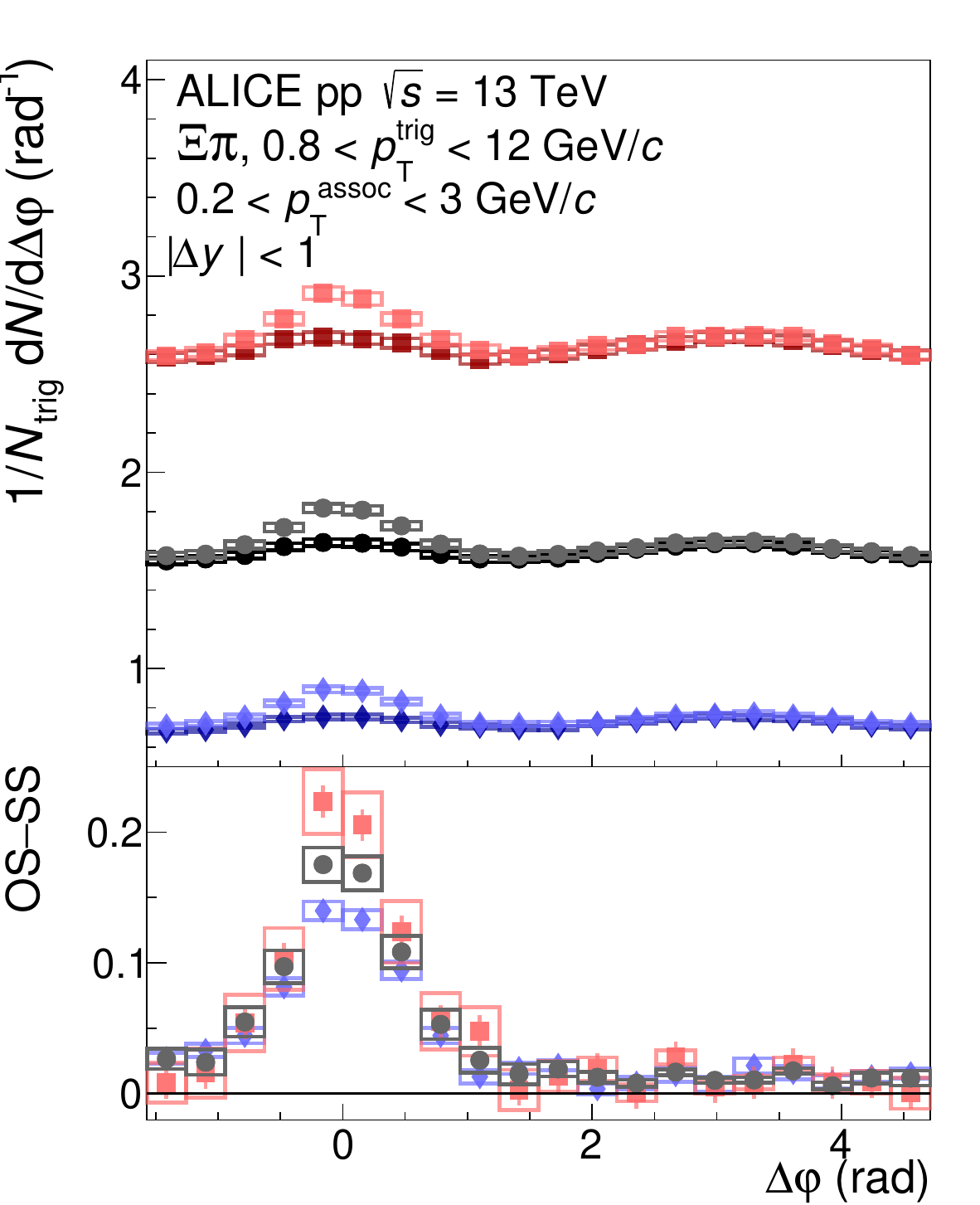}~\includegraphics[width = 0.33\textwidth]{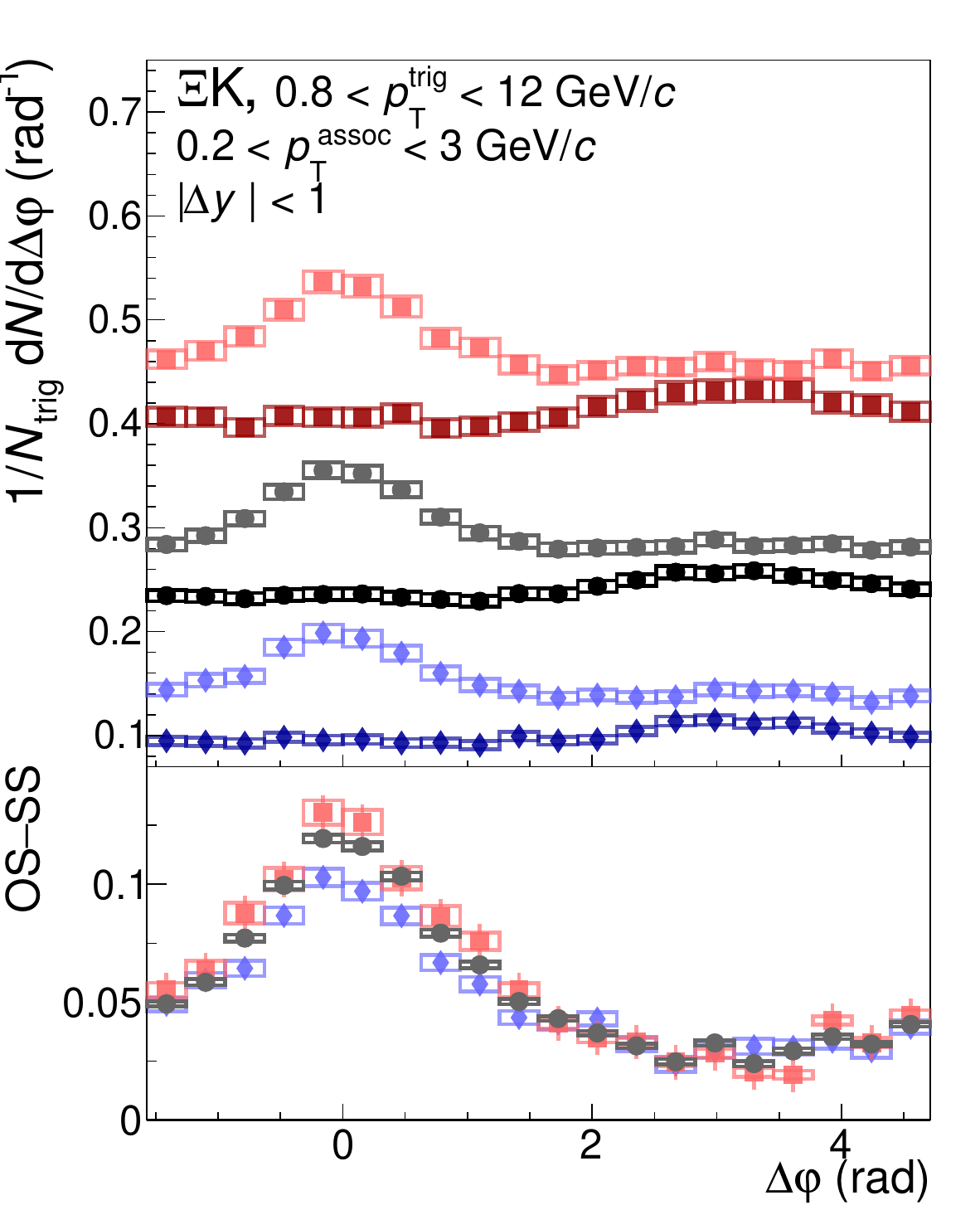}~\includegraphics[width = 0.33\textwidth]{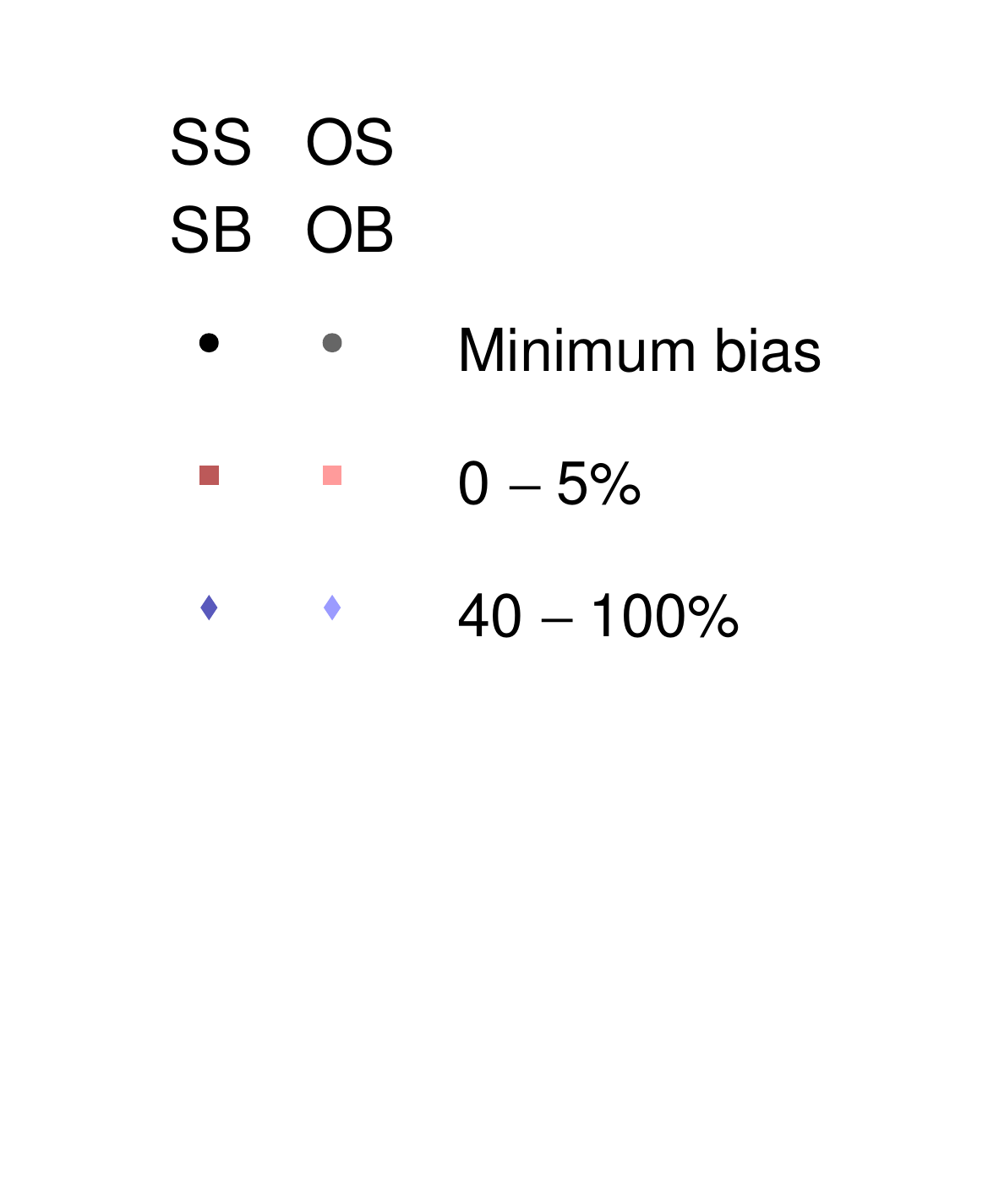}\\
    \includegraphics[width = 0.33\textwidth]{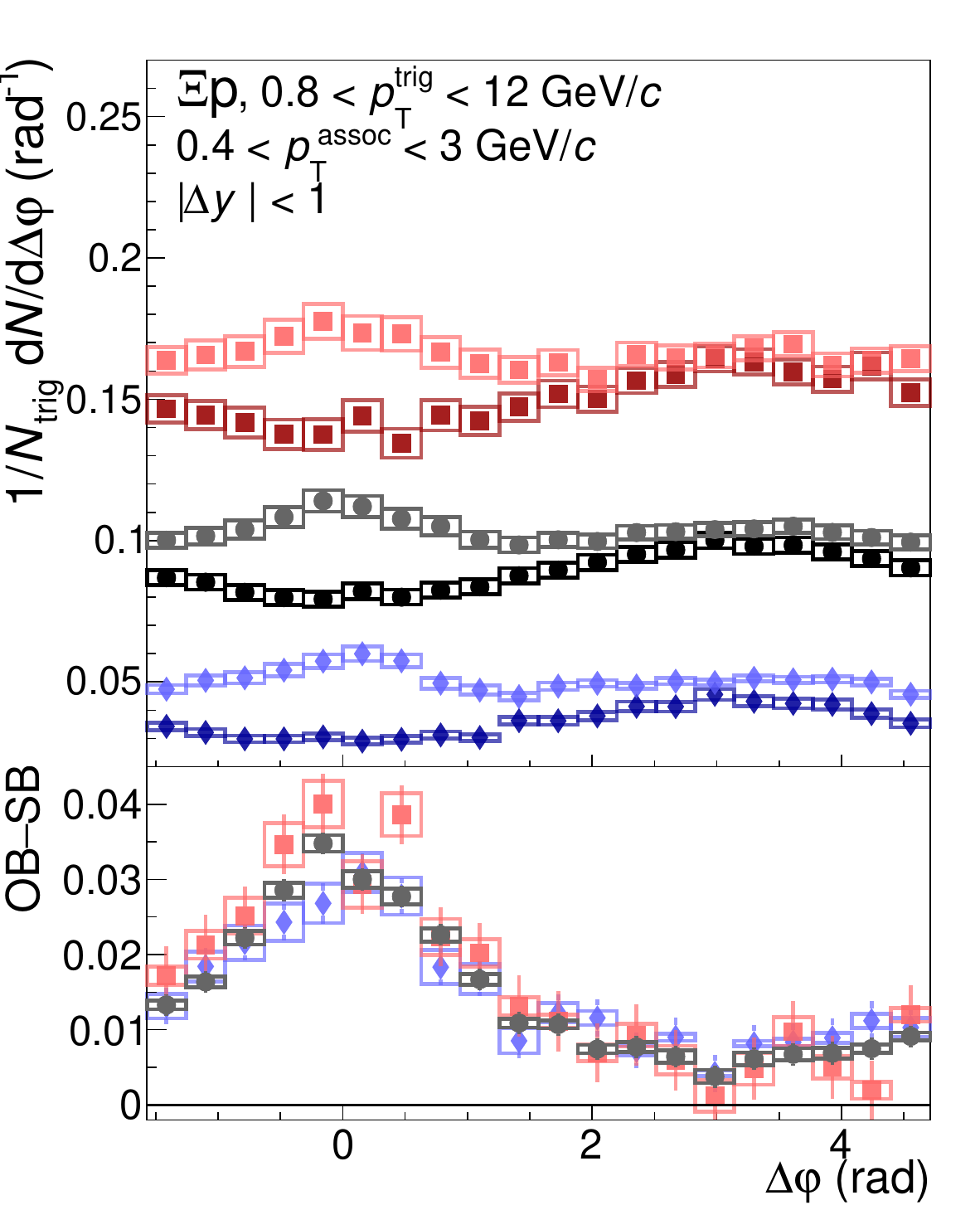}~\includegraphics[width = 0.33\textwidth]{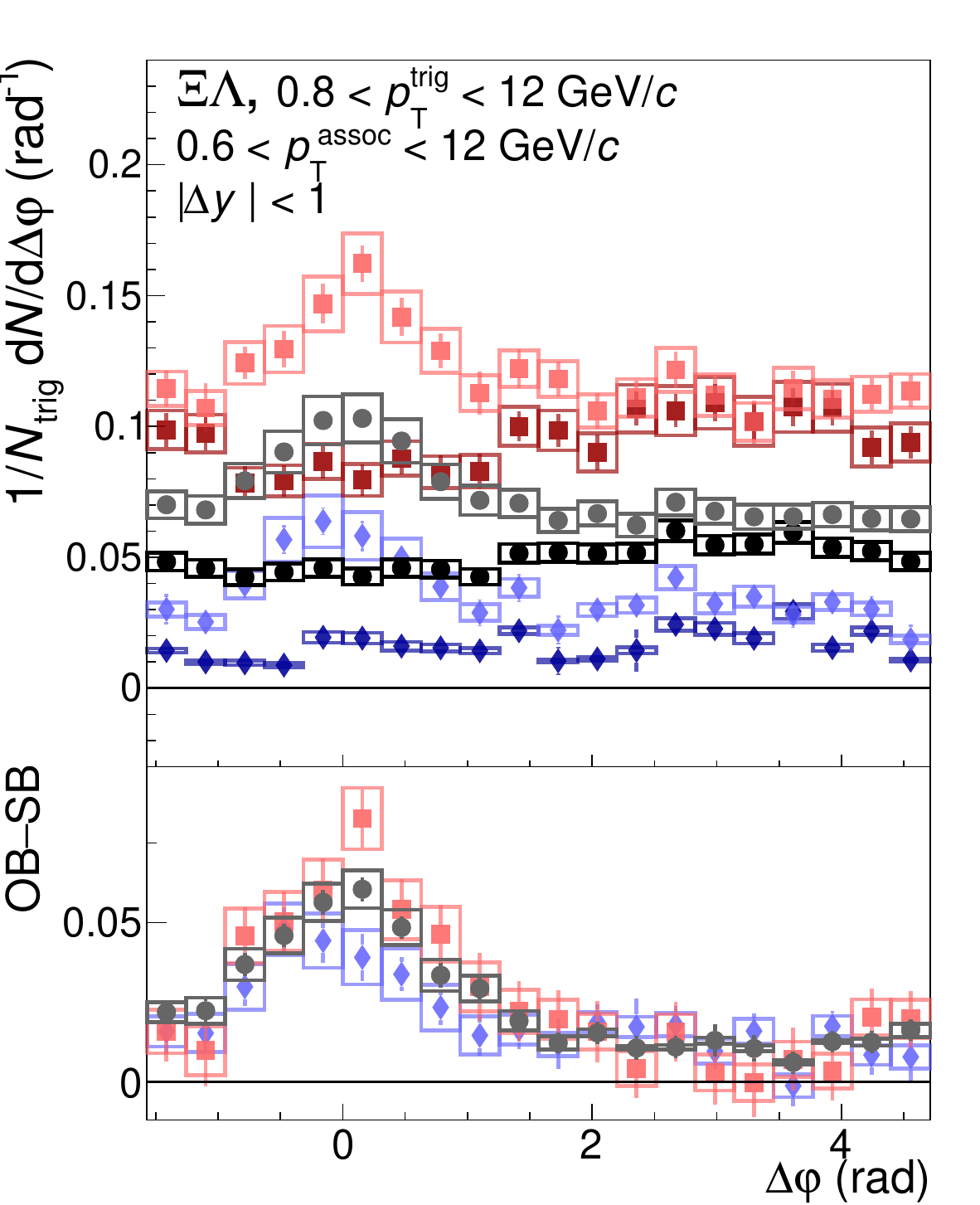}~\includegraphics[width = 0.33\textwidth]{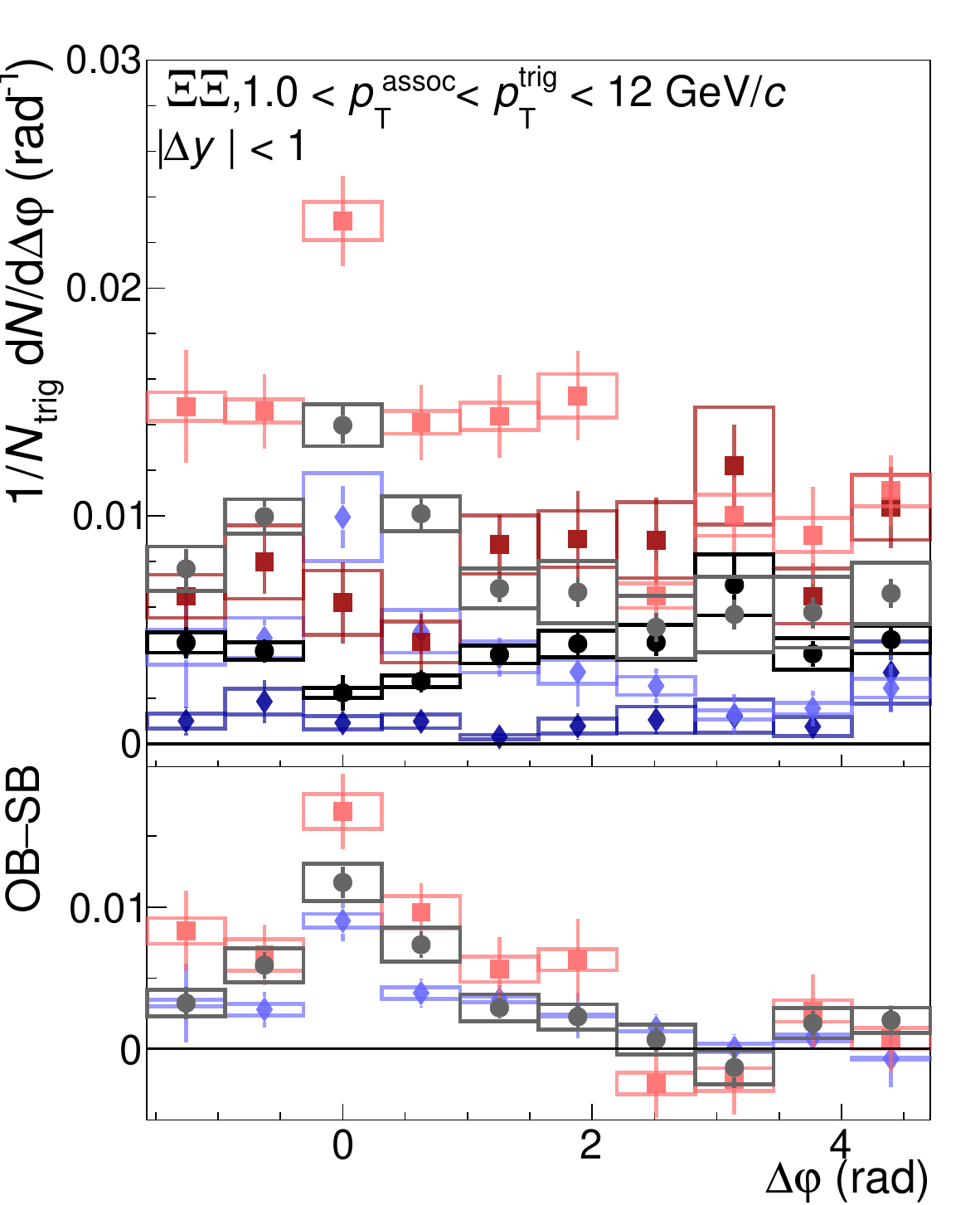}
    \caption{The $\Xi\pi$ (top left), $\Xi$K (top centre), $\Xi$p (bottom left), $\Xi\Lambda$ (bottom centre), and $\Xi\Xi$ (bottom right) correlation functions are shown for minimum bias (black), high-multiplicity ($0-5\%$, red), and low-multiplicity ($40-100\%$, blue) events, projected onto $\Delta\varphi$ ($|\Delta y| < 1$). In the top panels, opposite-sign correlations are shown in light markers, the same-sign correlations are shown with darker markers.  In the bottom panels, the OS--SS or OB--SB difference is shown in each multiplicity interval.  Statistical and systematic uncertainties are represented by bars and boxes, respectively.}
    \label{fig:multdepPhi}
\end{figure}

\subsection{Multiplicity dependence\label{sec:multdep}}

The per-trigger yields, OS--SS differences, and integrated yields were also studied as a function of the event multiplicity.  The multiplicity dependence of $\Xi-$hadron correlations may provide better discriminatory power than the multiplicity-inclusive results between the various pictures of the strangeness production mechanisms: In a statistical thermal model, there is expected to be a lifting of canonical strangeness suppression with increasing multiplicity, which would lead to weaker $\Xi\Lambda$ and $\Xi\Xi$ correlations.  A similar outcome could be observed in a core--corona model where the core fraction is anticipated to increase with increasing multiplicity.  In a string-breaking model such as \pythia{}, on the other hand, the production mechanism for strangeness production is expected to be independent of multiplicity (a consequence of jet universality).  However, with the addition of the junction and rope mechanisms, which both arise in high-string-density regions and thus play a larger role in high-multiplicity collisions, there may be a change in the dominant mechanism for producing baryons, particularly strange baryons, as a function of multiplicity.  

\begin{figure}[tbp]
    \includegraphics[width = 0.33\textwidth]{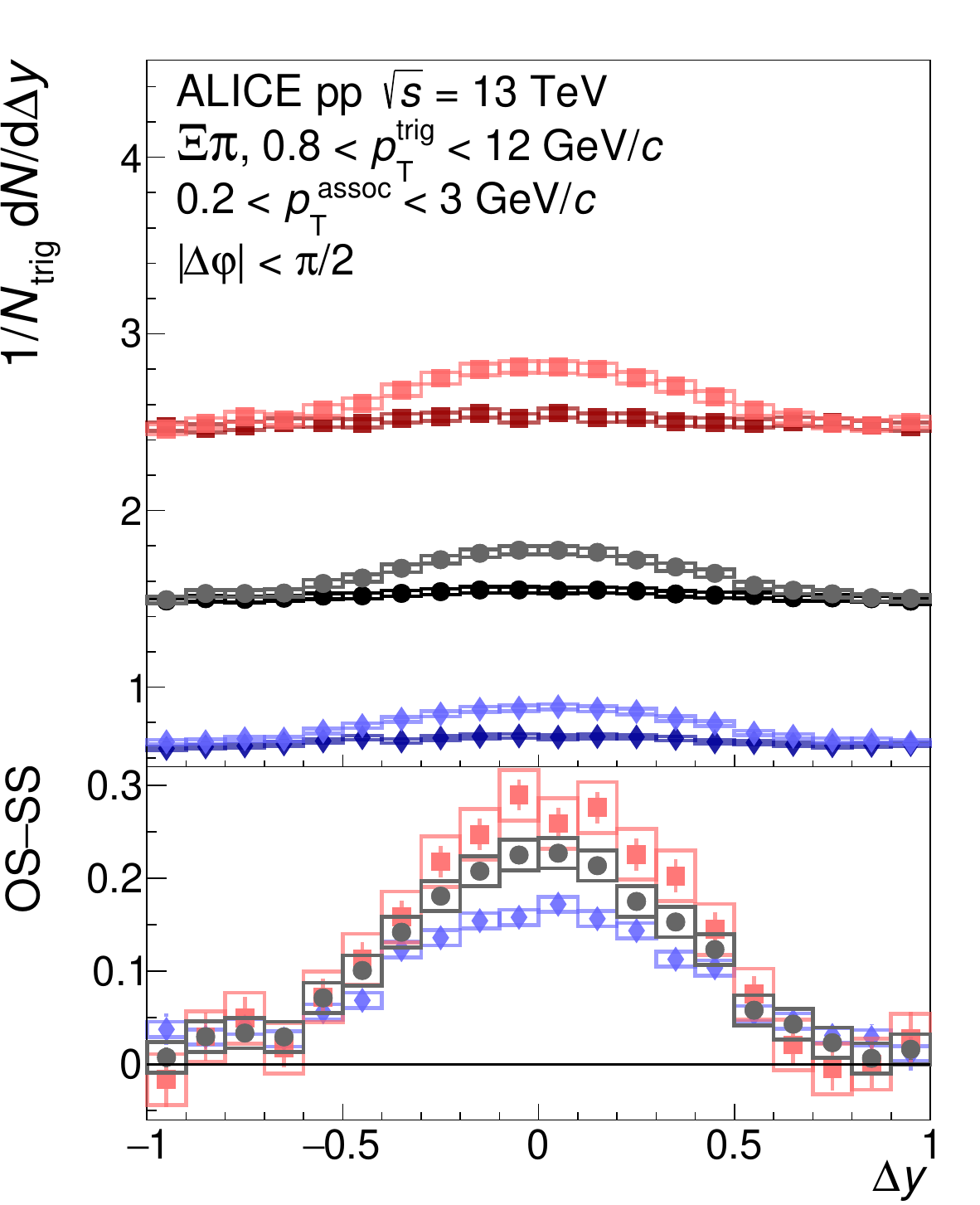}~\includegraphics[width = 0.33\textwidth]{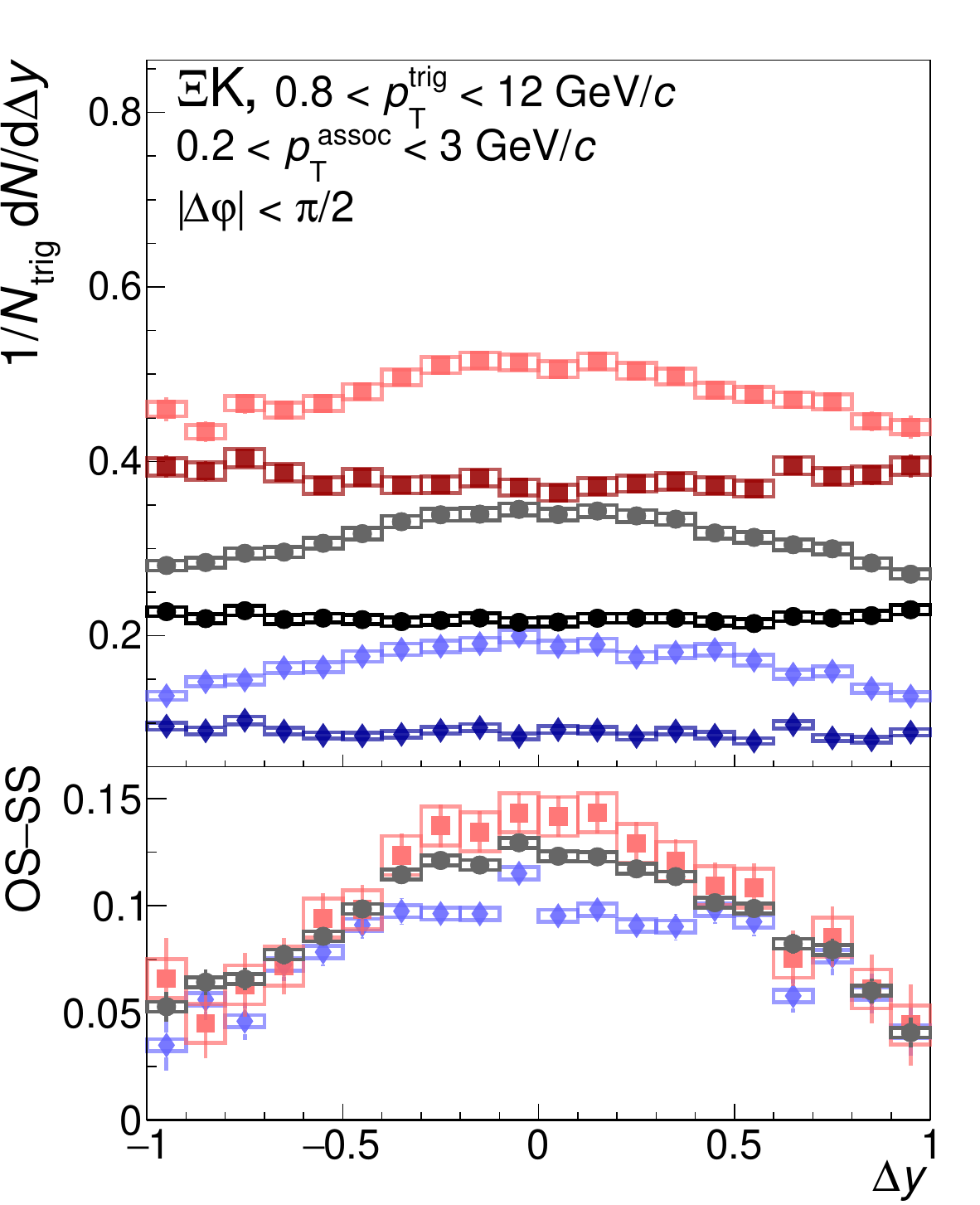}~\includegraphics[width = 0.33\textwidth]{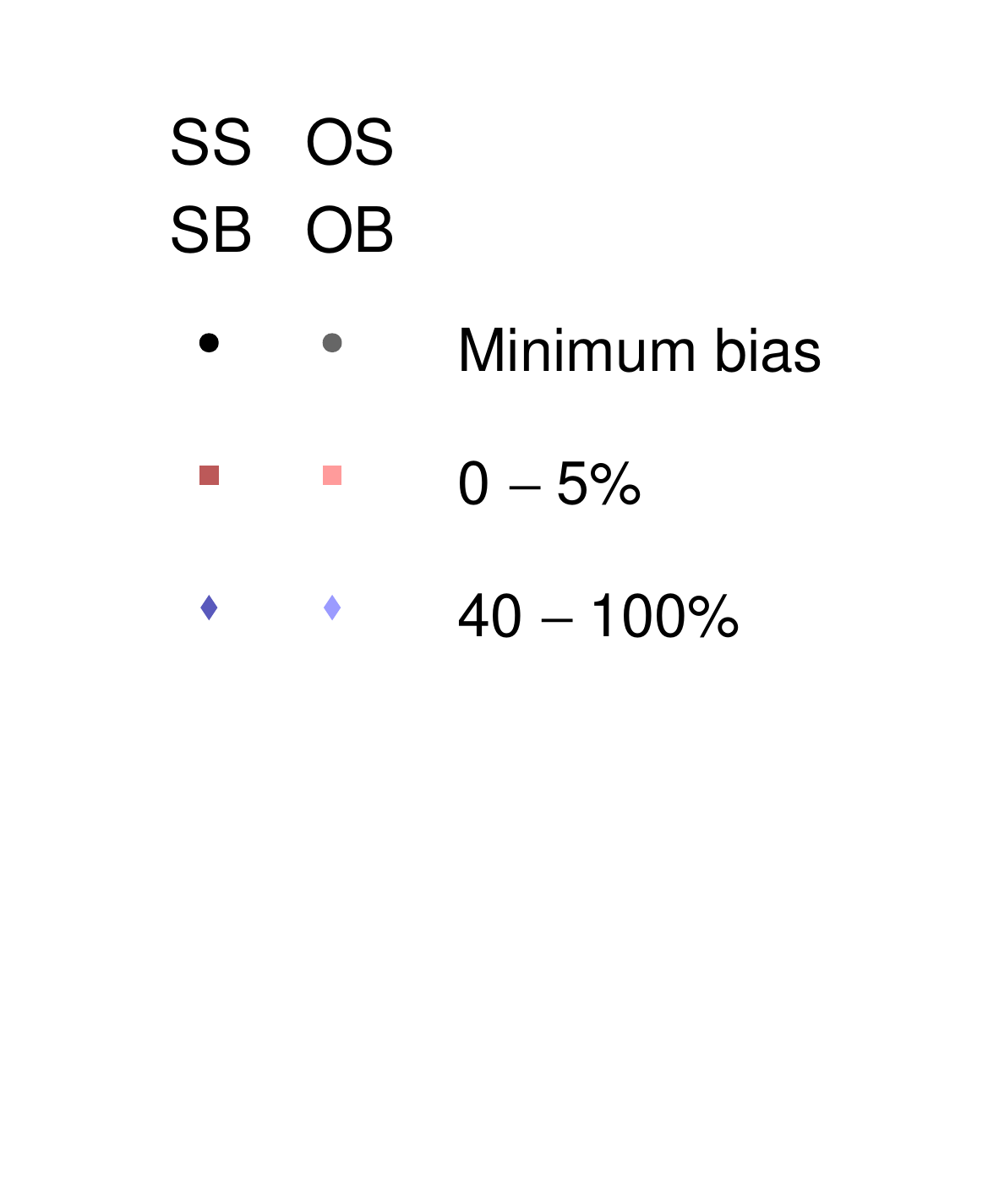}\\
    \includegraphics[width = 0.33\textwidth]{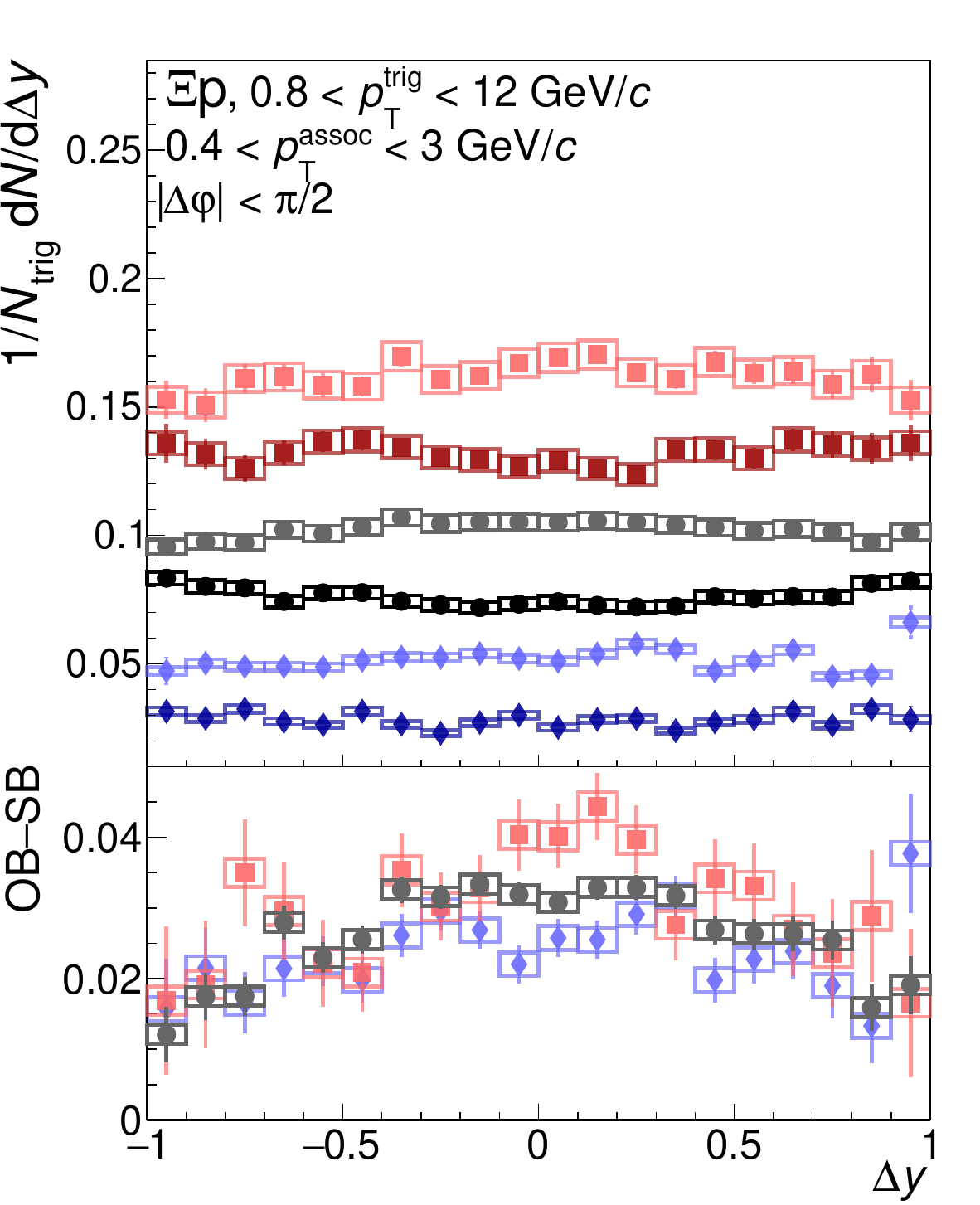}~\includegraphics[width = 0.33\textwidth]{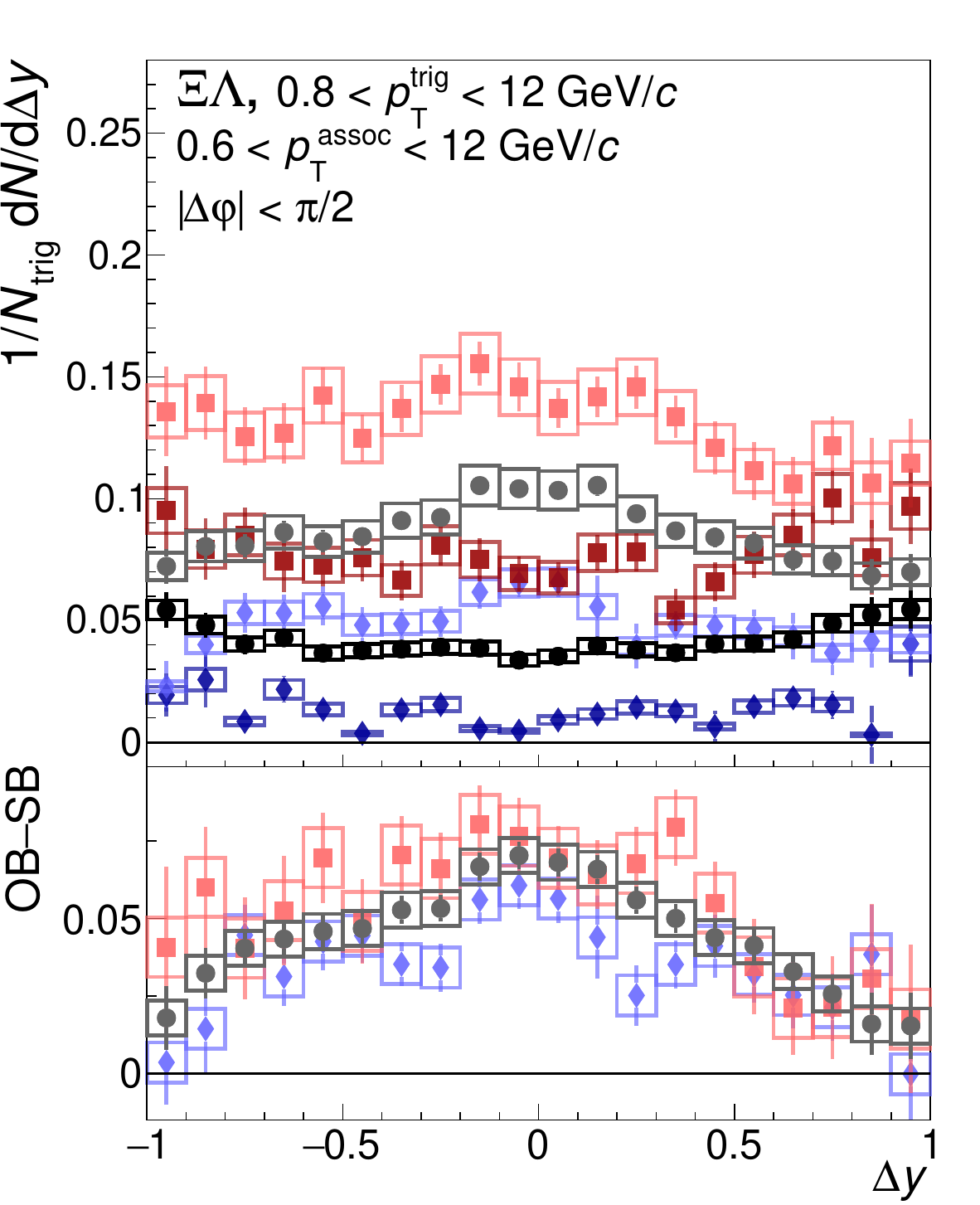}~\includegraphics[width = 0.33\textwidth]{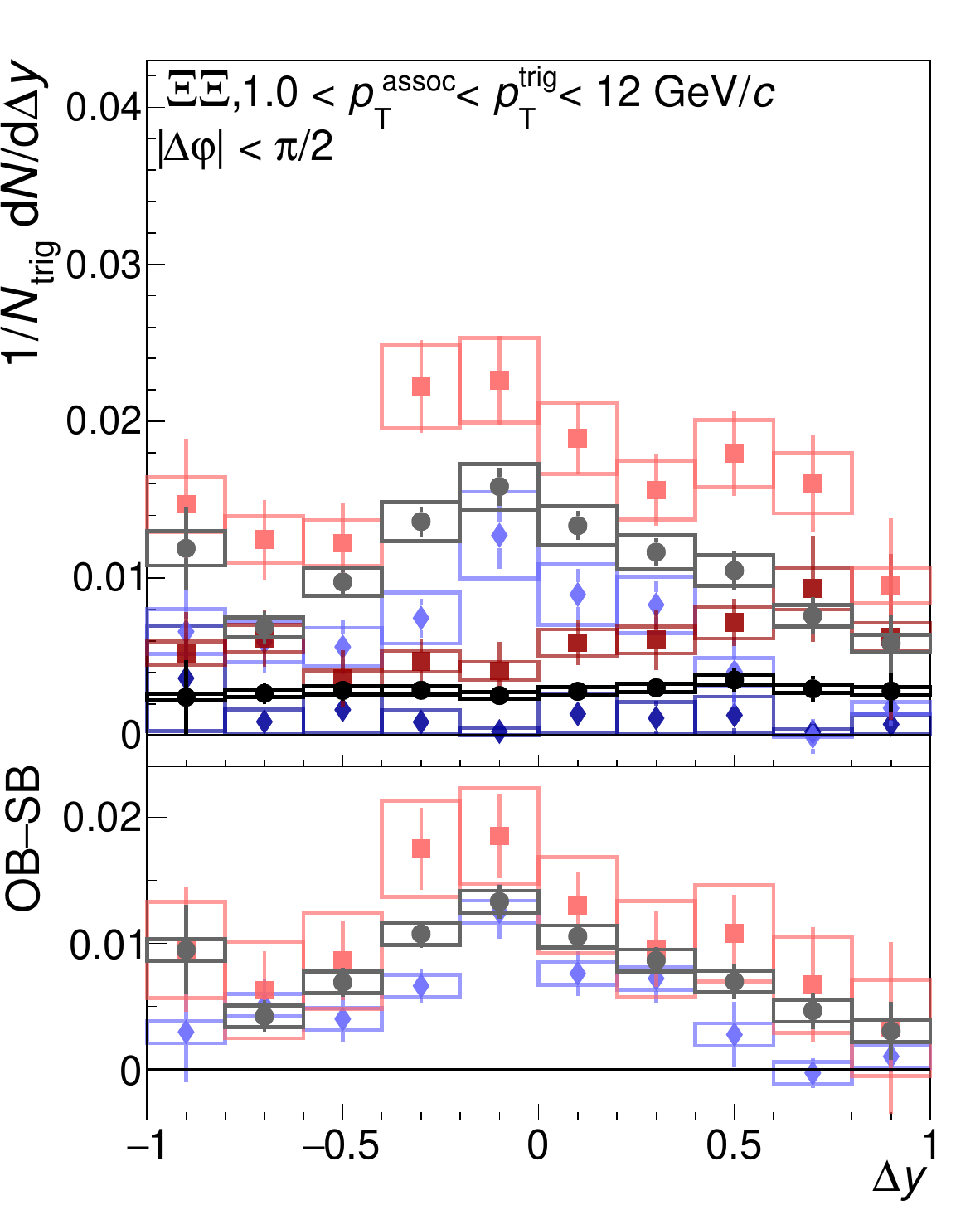}
    \caption{The $\Xi\pi$ (top left), $\Xi$K (top centre), $\Xi$p (bottom left), $\Xi\Lambda$ (bottom centre), and $\Xi\Xi$ (bottom right) correlation functions are shown for minimum bias (black), high-multiplicity ($0-5\%$, red), and low-multiplicity ($40-100\%$, blue) events, projected onto $\Delta y$ on the near side ($|\Delta\varphi| < \pi/2$). In the top panels, opposite-sign correlations are shown in light markers, the same-sign correlations are shown with darker markers.  In the bottom panels, the OS--SS or OB--SB difference is shown in each multiplicity interval.  Statistical and systematic uncertainties are represented by bars and boxes, respectively.}
    \label{fig:multdepY}
\end{figure}

\begin{figure}[tbp]
    \begin{center}
    \includegraphics[width = 0.49\textwidth]{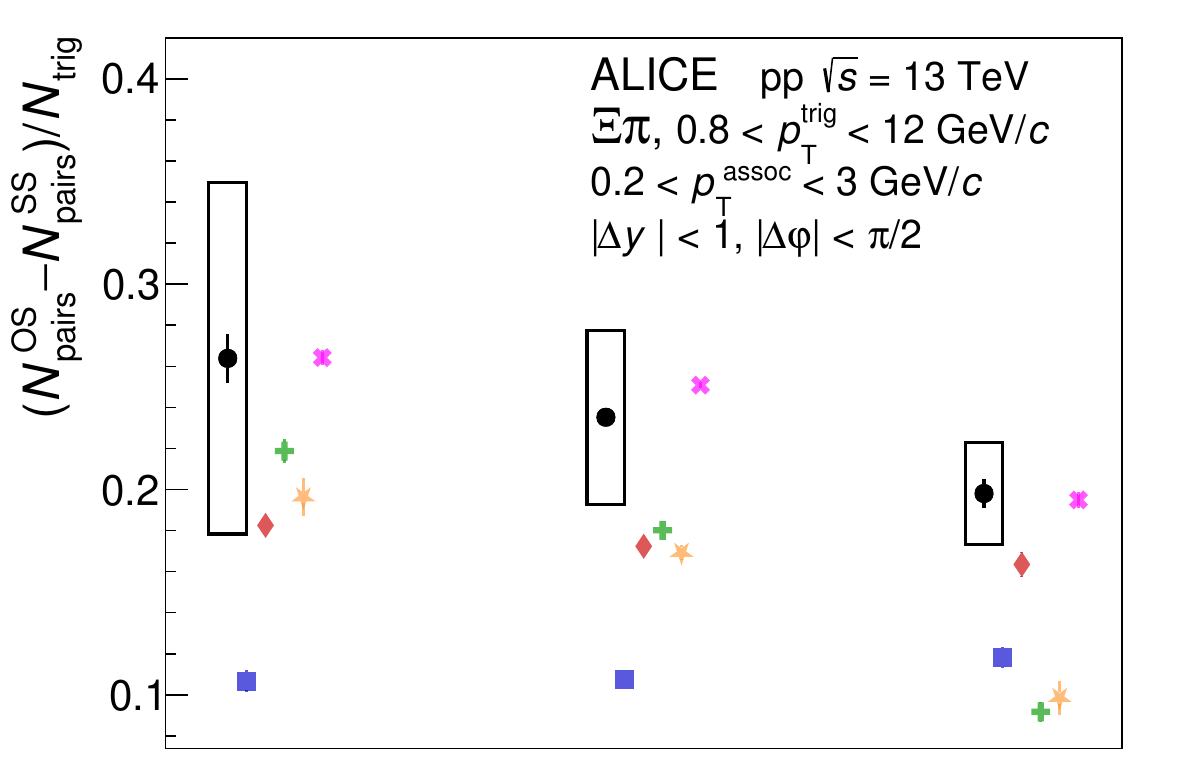}~\includegraphics[width = 0.49\textwidth]{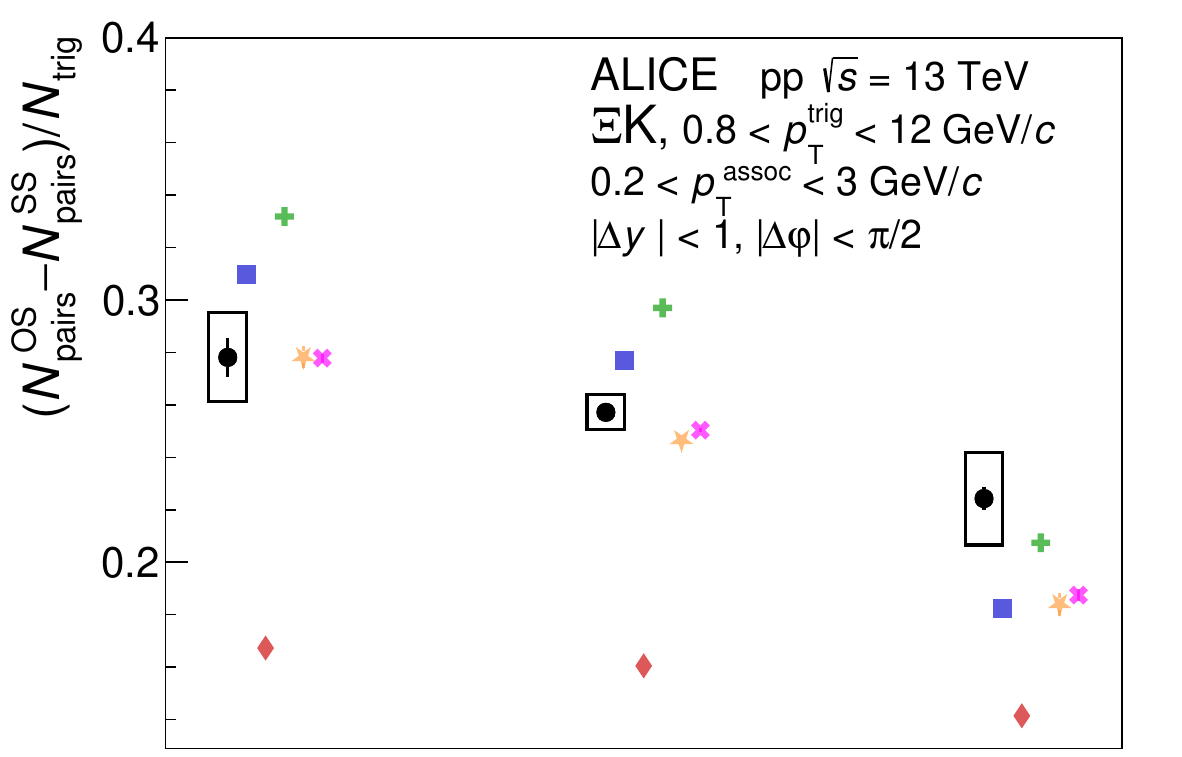}\\
    \includegraphics[width = 0.49\textwidth]{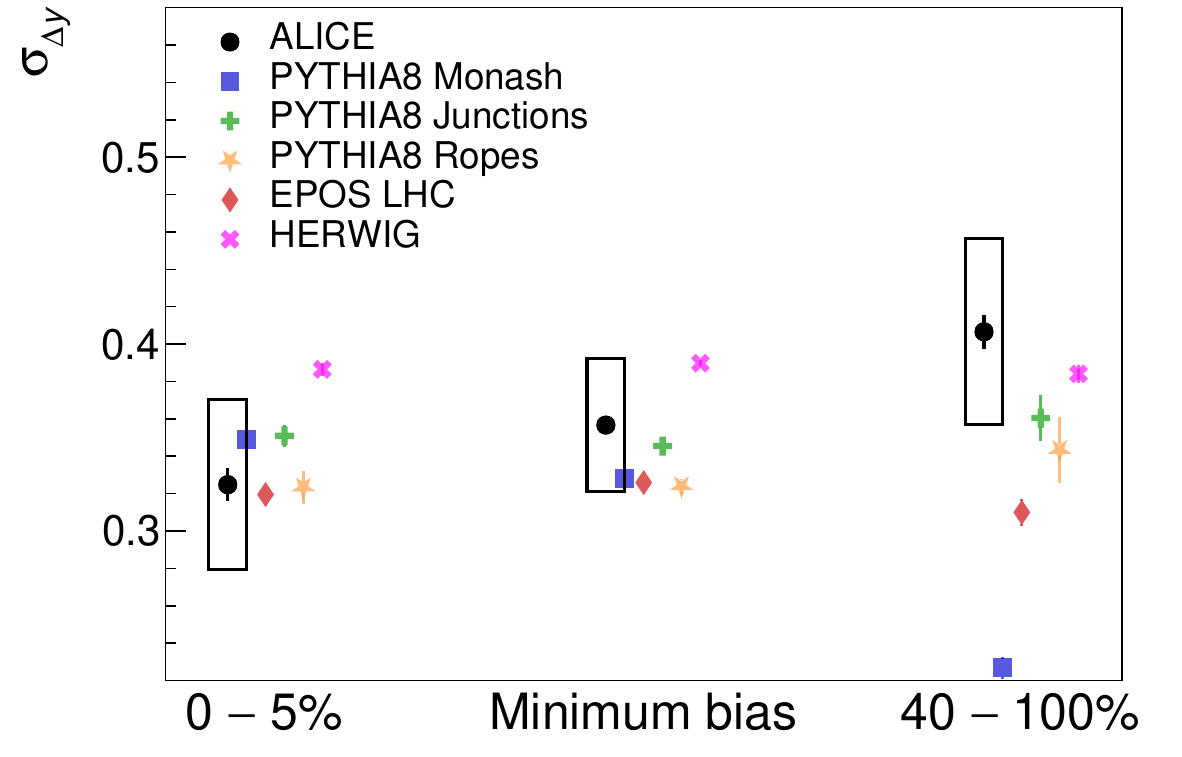}~\includegraphics[width = 0.49\textwidth]{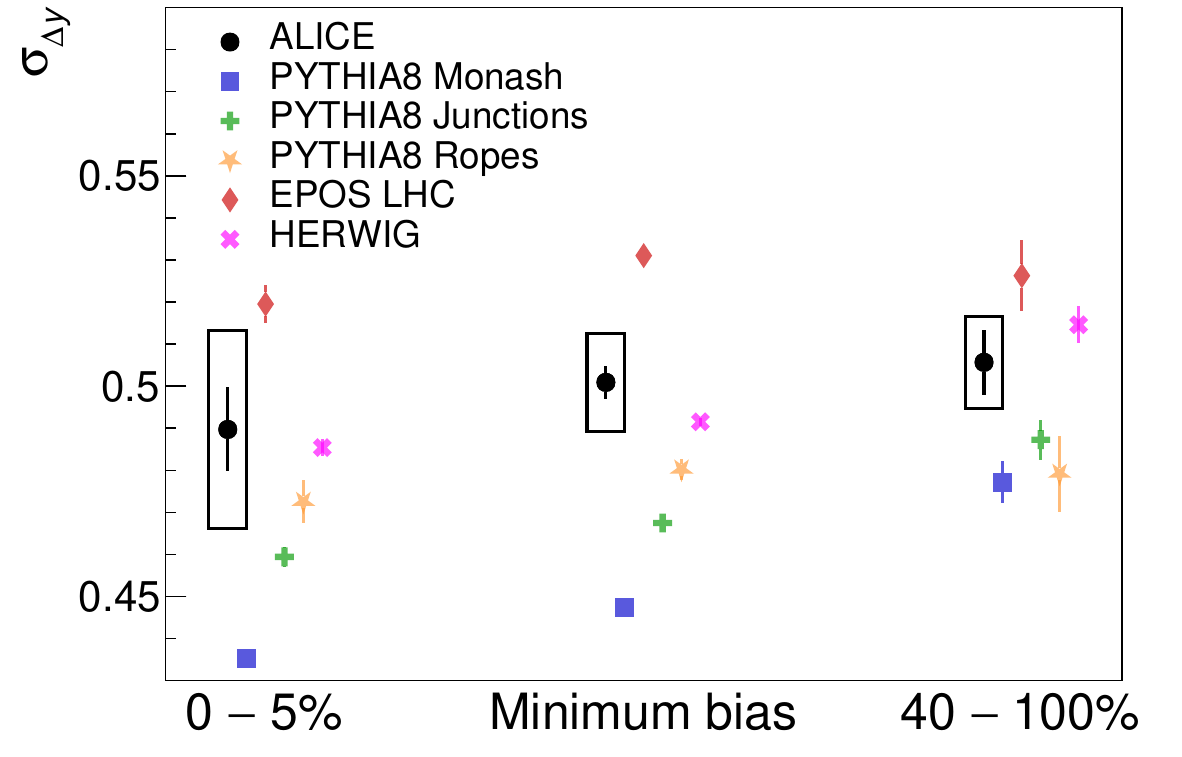}
    \end{center}
    \caption{The integrated OS--SS near-side yields (top) and near-side RMS widths in $\Delta y$ (bottom) are shown for $\Xi\pi$ (left) and $\Xi\mathrm{K}$ (right) correlations as a function of multiplicity.  Statistical and systematic uncertainties are represented by bars and boxes, respectively.  The ALICE data are compared with the following models: \pythia{} Monash tune (blue), \pythia{} with junctions enabled (green), \pythia{} with junctions and ropes (yellow), \eposlhc{} (red), and \herwig{} (pink). The statistical uncertainties on the model predictions, denoted by vertical bars, are smaller than the marker sizes in most cases.}
    \label{fig:xipi_xik_multdep}
\end{figure}

\begin{figure}[tbp]
    \begin{center}
    \includegraphics[width = 0.49\textwidth]{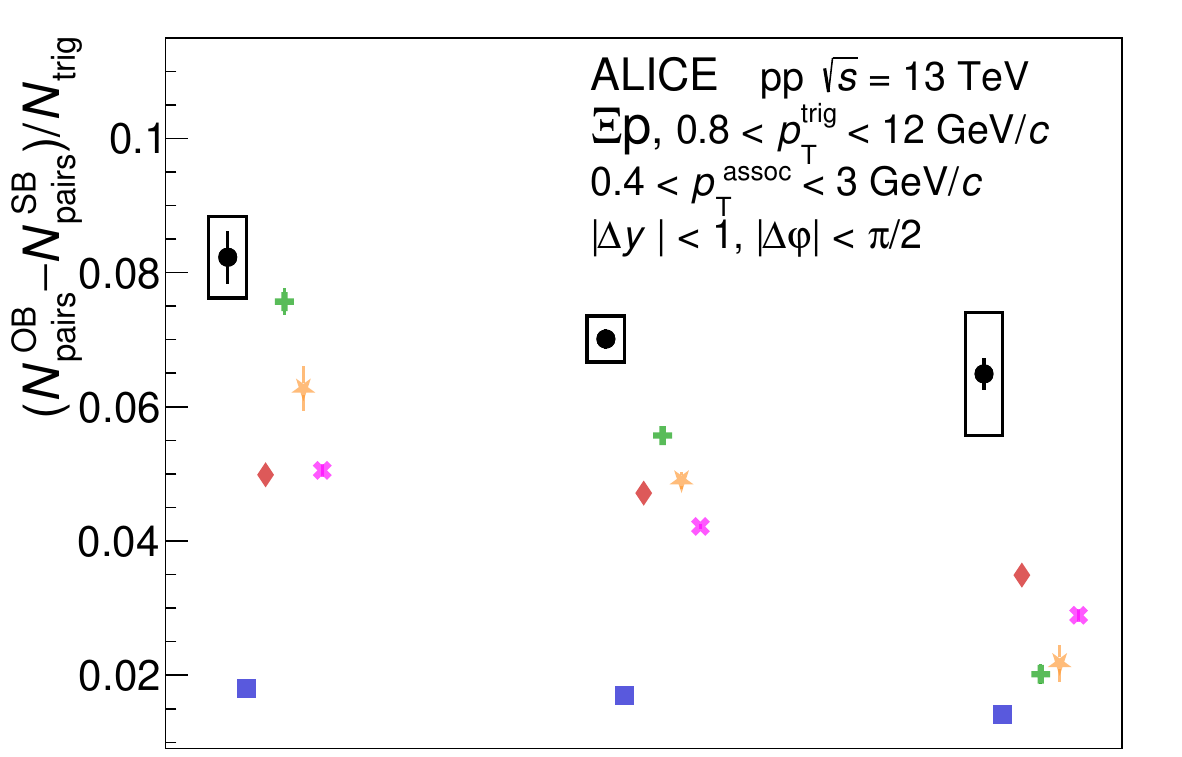}~\includegraphics[width = 0.49\textwidth]{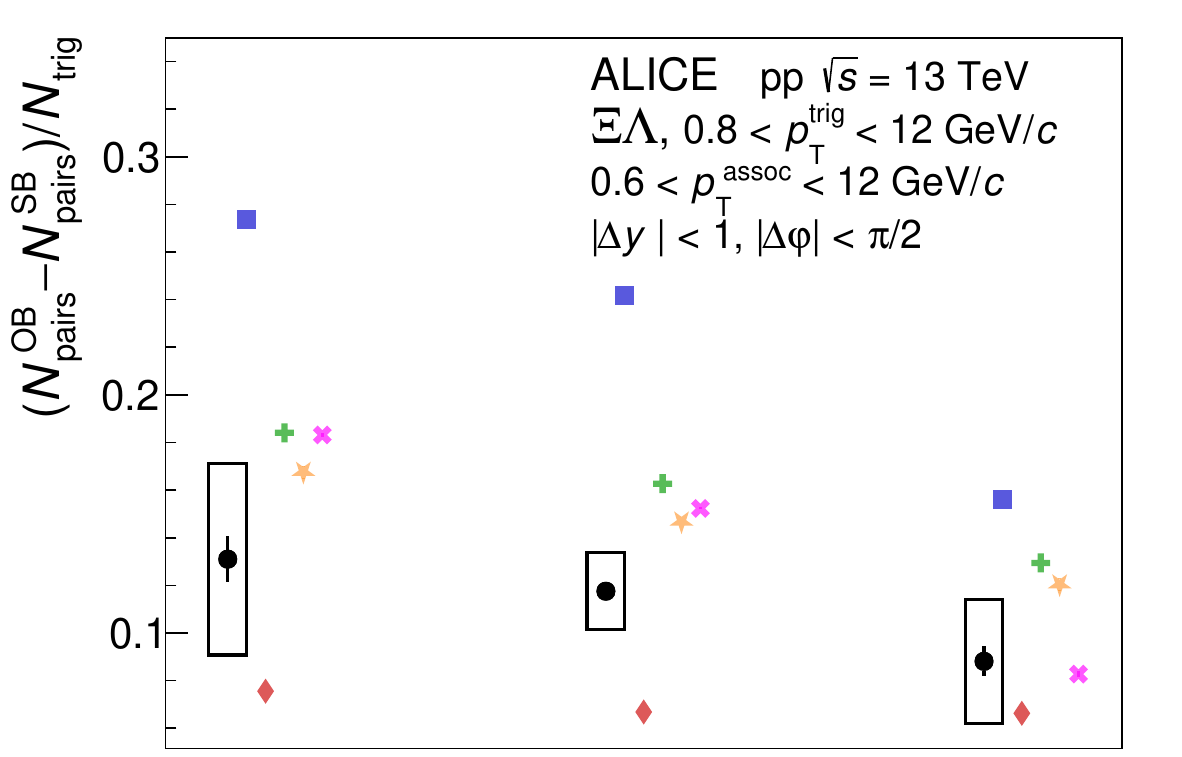}\\
    \includegraphics[width = 0.49\textwidth]{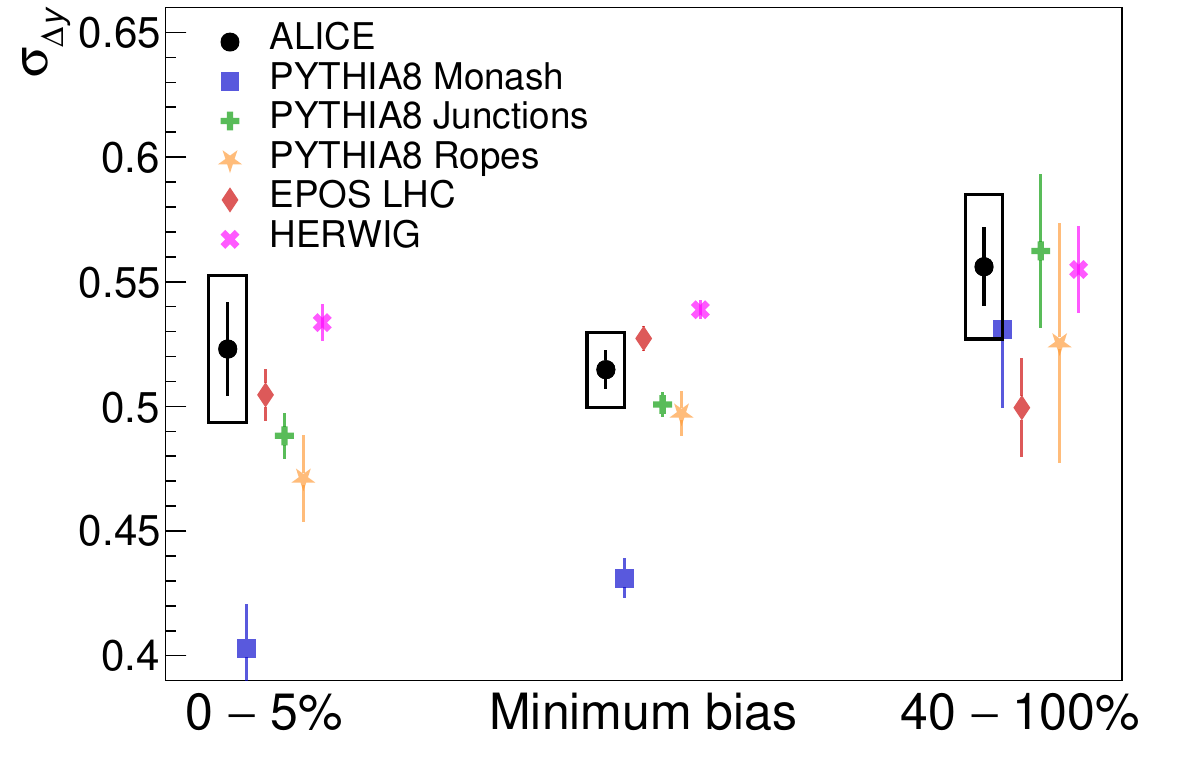}~\includegraphics[width = 0.49\textwidth]{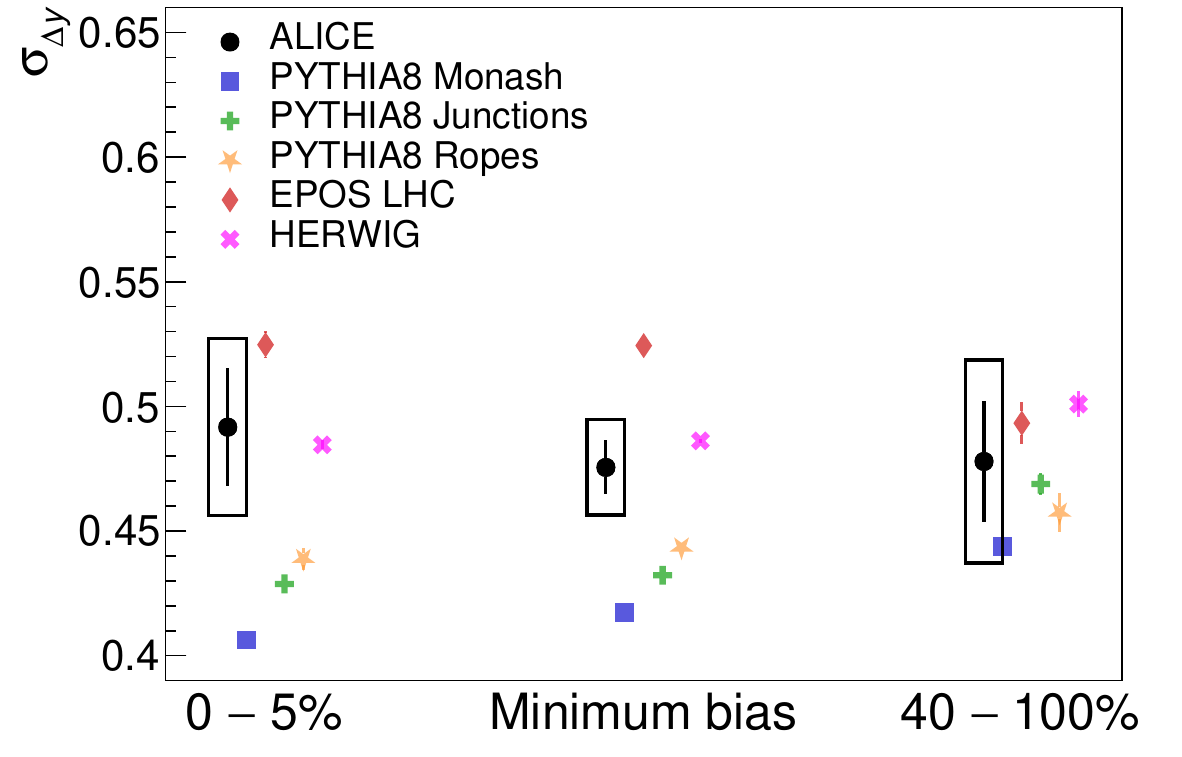}
    \end{center}
    \caption{The integrated OB--SB near-side yields (top) and near-side RMS widths in $\Delta y$ (bottom) are shown for $\Xi\mathrm{p}$ (left) and $\Xi\Lambda$ (right) correlations as a function of multiplicity.  Statistical and systematic uncertainties are represented by bars and boxes, respectively.  The ALICE data are compared with the following models: \pythia{} Monash tune (blue), \pythia{} with junctions enabled (green), \pythia{} with junctions and ropes (yellow), \eposlhc{} (red), and \herwig{} (pink). The statistical uncertainties on the model predictions, denoted by vertical bars, are smaller than the marker sizes in most cases.}
    \label{fig:xip_xilambda_multdep}
\end{figure}

\begin{figure}[tbp]
    \begin{center}
    \includegraphics[width = 0.49\textwidth]{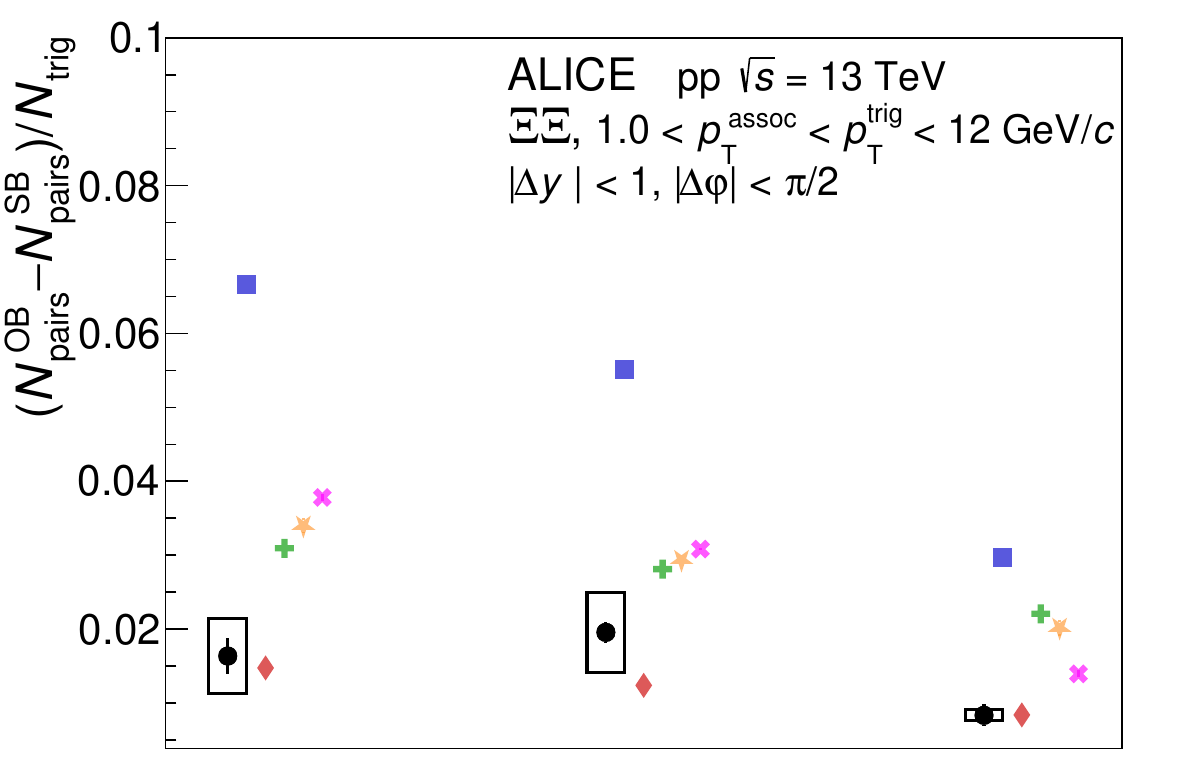}\\
    \includegraphics[width = 0.49\textwidth]{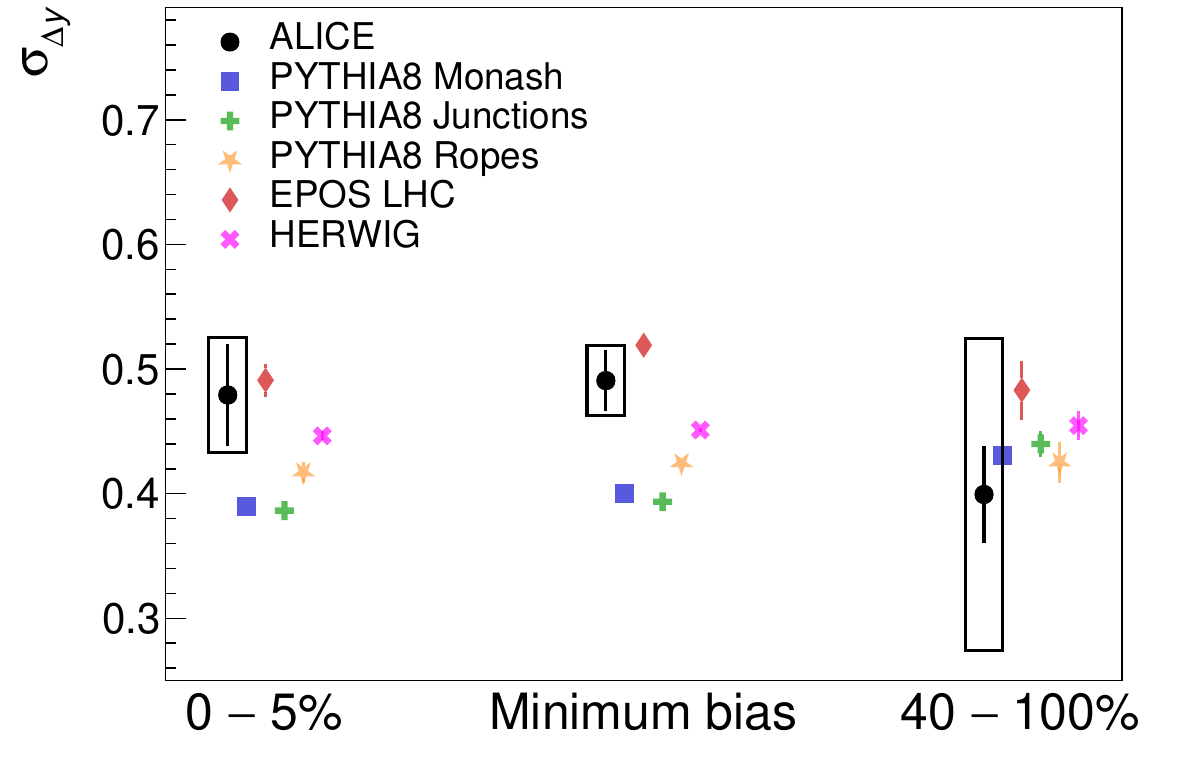}
    \end{center}
    \caption{The integrated OB--SB near-side yields (top) and near-side RMS widths in $\Delta y$ (bottom) are shown for $\Xi\Xi$ correlations as a function of multiplicity.  Statistical and systematic uncertainties are represented by bars and boxes, respectively.  The ALICE data are compared with the following models: \pythia{} Monash tune (blue), \pythia{} with junctions enabled (green), \pythia{} with junctions and ropes (yellow), \eposlhc{} (red), and \herwig{} (pink). The statistical uncertainties on the model predictions, denoted by vertical bars, are smaller than the marker sizes in most cases.}
    \label{fig:xixi_multdep}
\end{figure}

Figures~\ref{fig:multdepPhi} and~\ref{fig:multdepY} show the opposite-sign and same-sign $\Xi$--hadron correlation functions projected onto $\Delta\varphi$ and $\Delta y$, respectively, for low (40--100\%) and high (0--5\%) multiplicity events compared to the multiplicity-integrated (minimum bias) result.  The relevant systematic uncertainties described in Sec.~\ref{sec:systuncert} were re-evaluated in each multiplicity bin (note that the systematic uncertainties are highly correlated with multiplicity, for each particle species).  By definition, the level of the underlying event increases with multiplicity.   Furthermore, it appears that the near-side peak becomes slightly narrower for all the particle species with increasing event activity.  The near-side dip observed for baryon--baryon (and antibaryon--antibaryon) correlations does not demonstrate a noticeable dependence on multiplicity.  

More information can be gained by considering the charge-dependent (OS--SS, OB--SB) correlations, also shown in Figures~\ref{fig:multdepPhi} and~\ref{fig:multdepY}.  The integrated OS--SS and OB--SB near-side yields, as well as the near-side $\Delta y$ root-mean-square (RMS) widths, are compared to Monte Carlo models as a function of multiplicity for each particle species in Figures~\ref{fig:xipi_xik_multdep}--\ref{fig:xixi_multdep}.  A small multiplicity-dependence is observed for all particle species, with the charge-dependent near-side peak being taller and narrower in high-multiplicity events and shorter and broader in low-multiplicity events, compared to minimum bias.  This charge focusing with increasing multiplicity is consistent with the presence of radial flow~\cite{Voloshin:2003ud,ALICE:2018pal}.  

For the strange associated hadrons (K and $\Lambda$), the default \pythia{} Monash tune shows a sizeable increase of the charge-dependent near-side yield with higher multiplicity; this apparent violation of jet universality is likely due to the colour reconnection mechanism (which produces a radial flow-like effect).  Interestingly, when the rope or junction topologies are enabled, this significant multiplicity-dependence of the yields is instead observed for associated pions, kaons, and protons (and less pronounced for $\Lambda$).  This observation is consistent with expectations for junctions, which allows for the baryon number in the $\Xi$ to be more often balanced by antiprotons instead of anti-$\Lambda$ baryons.  The predictions for junctions and ropes (which include junctions) are not majorly qualitatively different, which indicates that these integrated yields may not be the ideal observable for distinguishing the effects of rope configurations.  \herwig{}, with its different string breaking model and cluster hadronisation mechanism, also shows an increasing OS--SS yield with multiplicity for all the species.  However, such multiplicity-dependent behaviour is less pronounced in the data, within the statistical and systematic uncertainties of the measurements, compared to both \pythia{} and \herwig{}.  On the other hand, \eposlhc{} shows very little multiplicity-dependence of the yields, which is also not consistent with experimental observations.  The multiplicity-dependence of the near-side widths is predicted to be minor by the Monte Carlo models, with a slight narrowing in high-multiplicity events for most particle species.  While the models do not quantitatively agree with the experimental data in all cases, in general the trends are similar.  These observations demonstrate that diquark breaking (the dominant baryon production mechanism within the standard Lund string model), and global strangeness conservation (as in \eposlhc{}) are not enough to describe all of the experimental behaviour, and additional mechanisms are needed.  However, the data are also not indicative of the turn-on of new dominant production mechanisms with multiplicity.  

\section{Conclusions}
\label{sec:conclusions}

The angular correlations between doubly-strange charged $\Xi$ baryons and identified hadrons ($\pi$, K, p, $\Lambda$, and $\Xi$) have been measured in \pp{} collisions by the ALICE Collaboration and compared to Monte Carlo models.  These correlations, and especially the difference between unlike-sign and like-sign correlations (for a general charge such as electric charge, strangeness, or baryon number), are particularly sensitive to the distribution of quantum numbers in the event, which gives information on the particle production mechanisms, subsequent diffusion, and hadronisation.  The correlation functions are also compared between high- and low-multiplicity \pp{} collisions, to search for potential indications of thermal strangeness production, deconfinement, or global instead of local strangeness conservation.  

The $\Xi\pi$ and $\Xi\mathrm{K}$ correlations show that the size of the underlying event, which gives rise to the pedestal below the near- and away-side jet peaks, is relatively well modelled by \pythia{}, likely because the Monte Carlo has been tuned to the experimentally-measured spectra.  However, the analysis of two-particle correlation functions yields more detailed and discriminatory power than the single-particle spectra alone.  The near-side correlation peak observed in the opposite-strangeness minus same-strangeness correlations is a signature of locally-correlated s$\overline{\rm s}$ production, which is observed in \pythia{} and \herwig{} but not in \eposlhc{}, a core--corona model where strangeness is only conserved globally.  However, \pythia{} and \herwig{} tend to predict stronger and narrower correlations than those measured in the experimental data, indicating that the effects of string-breaking are too large and that the diffusion of charges may be too small in these Monte Carlo models.  Finally, a near-side dip in same-baryon-number correlations demonstrates the difficulty of producing multiple baryons (or multiple antibaryons) close in phase space; while this feature was not captured by the default \pythia{} Monash tune, extended versions of \pythia{} which include colour ropes and baryon junctions are able to describe the data more accurately.  The multiplicity dependence of the correlation functions does not indicate any significant turn-on of new particle production mechanisms in high-multiplicity collisions, however, it is interesting to note that while \eposlhc{} does not describe local conservation of quantum numbers, it does quantitatively agree with experimental estimates of the charge balancing globally.  These novel results can be used to further refine and tune models of strangeness and baryon number production in hadronic collisions.


\newenvironment{acknowledgement}{\relax}{\relax}
\begin{acknowledgement}
\section*{Acknowledgements}

The ALICE Collaboration would like to thank all its engineers and technicians for their invaluable contributions to the construction of the experiment and the CERN accelerator teams for the outstanding performance of the LHC complex.
The ALICE Collaboration gratefully acknowledges the resources and support provided by all Grid centres and the Worldwide LHC Computing Grid (WLCG) collaboration.
The ALICE Collaboration acknowledges the following funding agencies for their support in building and running the ALICE detector:
A. I. Alikhanyan National Science Laboratory (Yerevan Physics Institute) Foundation (ANSL), State Committee of Science and World Federation of Scientists (WFS), Armenia;
Austrian Academy of Sciences, Austrian Science Fund (FWF): [M 2467-N36] and Nationalstiftung f\"{u}r Forschung, Technologie und Entwicklung, Austria;
Ministry of Communications and High Technologies, National Nuclear Research Center, Azerbaijan;
Conselho Nacional de Desenvolvimento Cient\'{\i}fico e Tecnol\'{o}gico (CNPq), Financiadora de Estudos e Projetos (Finep), Funda\c{c}\~{a}o de Amparo \`{a} Pesquisa do Estado de S\~{a}o Paulo (FAPESP) and Universidade Federal do Rio Grande do Sul (UFRGS), Brazil;
Bulgarian Ministry of Education and Science, within the National Roadmap for Research Infrastructures 2020-2027 (object CERN), Bulgaria;
Ministry of Education of China (MOEC) , Ministry of Science \& Technology of China (MSTC) and National Natural Science Foundation of China (NSFC), China;
Ministry of Science and Education and Croatian Science Foundation, Croatia;
Centro de Aplicaciones Tecnol\'{o}gicas y Desarrollo Nuclear (CEADEN), Cubaenerg\'{\i}a, Cuba;
Ministry of Education, Youth and Sports of the Czech Republic, Czech Republic;
The Danish Council for Independent Research | Natural Sciences, the VILLUM FONDEN and Danish National Research Foundation (DNRF), Denmark;
Helsinki Institute of Physics (HIP), Finland;
Commissariat \`{a} l'Energie Atomique (CEA) and Institut National de Physique Nucl\'{e}aire et de Physique des Particules (IN2P3) and Centre National de la Recherche Scientifique (CNRS), France;
Bundesministerium f\"{u}r Bildung und Forschung (BMBF) and GSI Helmholtzzentrum f\"{u}r Schwerionenforschung GmbH, Germany;
General Secretariat for Research and Technology, Ministry of Education, Research and Religions, Greece;
National Research, Development and Innovation Office, Hungary;
Department of Atomic Energy Government of India (DAE), Department of Science and Technology, Government of India (DST), University Grants Commission, Government of India (UGC) and Council of Scientific and Industrial Research (CSIR), India;
National Research and Innovation Agency - BRIN, Indonesia;
Istituto Nazionale di Fisica Nucleare (INFN), Italy;
Japanese Ministry of Education, Culture, Sports, Science and Technology (MEXT) and Japan Society for the Promotion of Science (JSPS) KAKENHI, Japan;
Consejo Nacional de Ciencia (CONACYT) y Tecnolog\'{i}a, through Fondo de Cooperaci\'{o}n Internacional en Ciencia y Tecnolog\'{i}a (FONCICYT) and Direcci\'{o}n General de Asuntos del Personal Academico (DGAPA), Mexico;
Nederlandse Organisatie voor Wetenschappelijk Onderzoek (NWO), Netherlands;
The Research Council of Norway, Norway;
Commission on Science and Technology for Sustainable Development in the South (COMSATS), Pakistan;
Pontificia Universidad Cat\'{o}lica del Per\'{u}, Peru;
Ministry of Science and Higher Education, National Science Centre and WUT ID-UB, Poland;
Korea Institute of Science and Technology Information and National Research Foundation of Korea (NRF), Republic of Korea;
Ministry of Education and Scientific Research, Institute of Atomic Physics, Ministry of Research and Innovation and Institute of Atomic Physics and Universitatea Nationala de Stiinta si Tehnologie Politehnica Bucuresti, Romania;
Ministry of Education, Science, Research and Sport of the Slovak Republic, Slovakia;
National Research Foundation of South Africa, South Africa;
Swedish Research Council (VR) and Knut \& Alice Wallenberg Foundation (KAW), Sweden;
European Organization for Nuclear Research, Switzerland;
Suranaree University of Technology (SUT), National Science and Technology Development Agency (NSTDA) and National Science, Research and Innovation Fund (NSRF via PMU-B B05F650021), Thailand;
Turkish Energy, Nuclear and Mineral Research Agency (TENMAK), Turkey;
National Academy of  Sciences of Ukraine, Ukraine;
Science and Technology Facilities Council (STFC), United Kingdom;
National Science Foundation of the United States of America (NSF) and United States Department of Energy, Office of Nuclear Physics (DOE NP), United States of America.
In addition, individual groups or members have received support from:
Czech Science Foundation (grant no. 23-07499S), Czech Republic;
European Research Council, Strong 2020 - Horizon 2020 (grant nos. 950692, 824093), European Union;
ICSC - Centro Nazionale di Ricerca in High Performance Computing, Big Data and Quantum Computing, European Union - NextGenerationEU;
Academy of Finland (Center of Excellence in Quark Matter) (grant nos. 346327, 346328), Finland.

\end{acknowledgement}

\bibliographystyle{utphys}   
\bibliography{bibliography}

\newpage
\appendix
\section{Selection criteria on $\Lambda$ and $\Xi$ candidates\label{sec:cuts}}
The criteria used to select $\Lambda(\overline{\Lambda})$ and $\Xi^-(\overline{\Xi}^+)$ candidates are shown in Tables~\ref{tab:Lambda_cuts} and~\ref{tab:Xi_cuts}, respectively.  
\begin{table}[h!]
\caption{Criteria used for the selection of $\Lambda$ candidates. For the \pt-dependent criteria, \pt{} is in GeV$/c$.}
\label{tab:Lambda_cuts}
\centering
\small
\begin{tabular}{lc}
\hline
$\mathbf{\textit{V}^0}$ \textbf{selection criteria}\\
\hline
Transverse momentum (GeV/$c$) & $0.6 < p_\mathrm{T} < 12$\\
Pseudorapidity & $|\eta| < 0.72$\\
$V^0$ radius (cm) & $\max(0.2,-1.1+1.2(\pt-0.35)^{0.56})<r_\mathrm{T}^\mathrm{V0}<83\pt-22$\\
DCA $V^0$ daughters (cm) & $\mathrm{DCA_{d-d}}<1.0$\\
Cosine of pointing angle & $\cos(\mathrm{PA})>0.995$\\
$\mathrm{K^0_s}$ rejection (\MeVmass)& $|m_{\pi\pi}-m_{\mathrm{K^0_S}}|>10$\\
\hline
\textbf{Daughter track selection criteria} &\\
\hline
General & either ITS refit or TOF hit for at least one daughter track,\\
& not included in $\Xi$ reconstruction\\
Transverse momentum (GeV/$c$) & $0.15 < p_\mathrm{T} < 20$\\
Pseudorapidity & $|\eta| < 0.8$\\
PID selection & $|n\sigma_{\rm TPC}| < 4$\\
Pion daughter DCA (cm) & $\mathrm{DCA_{d-PV}}>0.10$\\
Proton daughter DCA (cm) & $\mathrm{DCA_{d-PV}}>0.03$\\
\hline
\end{tabular}
\end{table}

\begin{table}[h!]
\caption{Criteria used for the selection of $\Xi$ candidates. Throughout this table, \pt\,is in GeV$/c$.}
\label{tab:Xi_cuts}
\centering
\small
\begin{tabular}{lc}
\hline
\textbf{Cascade selection criteria}\\
\hline
Transverse momentum (GeV/$c$) & $0.8 < p_\mathrm{T} < 12$\\
Pseudorapidity & $|\eta| < 0.72$\\
Cascade DCA (cm) & $\mathrm{DCA_{casc-PV}}<\min(2.0,0.007+1.34\,(p_\mathrm{T}-0.45)^{0.68})$\\
Cascade radius (cm) & $0.57+0.09(\,p_\mathrm{T}-0.45)^{0.81} < r_\mathrm{T}^\mathrm{casc} < 9+27\,(p_\mathrm{T}-0.45)^{1.7}$\\
Cosine of pointing angle & $\cos(\mathrm{PA}) > \max(0.993, 0.9983-3.2\times 10^{-3}(p_\mathrm{T}-0.45)^{-2.25})$\\
\hline
$\mathbf{\textit{V}^0}$ \textbf{selection criteria}\\
\hline
Invariant mass (\MeVmass) & $|m_{p\pi}-\mu_{\Lambda}|<2.6+2.5\,\pt$\\
DCA $V^0$ daughters (cm) & $\mathrm{DCA_{d-d}} < 1.5$\\
$V^0$ radius (cm) & $1.2 < r_\mathrm{T}^\mathrm{V0} < 16+57\,(p_\mathrm{T}-0.45)^{1.1}$\\
\hline
\textbf{Daughter track selection criteria} &\\
\hline
General & either ITS refit or TOF hit for at least one daughter track\\
Transverse momentum (GeV/$c$) & $0.15 < p_\mathrm{T} < 20$\\
Pseudorapidity & $|\eta| < 0.8$\\
PID selection & $|n\sigma_{\rm TPC}| < 4$\\
$V^0$ pion daughter DCA (cm) & $\mathrm{DCA_{d-PV}}> \max(0.03,-0.11+0.18\,(p_\mathrm{T}-0.45)^{-0.36})$\\
$V^0$ proton daughter DCA (cm) & $\mathrm{DCA_{d-PV}}> \max(0.03,-3.085+3.159\,(p_\mathrm{T}-0.45)^{-0.019})$\\
Bachelor DCA (cm) & $\mathrm{DCA_{bach-PV}}> 0.021+0.034\,(p_\mathrm{T}-0.45)^{-0.73}$\\
\hline
\end{tabular}
\end{table}

%
%

\newpage
\section{The ALICE Collaboration}
\label{app:collab}
\begin{flushleft} 
\small

S.~Acharya\,\orcidlink{0000-0002-9213-5329}\,$^{\rm 128}$, 
D.~Adamov\'{a}\,\orcidlink{0000-0002-0504-7428}\,$^{\rm 87}$, 
J.~Adolfsson\,\orcidlink{0000-0001-5651-4025}\,$^{\rm 76}$, 
G.~Aglieri Rinella\,\orcidlink{0000-0002-9611-3696}\,$^{\rm 33}$, 
M.~Agnello\,\orcidlink{0000-0002-0760-5075}\,$^{\rm 30}$, 
N.~Agrawal\,\orcidlink{0000-0003-0348-9836}\,$^{\rm 52}$, 
Z.~Ahammed\,\orcidlink{0000-0001-5241-7412}\,$^{\rm 136}$, 
S.~Ahmad\,\orcidlink{0000-0003-0497-5705}\,$^{\rm 16}$, 
S.U.~Ahn\,\orcidlink{0000-0001-8847-489X}\,$^{\rm 72}$, 
I.~Ahuja\,\orcidlink{0000-0002-4417-1392}\,$^{\rm 38}$
% ...

\section*{Affiliation Notes}

$^{\rm I}$ Deceased\\
$^{\rm II}$ Also at: Max-Planck-Institut fur Physik, Munich, Germany\\
$^{\rm III}$ Also at: Italian National Agency for New Technologies, Energy and Sustainable Economic Development (ENEA), Bologna, Italy\\
$^{\rm IV}$ Also at: Dipartimento DET del Politecnico di Torino, Turin, Italy\\
$^{\rm V}$ Also at: Yildiz Technical University, Istanbul, T\"{u}rkiye\\
$^{\rm VI}$ Also at: Department of Applied Physics, Aligarh Muslim University, Aligarh, India\\
$^{\rm VII}$ Also at: Institute of Theoretical Physics, University of Wroclaw, Poland\\
$^{\rm VIII}$ Also at: An institution covered by a cooperation agreement with CERN\\

\section*{Collaboration Institutes}

$^{1}$ A.I. Alikhanyan National Science Laboratory (Yerevan Physics Institute) Foundation, Yerevan, Armenia\\
$^{2}$ AGH University of Krakow, Cracow, Poland\\
$^{3}$ Bogolyubov Institute for Theoretical Physics, National Academy of Sciences of Ukraine, Kiev, Ukraine\\
$^{4}$ Bose Institute, Department of Physics  and Centre for Astroparticle Physics and Space Science (CAPSS), Kolkata, India\\
$^{5}$ California Polytechnic State University, San Luis Obispo, California, United States\\
$^{6}$ Central China Normal University, Wuhan, China\\
$^{7}$ Centro de Aplicaciones Tecnol\'{o}gicas y Desarrollo Nuclear (CEADEN), Havana, Cuba\\
$^{8}$ Centro de Investigaci\'{o}n y de Estudios Avanzados (CINVESTAV), Mexico City and M\'{e}rida, Mexico\\
$^{9}$ Chicago State University, Chicago, Illinois, United States\\
$^{10}$ China Institute of Atomic Energy, Beijing, China\\
$^{11}$ China University of Geosciences, Wuhan, China\\
$^{12}$ Chungbuk National University, Cheongju, Republic of Korea\\
$^{13}$ Comenius University Bratislava, Faculty of Mathematics, Physics and Informatics, Bratislava, Slovak Republic\\
$^{14}$ COMSATS University Islamabad, Islamabad, Pakistan\\
$^{15}$ Creighton University, Omaha, Nebraska, United States\\
$^{16}$ Department of Physics, Aligarh Muslim University, Aligarh, India\\
$^{17}$ Department of Physics, Pusan National University, Pusan, Republic of Korea\\
$^{18}$ Department of Physics, Sejong University, Seoul, Republic of Korea\\
$^{19}$ Department of Physics, University of California, Berkeley, California, United States\\
$^{20}$ Department of Physics, University of Oslo, Oslo, Norway\\
$^{21}$ Department of Physics and Technology, University of Bergen, Bergen, Norway\\
$^{22}$ Dipartimento di Fisica, Universit\`{a} di Pavia, Pavia, Italy\\
$^{23}$ Dipartimento di Fisica dell'Universit\`{a} and Sezione INFN, Cagliari, Italy\\
$^{24}$ Dipartimento di Fisica dell'Universit\`{a} and Sezione INFN, Trieste, Italy\\
$^{25}$ Dipartimento di Fisica dell'Universit\`{a} and Sezione INFN, Turin, Italy\\
$^{26}$ Dipartimento di Fisica e Astronomia dell'Universit\`{a} and Sezione INFN, Bologna, Italy\\
$^{27}$ Dipartimento di Fisica e Astronomia dell'Universit\`{a} and Sezione INFN, Catania, Italy\\
$^{28}$ Dipartimento di Fisica e Astronomia dell'Universit\`{a} and Sezione INFN, Padova, Italy\\
$^{29}$ Dipartimento di Fisica `E.R.~Caianiello' dell'Universit\`{a} and Gruppo Collegato INFN, Salerno, Italy\\
$^{30}$ Dipartimento DISAT del Politecnico and Sezione INFN, Turin, Italy\\
$^{31}$ Dipartimento di Scienze MIFT, Universit\`{a} di Messina, Messina, Italy\\
$^{32}$ Dipartimento Interateneo di Fisica `M.~Merlin' and Sezione INFN, Bari, Italy\\
$^{33}$ European Organization for Nuclear Research (CERN), Geneva, Switzerland\\
$^{34}$ Faculty of Electrical Engineering, Mechanical Engineering and Naval Architecture, University of Split, Split, Croatia\\
$^{35}$ Faculty of Engineering and Science, Western Norway University of Applied Sciences, Bergen, Norway\\
$^{36}$ Faculty of Nuclear Sciences and Physical Engineering, Czech Technical University in Prague, Prague, Czech Republic\\
$^{37}$ Faculty of Physics, Sofia University, Sofia, Bulgaria\\
$^{38}$ Faculty of Science, P.J.~\v{S}af\'{a}rik University, Ko\v{s}ice, Slovak Republic\\
$^{39}$ Frankfurt Institute for Advanced Studies, Johann Wolfgang Goethe-Universit\"{a}t Frankfurt, Frankfurt, Germany\\
$^{40}$ Fudan University, Shanghai, China\\
$^{41}$ Gangneung-Wonju National University, Gangneung, Republic of Korea\\
$^{42}$ Gauhati University, Department of Physics, Guwahati, India\\
$^{43}$ Helmholtz-Institut f\"{u}r Strahlen- und Kernphysik, Rheinische Friedrich-Wilhelms-Universit\"{a}t Bonn, Bonn, Germany\\
$^{44}$ Helsinki Institute of Physics (HIP), Helsinki, Finland\\
$^{45}$ High Energy Physics Group,  Universidad Aut\'{o}noma de Puebla, Puebla, Mexico\\
$^{46}$ Horia Hulubei National Institute of Physics and Nuclear Engineering, Bucharest, Romania\\
$^{47}$ HUN-REN Wigner Research Centre for Physics, Budapest, Hungary\\
$^{48}$ Indian Institute of Technology Bombay (IIT), Mumbai, India\\
$^{49}$ Indian Institute of Technology Indore, Indore, India\\
$^{50}$ INFN, Laboratori Nazionali di Frascati, Frascati, Italy\\
$^{51}$ INFN, Sezione di Bari, Bari, Italy\\
$^{52}$ INFN, Sezione di Bologna, Bologna, Italy\\
$^{53}$ INFN, Sezione di Cagliari, Cagliari, Italy\\
$^{54}$ INFN, Sezione di Catania, Catania, Italy\\
$^{55}$ INFN, Sezione di Padova, Padova, Italy\\
$^{56}$ INFN, Sezione di Pavia, Pavia, Italy\\
$^{57}$ INFN, Sezione di Torino, Turin, Italy\\
$^{58}$ INFN, Sezione di Trieste, Trieste, Italy\\
$^{59}$ Inha University, Incheon, Republic of Korea\\
$^{60}$ Institute for Gravitational and Subatomic Physics (GRASP), Utrecht University/Nikhef, Utrecht, Netherlands\\
$^{61}$ Institute of Experimental Physics, Slovak Academy of Sciences, Ko\v{s}ice, Slovak Republic\\
$^{62}$ Institute of Physics, Homi Bhabha National Institute, Bhubaneswar, India\\
$^{63}$ Institute of Physics of the Czech Academy of Sciences, Prague, Czech Republic\\
$^{64}$ Institute of Space Science (ISS), Bucharest, Romania\\
$^{65}$ Institut f\"{u}r Kernphysik, Johann Wolfgang Goethe-Universit\"{a}t Frankfurt, Frankfurt, Germany\\
$^{66}$ Instituto de Ciencias Nucleares, Universidad Nacional Aut\'{o}noma de M\'{e}xico, Mexico City, Mexico\\
$^{67}$ Instituto de F\'{i}sica, Universidade Federal do Rio Grande do Sul (UFRGS), Porto Alegre, Brazil\\
$^{68}$ Instituto de F\'{\i}sica, Universidad Nacional Aut\'{o}noma de M\'{e}xico, Mexico City, Mexico\\
$^{69}$ iThemba LABS, National Research Foundation, Somerset West, South Africa\\
$^{70}$ Jeonbuk National University, Jeonju, Republic of Korea\\
$^{71}$ Johann-Wolfgang-Goethe Universit\"{a}t Frankfurt Institut f\"{u}r Informatik, Fachbereich Informatik und Mathematik, Frankfurt, Germany\\
$^{72}$ Korea Institute of Science and Technology Information, Daejeon, Republic of Korea\\
$^{73}$ KTO Karatay University, Konya, Turkey\\
$^{74}$ Laboratoire de Physique Subatomique et de Cosmologie, Universit\'{e} Grenoble-Alpes, CNRS-IN2P3, Grenoble, France\\
$^{75}$ Lawrence Berkeley National Laboratory, Berkeley, California, United States\\
$^{76}$ Lund University Department of Physics, Division of Particle Physics, Lund, Sweden\\
$^{77}$ Nagasaki Institute of Applied Science, Nagasaki, Japan\\
$^{78}$ Nara Women{'}s University (NWU), Nara, Japan\\
$^{79}$ National and Kapodistrian University of Athens, School of Science, Department of Physics , Athens, Greece\\
$^{80}$ National Centre for Nuclear Research, Warsaw, Poland\\
$^{81}$ National Institute of Science Education and Research, Homi Bhabha National Institute, Jatni, India\\
$^{82}$ National Nuclear Research Center, Baku, Azerbaijan\\
$^{83}$ National Research and Innovation Agency - BRIN, Jakarta, Indonesia\\
$^{84}$ Niels Bohr Institute, University of Copenhagen, Copenhagen, Denmark\\
$^{85}$ Nikhef, National institute for subatomic physics, Amsterdam, Netherlands\\
$^{86}$ Nuclear Physics Group, STFC Daresbury Laboratory, Daresbury, United Kingdom\\
$^{87}$ Nuclear Physics Institute of the Czech Academy of Sciences, Husinec-\v{R}e\v{z}, Czech Republic\\
$^{88}$ Oak Ridge National Laboratory, Oak Ridge, Tennessee, United States\\
$^{89}$ Ohio State University, Columbus, Ohio, United States\\
$^{90}$ Physics department, Faculty of science, University of Zagreb, Zagreb, Croatia\\
$^{91}$ Physics Department, Panjab University, Chandigarh, India\\
$^{92}$ Physics Department, University of Jammu, Jammu, India\\
$^{93}$ Physics Program and International Institute for Sustainability with Knotted Chiral Meta Matter (SKCM2), Hiroshima University, Hiroshima, Japan\\
$^{94}$ Physikalisches Institut, Eberhard-Karls-Universit\"{a}t T\"{u}bingen, T\"{u}bingen, Germany\\
$^{95}$ Physikalisches Institut, Ruprecht-Karls-Universit\"{a}t Heidelberg, Heidelberg, Germany\\
$^{96}$ Physik Department, Technische Universit\"{a}t M\"{u}nchen, Munich, Germany\\
$^{97}$ Politecnico di Bari and Sezione INFN, Bari, Italy\\
$^{98}$ Research Division and ExtreMe Matter Institute EMMI, GSI Helmholtzzentrum f\"ur Schwerionenforschung GmbH, Darmstadt, Germany\\
$^{99}$ Saga University, Saga, Japan\\
$^{100}$ Saha Institute of Nuclear Physics, Homi Bhabha National Institute, Kolkata, India\\
$^{101}$ School of Physics and Astronomy, University of Birmingham, Birmingham, United Kingdom\\
$^{102}$ Secci\'{o}n F\'{\i}sica, Departamento de Ciencias, Pontificia Universidad Cat\'{o}lica del Per\'{u}, Lima, Peru\\
$^{103}$ Stefan Meyer Institut f\"{u}r Subatomare Physik (SMI), Vienna, Austria\\
$^{104}$ SUBATECH, IMT Atlantique, Nantes Universit\'{e}, CNRS-IN2P3, Nantes, France\\
$^{105}$ Sungkyunkwan University, Suwon City, Republic of Korea\\
$^{106}$ Suranaree University of Technology, Nakhon Ratchasima, Thailand\\
$^{107}$ Technical University of Ko\v{s}ice, Ko\v{s}ice, Slovak Republic\\
$^{108}$ The Henryk Niewodniczanski Institute of Nuclear Physics, Polish Academy of Sciences, Cracow, Poland\\
$^{109}$ The University of Texas at Austin, Austin, Texas, United States\\
$^{110}$ Universidad Aut\'{o}noma de Sinaloa, Culiac\'{a}n, Mexico\\
$^{111}$ Universidade de S\~{a}o Paulo (USP), S\~{a}o Paulo, Brazil\\
$^{112}$ Universidade Estadual de Campinas (UNICAMP), Campinas, Brazil\\
$^{113}$ Universidade Federal do ABC, Santo Andre, Brazil\\
$^{114}$ Universitatea Nationala de Stiinta si Tehnologie Politehnica Bucuresti, Bucharest, Romania\\
$^{115}$ University of Cape Town, Cape Town, South Africa\\
$^{116}$ University of Derby, Derby, United Kingdom\\
$^{117}$ University of Houston, Houston, Texas, United States\\
$^{118}$ University of Jyv\"{a}skyl\"{a}, Jyv\"{a}skyl\"{a}, Finland\\
$^{119}$ University of Kansas, Lawrence, Kansas, United States\\
$^{120}$ University of Liverpool, Liverpool, United Kingdom\\
$^{121}$ University of Science and Technology of China, Hefei, China\\
$^{122}$ University of South-Eastern Norway, Kongsberg, Norway\\
$^{123}$ University of Tennessee, Knoxville, Tennessee, United States\\
$^{124}$ University of the Witwatersrand, Johannesburg, South Africa\\
$^{125}$ University of Tokyo, Tokyo, Japan\\
$^{126}$ University of Tsukuba, Tsukuba, Japan\\
$^{127}$ Universit\"{a}t M\"{u}nster, Institut f\"{u}r Kernphysik, M\"{u}nster, Germany\\
$^{128}$ Universit\'{e} Clermont Auvergne, CNRS/IN2P3, LPC, Clermont-Ferrand, France\\
$^{129}$ Universit\'{e} de Lyon, CNRS/IN2P3, Institut de Physique des 2 Infinis de Lyon, Lyon, France\\
$^{130}$ Universit\'{e} de Strasbourg, CNRS, IPHC UMR 7178, F-67000 Strasbourg, France, Strasbourg, France\\
$^{131}$ Universit\'{e} Paris-Saclay, Centre d'Etudes de Saclay (CEA), IRFU, D\'{e}partment de Physique Nucl\'{e}aire (DPhN), Saclay, France\\
$^{132}$ Universit\'{e}  Paris-Saclay, CNRS/IN2P3, IJCLab, Orsay, France\\
$^{133}$ Universit\`{a} degli Studi di Foggia, Foggia, Italy\\
$^{134}$ Universit\`{a} del Piemonte Orientale, Vercelli, Italy\\
$^{135}$ Universit\`{a} di Brescia, Brescia, Italy\\
$^{136}$ Variable Energy Cyclotron Centre, Homi Bhabha National Institute, Kolkata, India\\
$^{137}$ Warsaw University of Technology, Warsaw, Poland\\
$^{138}$ Wayne State University, Detroit, Michigan, United States\\
$^{139}$ Yale University, New Haven, Connecticut, United States\\
$^{140}$ Yonsei University, Seoul, Republic of Korea\\
$^{141}$  Zentrum  f\"{u}r Technologie und Transfer (ZTT), Worms, Germany\\
$^{142}$ Affiliated with an institute covered by a cooperation agreement with CERN\\
$^{143}$ Affiliated with an international laboratory covered by a cooperation agreement with CERN.\\

\end{flushleft}

\end{document}